\documentclass[5p,number,sort&compress,times]{elsarticle}
\usepackage{amsmath,amsfonts}  
\usepackage{hyperref}
\usepackage{color}
\usepackage{wasysym}
\usepackage{array}
\usepackage{longtable}
\journal{Physica D}

\begin{document}

\begin{frontmatter}
\title{Moment-Closure Approximations for Discrete Adaptive Networks}

\author[mpi]{G.~Demirel\corref{cor1}}
\ead{guven@pks.mpg.de}
\author[mpi]{F.~Vazquez\fnref{fn1}}
\author[mpi]{G.A.~B\"{o}hme}
\author[bri]{T.~Gross}

\cortext[cor1]{Corresponding author}
\fntext[fn1]{Current address: Instituto de Fisica de Liquidos y Sistemas Biologicos (CCT-CONICET-La Plata, UNLP), Calle 59 Nro 789, 1900 La Plata, Argentina}
\address[mpi]{Max-Planck-Institute for the Physics of Complex Systems, N\"{o}thnitzer Stra\ss e 38, 01187 Dresden, Germany}
\address[bri]{Merchant Venturers School of Engineering, University of Bristol -- Woodland Road, Clifton BS8 1UB Bristol, United Kingdom}

\begin{abstract}
Moment-closure approximations are an important tool in the analysis of the dynamics on both static and adaptive networks. Here, we provide a broad survey over different 
approximation schemes by applying each of them to the adaptive voter model. While already the simplest schemes provide reasonable qualitative results, even very complex 
and sophisticated approximations fail to capture the dynamics quantitatively. We then perform a detailed analysis that identifies the emergence of specific correlations 
as the reason for the failure of established approaches, before presenting a simple approximation scheme that works best in the parameter range where all other approaches 
fail. By combining a focused review of published results with new analysis and illustrations, we seek to build up an intuition regarding the situations when existing 
approaches work, when they fail, and how new approaches can be tailored to specific problems.   
\end{abstract}


\end{frontmatter}

\section{INTRODUCTION}
Complex networks have been ubiquitously used to model problems from various disciplines \citep{Albert2002,Newman2003,Newman2006,Boccaletti2006,Dorogovtsev2008,Castellano2009}. 
Treating a complex system as a network, a set of discrete nodes and links, leads to a conceptual simplification that often allows subsequent analytical insight that provides 
a deep understanding. 

For many questions the networks of interest are not static entities but change in time due to the dynamics \emph{of} and \emph{on} the network. In the dynamics \emph{of} networks, 
the network itself is regarded as a dynamical system. Prominent examples are network growth models leading to specific topologies such as scale-free \citep{Barabasi1999} 
and small-world networks \citep{Watts1998}. The dynamics \emph{on} networks concerns dynamical processes such as epidemic spreading \citep{Pastor-Satorras2001} that occur 
on a given fixed network, where each node carries a state which evolves through interactions with its neighbors. 

If the dynamics on and of networks occur simultaneously and interdependently then the network topology coevolves with the states of the nodes and an adaptive network 
is formed \citep{Gross2008,Gross2009}. Adaptive networks have been used to model problems of opinion formation \citep{Holme2006,Vazquez2008,Nardini2008,Kimura2008,Zschaler2012,Durrett2012,Boehme2012}, 
epidemic spreading \citep{Gross2006,Shaw2008,Risau-Gusman2009,Shaw2010,Marceau2010,Graeser2011,Lagorio2011,Wang2011,Juher2012}, 
evolution of cooperation \citep{Zimmermann2000,Skyrms2000,Zimmermann2004,Pacheco2006,Segbroeck2009,Poncela2009,Szolnoki2009,Do2010,Zschaler2010,Segbroeck2011,Lee2011,Fehl2011,Santos2012}, 
synchronization \citep{Zhou2006,Sorrentino2008,Aoki2009,Assenza2011,Botella2012}, neuronal activity \citep{Bornholdt2000,Bornholdt2003,Levina2007,Levina2009,Jost2009,Meisel2009,Ren2010,Meisel2012,Droste2012}, 
collective motion \citep{Huepe2011,Couzin2011}, cartelisation of markets \citep{Peixoto2012}, and particle diffusion \citep{Kim2008} among others.

Network models in general and adaptive networks in particular provide a powerful framework to model, analyze, and eventually understand a wide range of self-organization phenomena. 
For instance \citet{Tomita2009} showed that a very simple adaptive network model can be used to produce a huge variety of different self-organizing structures including a self-replicating 
Turing machine.

While specific models can be studied by agent-based simulation, the numerical performance scales badly with the complexity of update rules in the model, which makes exploration 
of a wider range of models difficult. In particular those models where the update of the state or neighborhood of nodes depends on the states of multiple other neighboring nodes pose strong numerical demands. Additionally 
the general bad data locality of network simulations precludes efficient parallelization. This defines a strong need for analytical approaches, and, based on recent successes, 
highlights the exploration of dynamic networks with complex update rules as an area where analytical work could outpace and guide numerical exploration.

A direct microscopic description of dynamical networks generally constitutes a very high-dimensional dynamical system. While in some cases exact analytical results were nevertheless 
obtained (e.g.\ \citep{Atay2004,Do2010,Do2012}), there are presently no approaches that are generally applicable. Much of the theoretical progress therefore relies on the 
derivation of reasonably low-dimensional coarse-grained approximations to the full microscopic model.   

For networks in which the node can only assume states from a (small) discrete set of possibilities, approximation schemes are well established. These schemes are deeply rooted 
in physics and can be traced back to early work on the Ising model \citep{Bethe1935,Peierls1936}. In the networks literature there is presently a veritable zoo of different approximation 
schemes that build on similar principles but take different information into account. In the following we refer to these approaches as \emph{moment-closure approximations}. 
The common idea in all of these approaches is to derive evolution equations for the abundance of certain subgraphs in the network. One starts with writing an evolution equation 
for small subgraphs, such as single nodes, before writing equations for larger motifs -- a process that is reminiscent of classical moment expansions. The system of equations 
that is thus obtained generally depends on the abundance of other, typically larger, subgraphs that are not captured, and thus needs to be closed by estimating the abundance 
of these subgraphs -- the actual moment-closure approximation.       

Despite their underlying similarity moment-closure approximations proposed in the recent literature differ widely by the type and number of the subgraphs they capture. 
Generally speaking, capturing the dynamics of more subgraphs, leads to better approximations at the cost of having to deal with a larger system of equations 
(see \citep{Kimura2008,Marceau2010,Matsuda1992,Bauch2005,Szabo2004,Peyrard2008}). In practice some recently proposed schemes are successfully applied which only capture 
one or two subgraphs, while others capture thousands or millions of subgraphs. 

For the analysis of adaptive networks, but also certain types of dynamics on static networks, moment-closure approximations are presently the most commonly applied theoretical 
tool. In adaptive networks they were used for instance to study epidemics \citep{Gross2006,Shaw2008,Gross2008a,Risau-Gusman2009,Shaw2010,Marceau2010,Graeser2011,Jolad2011,Wang2011,Juher2012,Wieland2012,Wieland2012b,Demirel2012b}, 
collective motion \citep{Huepe2011,Couzin2011}, evolution of cooperation \citep{Zschaler2010,Graeser2009,Fu2009}, and social opinion formation \citep{Vazquez2008,Kimura2008,Zanette2006,Demirel2011}.

Despite the abundance of examples there is so far little intuition on when particular approximation schemes work and when they fail. This is most notable when considering the 
adaptive SIS model \citep{Gross2006} and the adaptive voter model \citep{Nardini2008,Vazquez2008,Kimura2008}. Both of these models are adaptive network models of similar 
complexity, and, depending on personal taste, either can be considered as the most simple non-trivial adaptive network. However, for the adaptive SIS model, the dynamics can be faithfully 
captured already by simple approximation schemes \citep{Gross2006}, with more sophisticated approaches leading expectedly to a further improvement\citep{Marceau2010,Wieland2012,Wieland2012b}. 
By contrast, for the adaptive voter model, simple approximation schemes only provide unsatisfactory results \citep{Vazquez2008,Kimura2008} and, as we show here, more sophisticated approaches can actually perform worse.  

In the present paper, we aim to offer an in-depth analysis of the performance and the failure of different approximation schemes. For the purpose of illustration we focus on the 
adaptive voter model as it provides a mathematically simple, yet challenging example system. To this system we apply the major approximation schemes proposed in the recent literature. 
Thereby, we build up an intuition of the advantages and disadvantages of the respective schemes.

A second goal of the present paper is to provide researchers entering the field with ``worked examples" for the major approximation schemes. We present these examples in strongly abbreviated form
in the main text, while providing all calculations in full detail in the appendices.

The paper is structured as follows: We start in Sec.~\ref{sec:models} by introducing the adaptive voter model. In Sec.~\ref{sec:homogeneous} we study the so-called homogeneous 
approximations, which result in relatively low dimensional equation systems. For these we explore in particular the effect of the order of approximation. Then, in Sec.~\ref{sec:heterogeneous} 
we discuss different heterogeneous approximations, which yield high-dimensional equation systems, but surprisingly do not significantly improve the performance of the approximation. 
Finally, in Sec.~\ref{sec:percolation} we introduce a slightly different expansion that captures very similar information but works exactly in those parameter ranges where 
other approximations fail. A summary and discussion in Sec.~\ref{sec:discussion} concludes the paper.


\section{ADAPTIVE VOTER MODEL}\label{sec:models} 
The voter model considers the competition of equally attractive and mutually exclusive opinions (say A and B) in a population of interacting agents. The agents are represented 
by nodes that have an internal binary state variable, indicating the opinion held by the corresponding agents. The state is updated dynamically in time due to social interactions, 
occurring between linked agents. 

In the original non-adaptive voter model \citep{Holley1975} the underlying interaction topology is static. At each time step, a pair of nodes connected by a link is selected. 
If they share the same state, nothing happens. Otherwise, one of the two adopts the other's state. This model has been explored in many subsequent works and in particular 
the dependence of the time needed to reach consensus on the underlying topology and details of the update rule is well understood 
\citep{Sood2005,Sood2008,Vazquez2008a,Castellano2003,Vilone2004,Suchecki2005a,Castellano2005,Suchecki2005b,Castello2007a}.

\begin{figure}[ht!]
  \centering \includegraphics[width=3.5in,keepaspectratio]{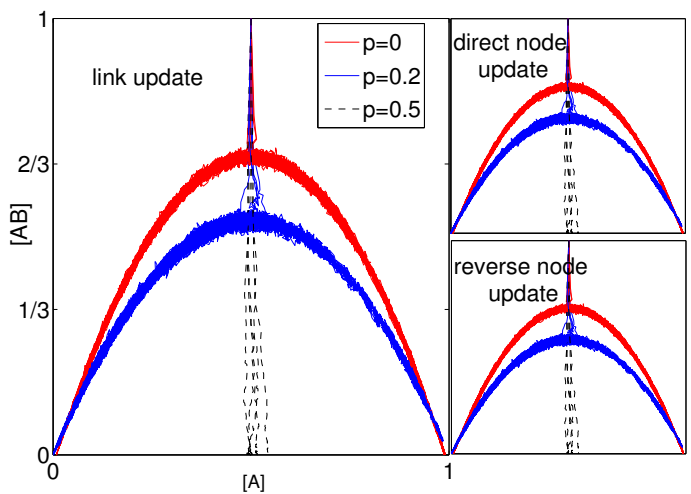}
  \caption{(Color online) Parabola of active states. Shown are five representative trajectories for each of three parameter values and three different update rules. The long-term behavior depends on the value of the rewiring rate $p$. If this rate exceeds a threshold $p^{*}$ then 
  the system quickly approaches a fragmented state in which the number of active links $[AB]$ vanishes, while the density of a given opinion (say, $[A]$) remains almost constant. If $p$ is below the threshold rewiring rate then the model approaches 
  a parabola of meta-stable active states on which it remains for a long time while undergoing a random walk in the opinion density that eventually leads to consensus. Parameters: $N=10^5$ and $\langle k \rangle=4$.}
  \label{fig:parabola}
\end{figure}

The adaptive variants of the model \citep{Holme2006,Vazquez2008,Nardini2008,Kimura2008} additionally incorporates a phenomenon known as social segregation, the tendency of humans 
to link preferentially to others holding similar opinions and distance themselves from those holding opposite opinions. Starting from a suitable initial condition, a pair of nodes 
connected by a link is selected at each time step. If the link is \emph{inert}, connecting nodes in the same state, then nothing happens. Otherwise, the link is \emph{active} and 
either (with probability $p$) one of the nodes rewires the link by detaching it from the other node and attaching to a random node that shares its opinion, or one of the nodes copies 
the other node's state.  

In the spirit of discrete event simulations, using the Gillespie algorithm, the probability $p$ can be thought of as an effective parameter capturing the rate of rewiring events, 
normalized by the total rate of update events happening in the network. We therefore call $p$ the rewiring rate. 

Previous investigations \citep{Holme2006,Nardini2008,Vazquez2008,Kimura2008,Boehme2011} 
have shown that there is a critical rewiring rate, $p^*$, at which the behavior of the model changes qualitatively. If $p>p^*$ then the system quickly approaches a fragmented state in 
which the network splits into two disconnected components that hold opinion A and B, respectively. Since no active links survive the dynamics freezes in this fragmented state.   
By contrast, if $p<p^*$ then the system remains active for a long time before eventually reaching a complete consensus on one of the opinions, which is likewise an absorbing state 
(see Fig.~\ref{fig:parabola}). 
 
We note many variants of the adaptive voter model using different update rules have been investigated in the literature \citep{Holme2006,Nardini2008,Vazquez2008,Kimura2008,Zanette2006,Gil2006,Benczik2008,Fu2008,Benczik2009,Sobkowicz2009,Zhong2010,Zschaler2012}.
Apart from models with more than two states\citep{Holme2006,Herrera2011,Boehme2012} and models with directed links\citep{Zschaler2012}, the most prominent difference is 
the precise procedure for selecting the link that is updated and the direction of the update (e.g.~which node adopts the others opinion) on this link. One distinguishes 
between \emph{direct node update}, \emph{reverse node update}, and \emph{link update} rules. In the two node update rules, one selects a random node (node X) then a random 
neighbor (node Y). Selecting the nodes in this way creates a slight bias in the selected nodes regarding the \emph{degree}, i.e. the number of neighbors of the respective nodes. 
Essentially, a node of higher degree has a proportionally higher probability of being found by following a link, and conversely the nodes found in this way have a higher degree.
Therefore, the average degree of node Y is higher than the average degree of node X \citep{Newman2003}. In models using direct node update, X retains the link in rewiring events 
and adopts Y's state in update events, whereas in models using reverse node update Y retains the link in rewiring events and copies X's opinion in update events. By contrast, 
in the link update rule one randomly selects a link, such that both selected nodes have identical statistical properties. While the precise update rule can have a significant impact 
on convergence times \citep{Nardini2008}, the qualitative features of the models, described above remain unchanged (see Fig.~\ref{fig:parabola}).

In the remainder of this paper we apply different moment-closure approximations for capturing the dynamics of the voter model. Specifically, we compare the ability of the 
different approaches to predict the density of active links in the state when both opinions are equally abundant.  

The specific model that we consider throughout most of the paper is as follows: We start with an Erd\H{o}s-Renyi random graph with $N$ nodes and $L$ links, such that the mean 
degree is $\langle k \rangle = 2 L / N$. Initial opinions are assigned randomly with equal probability. Therefore, the initial abundances obey $N_{A}=N_{B}=N/2$. The network is then 
evolved by a \emph{link} update rule: At each time step an active link is selected at random. With probability $p$ one of the nodes rewires the link and connects to a random node of same 
state. Otherwise, that is with the complementary probability $1-p$, one of the nodes adopts the other's state. The respective node that retains the link or adopts the other's opinion is selected randomly with equal probability. 
The model is simulated according to these rules until either fragmentation occurs or an active state is reached where the density of active links remains approximately constant 
over an intermediate time scale.  


\section{HOMOGENEOUS APPROXIMATIONS} \label{sec:homogeneous}
For classifying the different approximation schemes that have been proposed, it is useful to distinguish between homogeneous and heterogeneous approximations. While all 
approximations attempt to capture the dynamics of certain subgraphs, they differ in the way in which subgraphs are identified: Homogeneous approximations 
\citep{Vazquez2008,Nardini2008,Kimura2008,Gross2006,Shaw2008,Gross2008a,Vazquez2008a,Zschaler2010,Demirel2011,Gleeson2012} classify subgraphs according to states of the nodes and 
the internal topology in the subgraph, whereas heterogeneous approximations additionally take the degree of the nodes in the subgraph into account
\citep{Pastor-Satorras2001,Marceau2010,Pastor-Satorras2001a,Moreno2002,Sood2005,Sood2008,Pugliese2009,Noel2009,Gleeson2011,Durrett2012,Demirel2012b}.  

Homogeneous moment-closure approximations have been used in the past two decades to explain dynamics on static networks \citep{Matsuda1992,Bauch2005,Keeling1999,House2011,Gleeson2012} 
and have more recently been applied to adaptive models of epidemic spreading \citep{Gross2006,Shaw2008,Risau-Gusman2009,Fu2009,Shaw2010,Juher2012,Zanette2008}, opinion formation 
\citep{Vazquez2008,Kimura2008,Nardini2008,Zanette2006,Demirel2011}, cooperation games \citep{Zschaler2010,Graeser2009,Fu2009}, and collective motion \citep{Huepe2011,Couzin2011}.

The central idea of homogeneous approximations is to capture the dynamics of the network by writing balance equations for the density or abundance of a certain set of labeled subgraphs 
(motifs),  which are called \emph{network moments}. The system of differential equations of network moments constitutes the moment expansion which is then closed by the so-called moment-closure 
approximation.

\subsection{Moment Expansion}
Writing the rate equation for a network moment necessitates calculating the rates of all possible processes that result in either the formation or destruction of the 
respective subgraph. For the adaptive voter model, the moment expansion for small subgraphs (nodes and links) has already been developed in \citet{Vazquez2008} and 
independently for an identical model by \citet{Kimura2008}:
\begin{eqnarray} \label{eq:ode_link_triplet}\renewcommand{\arraystretch}{2}
\displaystyle \frac{\rm d}{\rm dt} [A] & = & 0, \nonumber \\
\displaystyle \frac{\rm d}{\rm dt} [AA]  & = & \displaystyle \frac{1}{2}[AB] + \frac{(1-p)}{2} \big(2 [ABA] - [AAB] \big), \nonumber \\
\displaystyle \frac{\rm d}{\rm dt} [BB] &=& \displaystyle \frac{1}{2}[AB] + \frac{(1-p)}{2}\big(2 [BAB] - [ABB]\big),
\end{eqnarray}
where $[X]$ denotes the density of nodes with state X, $[XY]$ denotes the density of links between nodes of state X and Y, and $[XYZ]$ denotes the density of triplets 
constituted by a node of state Y in the center and its two neighbors of state X and Z, with $\text{X, Y, Z}\in\{\text{A,B}\}$. The densities are normalized with respect 
to the number of nodes $N$, i.e. $[\Omega]=N_{\Omega}/N$ where we used $\Omega$ as a placeholder for an arbitrary subgraph (e.g. A-node, AB-link, AAB-triplet), such that 
$N_{\Omega}$ is the total abundance of that subgraph in the network.

The first equation in Eq.~(\ref{eq:ode_link_triplet}) states that the density of nodes of state A (and equivalently B) is conserved in the thermodynamic limit. This is peculiar to the voter model and is a direct
consequence of symmetry in state adoption. In an update step, the A-node adopts state B with probability $(1-p)/2$ and equivalently the B-node adopts state A with the same probability 
$(1-p)/2$ leading to a vanishing net drift in the deterministic limit.

The second equation captures the change in the density of AA-links $[AA]$. If a B-node adopts state A on an AB-link (at rate $(1-p)[AB]/2$) or an A-node rewires an AB-link 
and forms a link to another A-node (at rate $p[AB]/2$), an AA-link is directly created in the update. The total rate of this direct creation of AA-Links is thus $[AB]/2$. 
Additionally, AA-links can be created indirectly. Consider a B-node that is connected to two A-neighbors such that the three nodes form an ABA-triplet. When this B-node 
adopts state A, instead of one, two AA-links are formed. While the rate of adoption events creating A-nodes is $(1-p)[AB]/2$ the expected additional links created indirectly 
in such adoption events are $2[ABA]/[AB]$. Thus the total rate for the indirect creation of AA-links is $(1-p)[ABA]$. Finally, AA-links can be destroyed indirectly if one 
of the A-nodes forming an AA-link adopts state B due to an interaction with another B-node, which happens at rate $(1-p)[AAB]/2$. The rate equation for $[BB]$ is obtained by 
interchanging $A$ and $B$ in the second equation. 

The system of equations (\ref{eq:ode_link_triplet}) is not closed as it contains the (unknown) densities of larger subgraphs. This is a general property of moment expansions. 
For a system where the update rules directly affect subgraphs of diameter $d$ (here, 1) the evolution equation for a subgraph of diameter $l$ generally contains subgraphs of 
diameter up to $d+l$ due to indirect effects.   

For a more precise discussion it is useful to define the \emph{order} of a network moment as the number of links contained in the corresponding subgraph. For instance, 
the moments $[X]$, $[XY]$ and $[XYZ]$ have order zero, one and two respectively. Accordingly, one can define the order of a network model as the largest order of subgraphs 
used in definition of the network update process, e.g. the voter model has order one since it involves a link at most in the definition.

Let us denote the model order as $o_{M}$, the set of node states as $S$ ($S=\{A,B\}$ in the voter model), and the set of network moments of order $o$ as $[\Omega_{o}]$, 
e.g. $[\Omega_{0}]=\{[X]:X\in S\}$, $[\Omega_{1}]=\{[XY]:X,Y\in S\}$, and $[\Omega_{2}]=\{[XYZ]:X,Y,Z\in S\}$. Then,
\begin{equation}\label{eq:gen}\renewcommand{\arraystretch}{2}
\displaystyle \frac{\rm d}{\rm dt} [\omega_{o}]  = f_{\omega_{o}}\left([\Omega_{0}],[\Omega_{1}],...,[\Omega_{o+o_{M}}]\right),
\end{equation}
where $[\omega_{o}]\in[\Omega_{o}]$ and $f_{\omega_{o}}$ is a $\mathbb{R}$-valued function that is moment specific. 

We note that there are variants of moment expansions that differ in the definition of the set $\Omega_{o}$. One intuitive approach is to include all possible subgraphs 
that contain $o$ links. However, more appropriate basis of subgraphs can be constructed by considering the specific properties of the model. For instance, sparse random 
graphs tend to be locally tree-like. If a model does not have any particular update rules that actively create small cycles, cyclic motifs are exceedingly rare and can 
thus in general be safely ignored. 

\begin{figure}[ht!]
  \centering
  \includegraphics[width=3.5in,keepaspectratio]{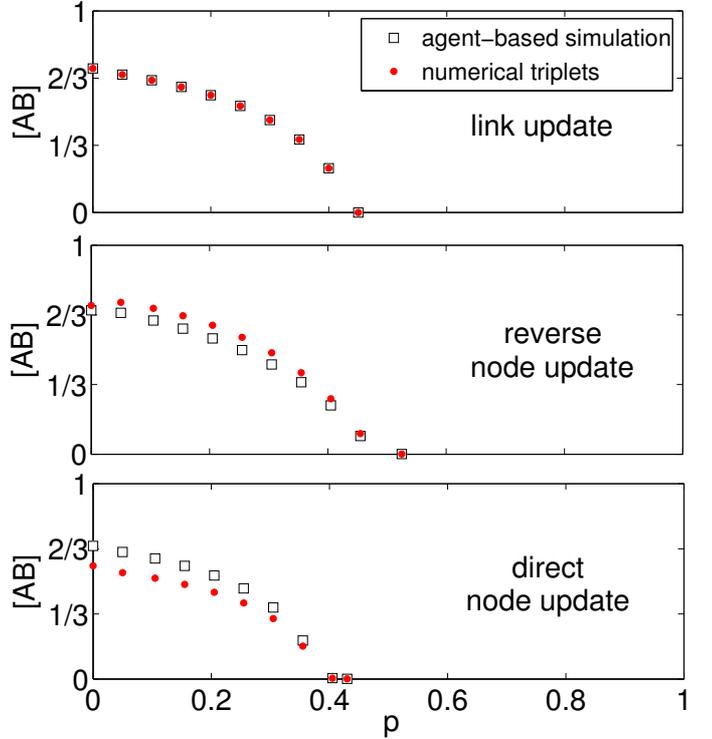}
  \caption{(Color online) Validity of the moment expansion. The quasi-stationary density $[AB]$ at $[A]=[B]=1/2$ from agent-based simulations is compared with $[AB]$ computed 
  analytically from the moment expansion (Eq.~(\ref{eq:tripletsollink}) for link and reverse node update (\ref{sec:appendix D}), and see \ref{sec:appendix E} for direct node update). 
  In this plot, the moment-closure approximation is avoided and the analytical results use the triplet densities $[AAB]$ and $[ABA]$ measured in the numerical simulation. 
  Parameters: $N=10^5$, $\langle k \rangle = 4$.}
 \label{fig:expansion_validity}
\end{figure}

In order to test the validity of the moment expansion Eq.~(\ref{eq:ode_link_triplet}) for the adaptive voter model, we solve for the density $[AB]$ in the same equation 
at the steady state, which yields 
\begin{equation}\label{eq:tripletsollink}
[AB] \ = \ (1-p) \left( [AAB] - 2 [ABA]\right).
\end{equation}
To assess the validity of the expansion independently of the subsequent closure approximation we compute the expected number of active links for the respective triplet 
densities $[AAB]$ and $[ABA]$ observed in an agent-based simulation. The resulting diagram, Fig.~\ref{fig:expansion_validity}, shows an almost perfect match for the active link 
density $[AB]$ with the numerical results for the model with link update. The match for node update rules is less good, in particular for low rewiring rates, but becomes better 
as we approach the transition point where the active links vanish and the network fragments.    

Let us discuss the bad performance of the expansion for the node update rules in a little more detail. The moment expansion can only be exactly valid in the thermodynamic limit. 
However, the link update model shows that for the network sizes considered here finite size effects can be neglected. The only other explanation is that the simulated system 
contains some correlations that are not picked up by the moment expansion. Because the effect of long-ranged correlations should have been captured. However, in the expansion 
we are implicitly assuming that the triplets are statistically distributed within the network, which is approximately true for the link update rule at low rewiring rates. 
By contrast in the node update rules the bias can induce additional correlations. This effect is counteracted by the rewiring as this mixes the network and thus improves 
the prediction.       

As a side-note, let us remark that the beneficial effect of rewiring is a general property. In a given static network model a specific topological feature, such as a node of 
particular high degree or a densely clustered region, might exist that biases the system constantly in a specific direction. In an adaptive network such topological features 
may emerge for a limited time before being destroyed by topological dynamics. In the long run the effect of unlikely local configurations often averages out. In this sense 
adaptive networks are ensembles of themselves and can thus often be very well approximated unless correlations arise systematically that are not captured by the approximation 
scheme used. 

\subsection{Moment-Closure Approximation}
In the previous section we have seen that moment expansions lead to an infinite hierarchy of equations. So far we truncated this expansion after the first order and used 
numerical values for the densities of larger subgraphs. Because we generally develop the expansion to obtain an analytical solution, we cannot rely on numerics, but must estimate 
the density of large subgraphs using a suitable approximation, known as \emph{moment-closure approximations}.   

For illustration let us approximate the second order moment $[ABA]$ in terms of the zeroth order moment $[B]$ and the first order moment $[AB]$ by the so-called \emph{pair approximation}. 
An ABA-triplet comprises two adjacent AB-links sharing a common B-node. Therefore we first note that the density of single AB-links is $[AB]$. Next, we compute the probability 
that an \emph{additional} A-node is connected to the B-node in this link.
  
Since we reached the B-node by following a link, we can expect it to have a higher-than-average degree. Specifically, the degree of the B-node in a randomly selected AB-link 
follows the distribution $Q_{k}^{B}=kP_{k}^{B}/([B]\langle k_{B} \rangle)$, where $P_{k}^{B}$ denotes the probability that a randomly selected B-node has degree $k$ and 
$\langle k_{B} \rangle$ is the mean degree of a randomly selected B-node. The expected number of \emph{additional} links of a B-node in an AB-link is thus  
$\langle q_{B} \rangle = \sum_{k}{(k-1)Q_{k}^{B}}$, which is also known as the mean excess degree \citep{Newman2003} (of B-nodes). 

A key assumption that we have to make at this point is that the AB-links are uncorrelated, except for the effect of the node degree described above. Using this assumption, 
each of the additional links of the B-node is an AB-link with probability $[AB]/([AB]+2[BB])$. Using $[AB]+2[BB] = \langle k_{B} \rangle [B]$, we find  
\begin{equation}\label{eq:linkpairappr}
2[ABA] \simeq \kappa_{B} \frac{[AB]^{2}}{[B]},
\end{equation}
where $\kappa_{B} = \langle q_{B} \rangle / \langle k_{B} \rangle$. 

Similarly,
\begin{equation}\label{eq:linkpairappr2}
[ABB] \simeq \kappa_{B} \frac{2[BB][AB]}{[B]} \text{ and } [BBB] \simeq \kappa_{B} \frac{2[BB]^{2}}{[B]}.
\end{equation}

\begin{figure}[ht!]
  \centering
  \includegraphics[width=3.2in,keepaspectratio]{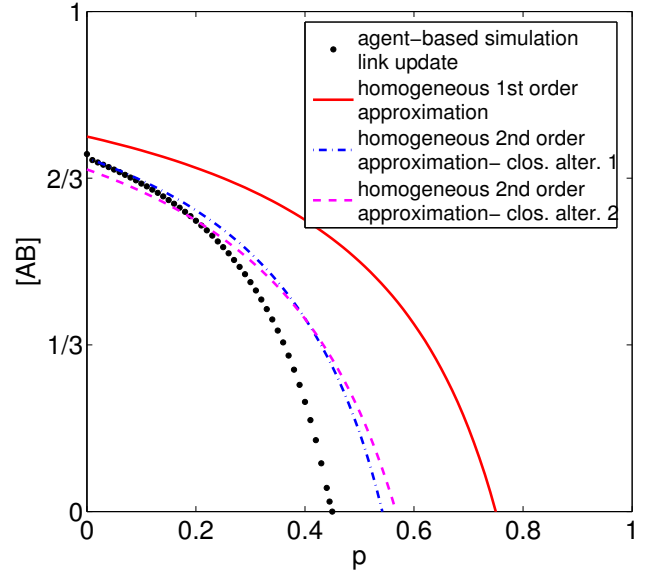}
  \caption{(Color online) Performance of the first and second-order homogeneous moment-closure approximations. As a function of the rewiring rate $p$, the density of active links $[AB]$ 
  from agent-based simulations is compared with results of the pair approximation, and two different triplet approximations $[(A)-BAA]=0.5[AAA][AAB]/[AA]$ and $[(A)-BAA]=0.5[AAB]^{2}/[AB]$ 
  (dashed, dash-dotted, respectively). Parameters: $N=10^5$, $\langle k \rangle = 4$.}
  \label{fig:performanceofclosures}
\end{figure}

The parameter $\kappa_B$ that appears in these equations is generally not known because the degree distribution is reshaped by the rewiring process (although see \citep{Zschaler2012} 
for an exception). Governing $\kappa_B$ are two counteracting effects, on the one hand we know that the degree of a node that is reached via a link is on average greater than the mean 
degree and on the other hand we have to subtract $1$ from this increased degree, because we are only interested in the number of additional links. On Erd\"{o}s-Renyi random graphs, 
these effects cancel exactly such that $\kappa_{B}=1$. Although the value of $\kappa_B$ can be significantly higher in networks with broad degree distributions, assuming $\kappa_{B}=1$ 
has yielded good results for models with a fairly wide degree distribution \citep{Gross2006}. 

Substituting Eqs.~(\ref{eq:linkpairappr}, \ref{eq:linkpairappr2}) into Eq.~(\ref{eq:ode_link_triplet}) and using the random-graph approximation $\kappa_{B}=1$ yields 
\begin{eqnarray} \label{eq:ode_pair_apprx}\renewcommand{\arraystretch}{2}
\displaystyle \frac{\rm d}{\rm dt} [A] & = & 0, \nonumber \\
\displaystyle \frac{\rm d}{\rm dt} [AA]  & = & \displaystyle \frac{1}{2} [AB] + \frac{(1-p)}{2} \left( \frac{[AB]^{2}}{[B]} - \frac{2[AA][AB]}{[A]} \right), \nonumber \\
\displaystyle \frac{\rm d}{\rm dt} [BB] &=& \displaystyle \frac{1}{2} [AB] + \frac{(1-p)}{2}\left(\frac{[AB]^{2}}{[A]} - \frac{2[BB][AB]}{[B]}\right). \nonumber \\ 
\end{eqnarray}

In order to test the performance of approximation we solve Eq.~(\ref{eq:ode_pair_apprx}) for the stationary density of active links  
\begin{equation}\label{eq:pairsollink2}
[AB] \ = \ \left( \frac{ \langle k \rangle (1-p)-1}{(1-p)}\right)[A]\left(1-[A]\right),
\end{equation}
which nicely captures the parabola shape of the states, shown in Fig.~\ref{fig:parabola}. Furthermore, considering the tip of the parabola $[A]=1/2$, we find
\begin{equation}\label{eq:pairsollink}
[AB] \ = \ \left( \frac{ \langle k \rangle (1-p)-1}{4(1-p)}\right),
\end{equation} 

A comparison of the Eq.~(\ref{eq:pairsollink}) with numerical results is shown in Fig.~\ref{fig:performanceofclosures}. The comparison shows that the approximation captures 
qualitative features of the model. The highest density of active links is found for $p=0$, then as $p$ is increased the density of active links declines and finally reaches zero 
at a finite rewiring rate $p^{*}$. However, the quantitative correspondence between the analytical and numerical results is very bad. In particular the pair approximation 
significantly overestimates the rewiring rate at which fragmentation occurs. 

\begin{figure}[ht!]
  \centering
  \includegraphics[width=3.0in,keepaspectratio]{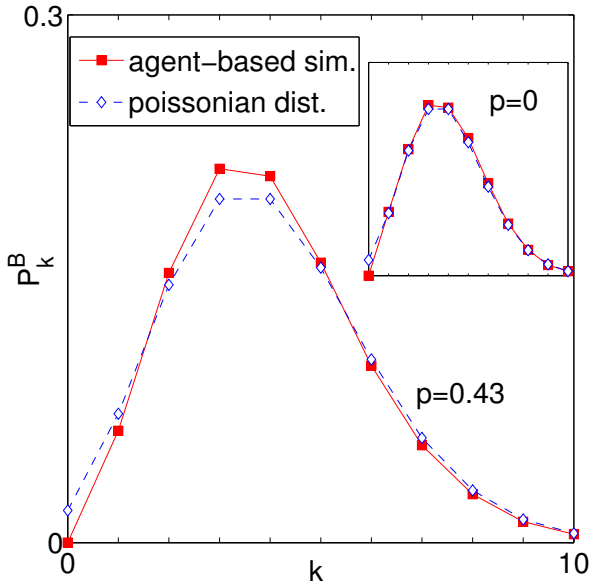}
 \caption{(Color online) Validity of the random-graph approximation. Comparison of the degree distribution of B-nodes, $P_{k}^{B}$, in agent-based simulations with a 
  Poisson distribution $P_k=e^{-\langle k \rangle} \langle k \rangle^k / k!$. The comparison shows that the degree distribution of the adaptive network remains almost exactly Poissonian, and hence the random-graph approximation 
  is valid to very good approximation. Parameters: $N=10^5$, $\langle k \rangle = 4$.}
  \label{fig:degreedist}
\end{figure}

Let us investigate the reason of the bad performance of the approximation in more detail. We have already argued above that the random-graph approximation $\kappa_B=1$ is 
probably harmless. This can be confirmed by comparing the degree distribution observed in simulations to the Poissonian distribution of a random graph (Fig.~\ref{fig:degreedist}). 
The comparison shows that the degree distribution stays very close to the Poissonian distribution. For such a close match the random-graph approximation is almost exact and 
cannot be the source of the major discrepancy observed in the results. 

Accepting the validity of the moment expansion and ruling out the random-graph approximation as a source of errors leaves us with only two further sources of errors: 
We have assumed that a) the actual density of large motifs can be replaced with its expectation value and b) that correlations between active links can be neglected 
when computing this expectation value.  

Let us first consider assumption a), which we call truncation assumption. Approximating any system by a lower-dimensional system is only possible if there is a time 
scale separation between slow low-order moments and fast higher-order moments \citep{Gross2008a}. The system then quickly converges to the slow manifold, characterized 
by the slow variables only \citep{Kuehn2011}. Therefore, dynamics of moments higher than some order are enslaved to the dynamics of lower ones and they can be expressed 
as algebraic functions of low order slow moments. In our moment expansion the higher-order moments are disproportionately more likely to be affected by updates. For instance 
a single rewiring event effects one link, but approximately $2 (\kappa_B \langle k \rangle)^2$ triplets. While a more detailed investigation of this point would probably be fruitful, 
we conclude that assumption a) is probably not the main source of error in the present approximation scheme.  

\begin{figure}[ht!]
  \centering
  \includegraphics[width=3.0in,keepaspectratio]{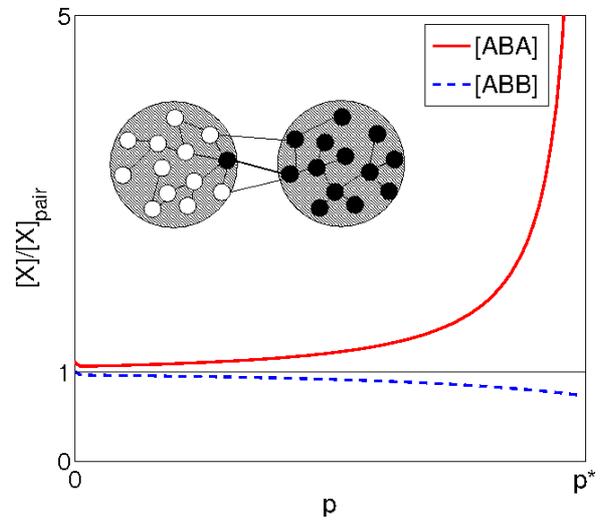}
  \caption{(Color online) Test of the pair approximation. Shown is the ratio between the observed number of triplets in agent-based simulations and the expected number based on the 
  observed number of nodes and links. The pair approximation is approximately valid for ABB triplets, whereas the error in the approximation of ABA triplets diverges as the system 
  approaches the fragmentation point $p^*$. The sketch in the inset explains this failure. Close to fragmentation many active links are created by very few nodes that are in the wrong cluster. 
  This induces a very high correlation between active links which is not capture by the pair approximation. Parameters: $N=10^5$, $\langle k \rangle = 4$.    
  }
  \label{fig:ABA}
\end{figure}
 
Accepting that the actual dynamical densities of higher-order moments can be replaced by their static expectation values, leaves us with the task of capturing the corresponding 
slow manifolds in a suitable functional form. Above, we derived such a functional form based on the assumption b), the absence of longer-ranged correlations. We can test this 
assumption by comparing the numbers of triplets observed in simulations to the expected values for uncorrelated active links. This comparison, shown in Fig.~\ref{fig:ABA}, 
indicates that the expectation for ABB-triplets is almost correct, while the error in the estimation of ABA-triplets diverges as the system approaches the fragmentation point.   

\begin{figure}[ht!]
  \centering
  \includegraphics[width=3.0in,keepaspectratio]{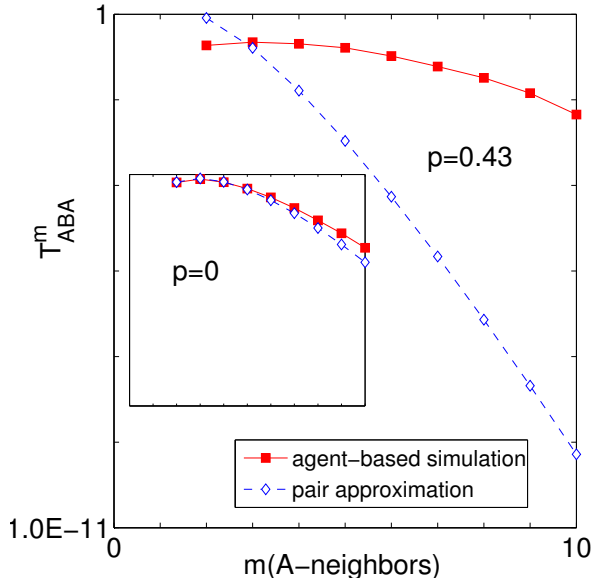}
  \caption{(Color online) Emergence of ABA-correlations close to fragmentation. Shown is $T_{ABA}^{m}$, the fraction of B-nodes with $m$ A-neighbors. Averages from agent-based simulations
  are compared with the expectation of an Erd\H{o}s-Renyi random graph without any second neighbor correlations (see \ref{sec:appendix F}). Close to the fragmentation transition, the A-neighbor distribution of 
  ABA-triplets  deviate from the pair approximation. Parameters: $N=10^5$, $\langle k \rangle = 4$.}
  \label{fig:fragmentedtopology}
\end{figure}

An intuitive explanation of the failure of the pair-approximation is shown in the inset of Fig.~\ref{fig:ABA}. The sketch shows a representation of a typical configuration close 
to fragmentation. The nodes have sorted into two large clusters, connected by a few remaining links. On those links occasionally opinion adoption events take place which introduce 
some ``wrong"  nodes into the clusters and thus create many active links. The majority of active links is thus located on few nodes. This creates both a disproportionately 
large number of ABA-triplets and constitutes a strong three-node correlation that is not captured by the pair approximation. 

The intuitive explanation above can be quantified by a numerical test shown in Fig.~\ref{fig:fragmentedtopology}. The results confirm that close to fragmentation most ABA-triplets 
occur on B-nodes having many A-neighbors.  

\subsection{Better homogeneous approximations}

The reasoning presented above identifies the actual closure approximation, essentially assuming the absence of longer-ranged correlations as the reason for the failure 
of the pair approximation close to the fragmentation transition. Let us therefore discuss ways in which the approximation can be improved. 

In principle every moment expansion should converge to the correct result if the order of the approximation is increased. Clearly despite its shortcoming the pair approximation 
is much better than the zeroth-order \emph{node approximation} that ignores all, even link level, correlations and assumes that the network is well-mixed in terms of node states 
\citep{Nardini2008}. This is analogous to the molecular field approximation of the Ising model \citep{Katsura1973}. The first-order approximation we used above is called the \emph{pair approximation} 
\citep{Vazquez2008,Kimura2008,Gross2006,Shaw2008,Zschaler2010,Demirel2011,Vazquez2008a}, which accounts for nearest-neighbor correlations but neglects higher order ones. 
This is analogous to the Bethe-Peierls approximation \citep{Bethe1935,Peierls1936,Burley1972,Katsura1973,Yedidia2005} in statistical physics. When the density of cycles is 
non-negligible, i.e. when the network has high clustering, the pair approximation is extended in order to account for the cyclic motifs, corresponding to Bethe-Kirkwood 
type approximations \citep{Law2003,Peyrard2008}. Approximations that make the closure at higher orders are analogous to high order Kikuchi approximations \citep{Kikuchi1951}. 

Increasing the order of the approximation is expected to yield better results \citep{Kimura2008,Matsuda1992,Bauch2005}, but convergence to the correct solution is not guaranteed 
to be fast or uniform. Moreover, increasing the order of the expansion creates a number of technical problems. First, the number of subgraphs increases very quickly with the order 
of the expansion. For instance, a third-order approximation to the adaptive voter model already consists of $29$ rate equations and $48$ estimated fourth-order moments. 
Second, uniquely enumerating the subgraphs and computing the correct prefactors that arise from symmetries in not completely trivial. Third, and perhaps most interestingly there 
are different mutually inconsistent possibilities for closing expansions beyond the pair level (see \ref{sec:appendix C}). A criterion on which closure should be used is an important open mathematical problem. 

In the context of the adaptive voter model a higher order closure has been used by \citet{Kimura2008}, and for the specific model studied here two different higher order closures 
are derived in \ref{sec:appendix B} and \ref{sec:appendix C}. 

The performance of higher level-closures is shown in Fig.~\ref{fig:performanceofclosures}. Although the triplet-level closures perform better than the pair-level closure, 
the prediction is still very bad close to the bifurcation point. We can explain this result by considering Fig.~\ref{fig:fragmentedtopology} again. In the triplet-level closure 
we have to estimate the density of four-node subgraphs. In this estimation we use the assumption that the four node subgraphs are uncorrelated. However, from the numerical results 
we know that many active links connect to nodes that have ten or more such links, which implies also a high correlation at the four-node-level. 

The reasoning above suggests that the order of the approximation will have to raised beyond the mean degree of the system to achieve a faithful result. Because of the technical difficulties 
described above this is clearly infeasible. However, note that by raising the expansion to this point we would be capturing much information that is clearly not important. Considering that 
the number of subgraphs rises combinatorially with the order of the expansion, say tenth-order expansion would include an enormous number of subgraphs including exceedingly rare ones, 
which clearly cannot be of importance. Consider furthermore that the tenth-order closure would also comprise 11-node chains, and thus captures correlations that are longer than the diameter 
of networks in reasonable simulations. This suggests, that much a better performance is achieved more cheaply if a tailored motif-basis is used that does not cover whole orders but selectively 
contains only those subgraphs that are otherwise hard to estimate - an idea to which we return in Sec.~\ref{sec:percolation}.  

A promising alternative approach is to use a relatively low order moment expansion, but use a more intelligent closure. For instance, \citet{Gross2008a} implements the approach of equation-free modeling 
\citep{Kevrekidis2004} to extract a proper closure term for an epidemic model automatically from very short bursts of simulation runs. This enabled a semi-analytical investigation 
where continuation software was used to explore the dynamics of the system.   
 
Another approach proposed in a recent paper \citep{Rogers2011} is to generate closures from a maximum entropy principle. This approach thus solves at least the problem of non-uniqueness 
of higher-order closures. However, in many cases the approach provides only implicit equations for the closure that do not seem to have a closed-form solution. This can be seen as an 
indication that explicit fully-consistent closure approximations for moment-expansions beyond the pair level might not exist at all. 

\section{HETEROGENEOUS APPROXIMATIONS}\label{sec:heterogeneous}
Presently it is widely believed that quite universally better results can be obtained by heterogeneous approximations that capture information on the degree of the nodes. Indeed, such approaches 
have yielded an improvement in several example systems \citep{Pugliese2009,Marceau2010,Gleeson2011,Durrett2012}.
 
In this section, we investigate two prominent heterogeneous moment-closure approximations. In the \emph{heterogeneous pair approximation} \citep{Pugliese2009}, links are grouped according to the state 
and the degree of the nodes at their ends. In the \emph{active neighborhood approximation} \citep{Noel2009,Marceau2010,Lindquist2011,Gleeson2011}, nodes are placed in compartments according to their state, 
degree and number of neighbors in a given state. 

Although the heterogeneous approaches can capture effects resulting from the heterogeneity of the degree distribution, they do not specifically address the complications identified above. Here, we test the 
performance of these approaches for the voter model, which reveals that they do not perform better than the homogeneous approximations, in this context.

\subsection{Heterogeneous Pair Approximation}\label{Het-pair-approx}
The heterogeneous pair approximation is based on writing a set of coupled rate equations for the density of active links $[AB]_{k,k'}$ between a node of degree $k$ and a node of degree $k'$. 
In networks with narrow degree distribution we expect these densities to be independent of $k$ and $k'$, whereas the same is not true for networks with broad degree distribution \citep{Pugliese2009}. 
Here we follow an approach which is an extension of the one developed in \citep{Vazquez2008a,Pugliese2009} for the case of adaptive networks. We consider the direct node update rule, since the derivation 
is easier in this case.

\begin{figure}[ht!]
\centering
  \includegraphics[width=3.0in,keepaspectratio]{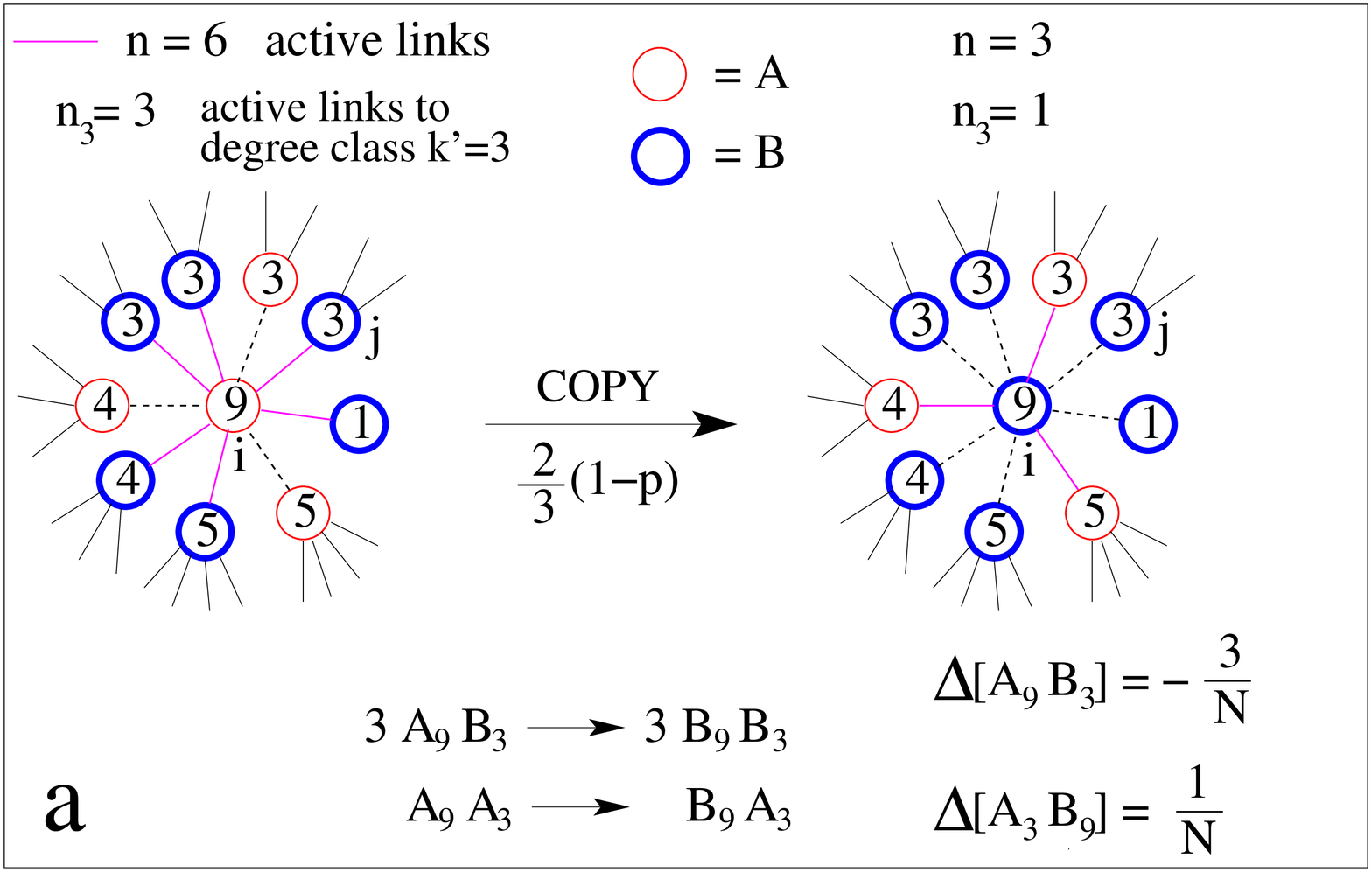}
  \includegraphics[width=3.0in,keepaspectratio]{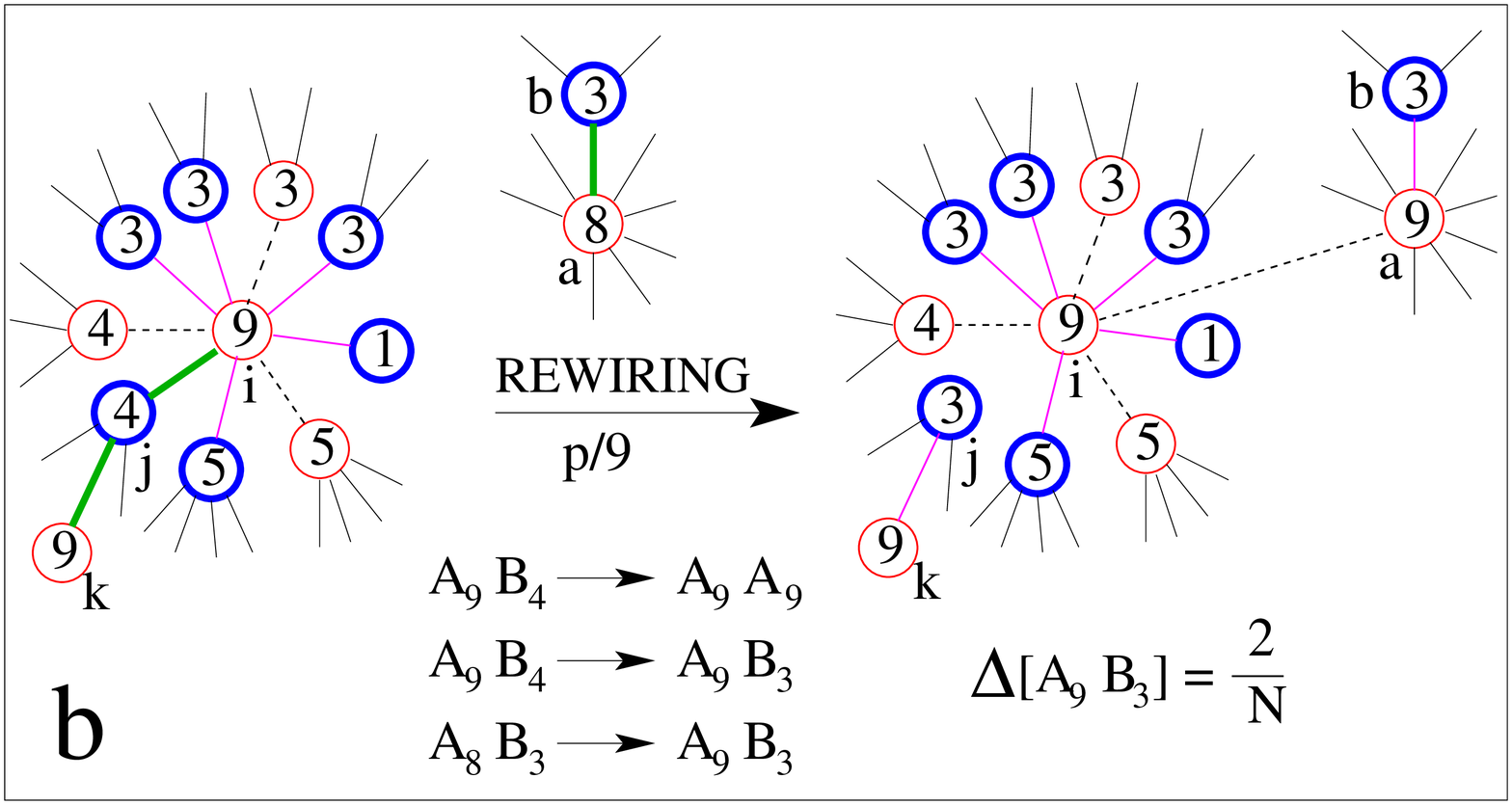}
  \includegraphics[width=3.0in,keepaspectratio]{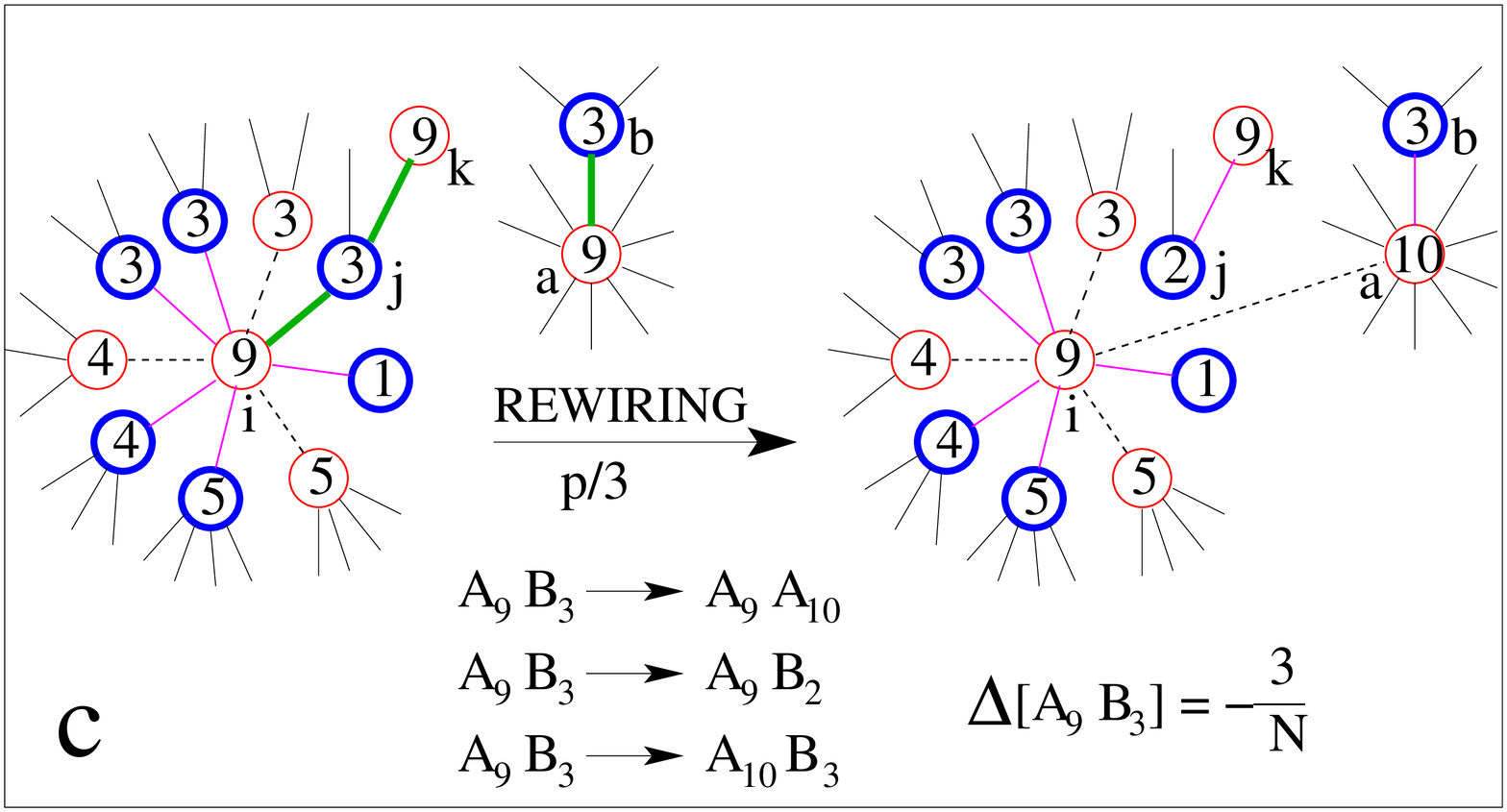}
  \caption{(Color online) Illustration of three update events in the direct voter dynamics that lead to a change in the density of active links $[A_9B_3]$ connecting nodes of degree $k=9$ and $k'=3$, when a node $i$ of degree $k=9$ and state 
	A is chosen. Initially (left), the number of neighbors of degree $k'=3$ is $\mathcal N_{3}=4$, from which $n_3=3$ are in the opposite state B, and the total number of active links is $n=6$.  After the update (right), some links 
	change its type.  At the bottom of each panel are indicated the transitions in link type that involve an $A_9 B_3$-link, and the associated changes in the density $[A_9 B_3]$.  (a) With probability $n(1-p)/k$ node $i$ copies the state of a randomly chosen B-neighbor, thus links attached to $i$ change from active to inert and vice-versa. (b) With probability $n_4 p/k$ an active link connected to one of the neighbors of degree 4 (node $j$ is this example) is chosen 
	and rewired to a node $a$ of class $(A,8)$.  (c) With probability $n_3 \,p/k$ an active link of type $A_9 B_3$ is chosen at random and rewired to a node $a$ of class $(A,9)$.  We note that this figure is an illustration of only some of
	the terms in Eq.(\ref{dABkk'dt}).}
  \label{fig:hetero-rates}
\end{figure}

We start by writing the active link density as $[AB]_{k,k'}=[A_k B_{k'}]+[B_k A_{k'}]$, where, for instance, $[A_k B_{k'}]$ denotes the density of links connecting a node of state A and degree $k$ with a node of state B and degree $k'$. Assuming a node with state A (node i) is chosen in an update event, then 

\begin{eqnarray}\label{dABkk'dt}
\hspace{-1cm} \left. \frac{d [A_k B_{k'}]}{dt} \right|_A &=&\sum_{l=1}^{\overline k}\frac{[A]_{l}}{1/N} \sum_{\mathcal N_1=0}^l ... \sum_{\mathcal N_{\overline k}=0}^l ~ \sum_{n_1=0}^{\mathcal N_1} ...\sum_{n_{\overline k}=0}^{\mathcal N_{\overline k}} \nonumber \\ 
&\times& M(\mathcal N_1,..,\mathcal N_{\overline k};l)~\prod_{m=1}^{\overline k} B(n_{m};\mathcal N_{m}) \nonumber \\
&\times& \Bigg\{ \frac{n}{l} (1-p) \left[ (\mathcal N_{k}- n_{k})  \delta_{l,k'} - n_{k'} \delta_{l,k} \right] -  \frac{n_{k'}}{l} p \,\delta_{l,k} \nonumber \\ 
&+& \frac{p}{l} \left[ n_{k'+1}\, N(A_k | B_{k'+1} A_l)- n_{k'}\, N(A_k | B_{k'} A_{l}) \right] \nonumber \\
&+& \frac{n}{l} p \left[ \frac{[A]_{k-1} N(B_{k'} | A_{k-1})}{[A]} - \frac{[A]_k N(B_{k'}|A_k)}{[A]} \right] \Bigg\} \frac{1}{N}, \nonumber \\
\end{eqnarray}

With probability $[A_{l}]$, node i belongs to class $(A,l)$. The product $M(\mathcal N_1,...,\mathcal N_{\overline k};l)\, \prod_{m=1}^{\overline k} B(n_{m};\mathcal N_{m})$ 
expresses the probability that the configuration around node i consists of $\mathcal N_{m}$ links to neighbors of the degree class $m$ ($m=1,..,\overline k$, with $\overline k$ 
the maximum degree) and $n_{m}$ of these neighbors have the opposite state B ($n_{m}=0,..,\mathcal N_{m}$). Here $B(n_{m};\mathcal N_{m})$ stands for the probability that $n_{m}$ 
of the $\mathcal N_m$ links to neighbors of class $m$ are active. 

We distinguish again between \emph{direct} and \emph{indirect} changes in $[A_k B_{k'}]$. A direct change takes place when node i is in either class $A_{k'}$ or $A_k$, 
giving the first two terms inside the brackets of Eq.~(\ref{dABkk'dt}), respectively. Node i adopts state B when it copies the state of a randomly chosen B-neighbor, which happens with probability $(1-p)n/k$, 
where $n=\sum_{m=1}^{\overline k} n_{m}$ is the number of active links. In these events, the corresponding changes in the density $[A_k B_{k'}]$ are $\Delta [A_k B_{k'}] = (\mathcal N_{k}- n_{k})/N$ 
and $\Delta [A_k B_{k'}] = - n_{k'}/N$ respectively (see Fig.~\ref{fig:hetero-rates}-a). 

An indirect change occurs due to an update on a neighboring node. The third term inside the brackets of Eq.~(\ref{dABkk'dt}) corresponds to the rewiring of an active link 
connected to a neighbor j of class $(B,k')$ (with probability $pn_{k'}/k$), that results in the loss of an $A_k B_{k'}$ link. 

Other indirect changes take place when node i is in a generic class $A_l$, and one of its links to a neighbor j is rewired, affecting the class of the links to the node j, 
and the class of a--b links, where a is the node that receives the rewired link and node b is a neighbor of node a (see Figs.~\ref{fig:hetero-rates}-b and c). 
In Fig.~\ref{fig:hetero-rates}-b we describe the situation in which node j is in class $(B,k'+1)$ (with probability $n_{k'+1}/k$).  Given that node j loses one link, it changes 
to class $(B,k')$, thus there is a gain of $N(A_k | B_{k'+1} A_k)$ links, represented in the fourth term of Eq.~(\ref{dABkk'dt}). The fifth term (see Fig.~\ref{fig:hetero-rates}-c) 
corresponds to a loss in a similar update, when node j is in class $(B,k')$. 

Finally, the last gain and loss terms belong to the case where any of the active links of node i is rewired (with probability $pn/k$) to a node a in class $(A,k-1)$ (see Fig.~\ref{fig:hetero-rates}-b), 
or in class $(A,k)$ (see Fig.~\ref{fig:hetero-rates}-c) respectively. Given that the link is rewired to an A-node chosen at random, the probabilities for these events are $[A_{k-1}]/[A]$ and $[A_{k}]/[A]$, 
respectively. In the former event, the $N(B_{k'}|A_{k-1})$ links of type $A_{k-1} B_{k'}$ attached to node $a$ change to type $A_k B_{k'}$, while in the latter event the $N(B_{k'}|A_k)$ links of type 
$A_k B_{k'}$ attached to node $a$ change to type $A_{k+1} B_{k'}$.

Given that $[A_k B_{k'}]$ may also change when a B-node is chosen, the evolution of $[A_k B_{k'}]$ is given by 
\begin{eqnarray*}
\left. \frac{d [A_k B_{k'}]}{dt} = \frac{d [A_k B_{k'}]}{dt} \right|_A + 
\left. \frac{d [A_k B_{k'}]}{dt} \right|_B.
\end{eqnarray*}

The second term on the right hand side can be obtained from Eq.~{\ref{dABkk'dt}, by interchanging A and $k$ by B and $k'$, respectively.
By carrying out the summations, we arrive at the rate equation for the evolution of $[AB]_{k,k'}$ (see \ref{sec:appendix G} for the derivation)

\begin{eqnarray}\label{d[AB]kk'dt}
\hspace{-1cm}\frac{d [AB]_{k,k'}}{dt} &=& (1-p) \left( Q_{k'} \frac{k-1}{k} [AB]_k + Q_k \frac{k'-1}{k'} [AB]_{k'} \right) \nonumber \\
&-& \Bigg\{ \frac{(1-p)}{2 \langle k \rangle [A] [B]} \left( \frac{k-1}{k} \frac{[AB]_k}{Q_k}  + \frac{k'-1}{k'} \frac{[AB]_{k'}}{Q_{k'}} \right)\nonumber \\
&& \ +\frac{1}{k} + \frac{1}{k'} \Bigg\} [AB]_{k,k'} \nonumber \\
&+& \frac{p}{4 \langle k \rangle [A] [B]} \Bigg\{ \frac{k' [AB]_{k,k'+1} \{AB\}_{k'+1}}{Q_{k'+1}} \nonumber \\ 
&& \ + \frac{k [AB]_{k+1,k'} \{AB\}_{k+1}}{Q_{k+1}}  \nonumber \\ 
&& \ - \left[ \frac{(k'-1) \{A B\}_{k'}}{Q_{k'}} + \frac{(k-1) \{A B\}_k}{Q_k} \right] [AB]_{k,k'} \nonumber \\
&& \ + 2 \langle k \rangle \left([AB]_{k-1,k'} + [A B]_{k,k'-1} - 2 [AB]_{k,k'} \right) \{AB\}\Bigg\}, \nonumber \\   
\end{eqnarray}
where $Q_{k}=kP_{k}/\langle k \rangle$ is the excess degree distribution, $\{A_{k} B\} \equiv \sum_{l=1}^{\overline k} [A_{k} B_{l}]/l$, 
$\{A B\}_{k} \equiv \{A_{k} B\}+\{A B_{k}\}$, $[AB]_k = \sum_{l=1}^{\overline k} ([A_k B_{l}] + [B_kA_{l}])$, and $\{A B \} \equiv \sum_{k=1}^{\overline k} \{A_{k} B\}$. 

\begin{figure}[ht!]
  \centering
  \includegraphics[width=3.2in,keepaspectratio]{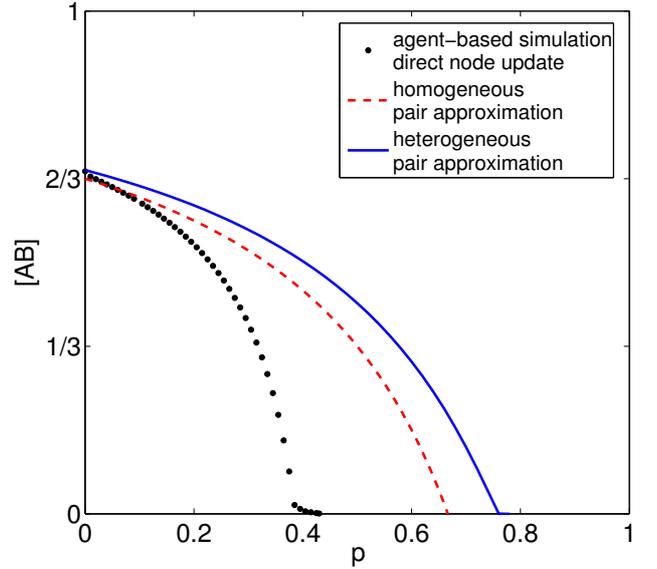}
  \caption{(Color online) Performance of heterogeneous pair approximation in comparison to homogeneous pair approximation and agent-based simulations for the direct node update rule.
	    Parameters: $N=10^5$, $\langle k \rangle = 4$.}
  \label{fig:transition_hetero}
\end{figure}

Equation~(\ref{d[AB]kk'dt}) together with the consistency conditions $[AB]_k = \sum_{l=1}^{\overline k} [AB]_{k,l}$ and $\{AB\}_k = \sum_{l=1}^{\overline k} [AB]_{k,l}/l$ 
form a closed system of coupled ordinary differential equations. Since the system is too complex to solve directly, we obtain the stationary solution by numerical integration 
starting from equiprobable $[A]_{0}=[B]_{0}=1/2$ initial states in a random graph with $P_k=e^{-\langle k \rangle} \langle k \rangle^k / k!$.  

In Fig.~\ref{fig:transition_hetero} we plot the global stationary density of active links $[AB]=\frac{1}{2} \sum_{k=1}^{\overline k} [AB]_k$ as a function of rewiring rate $p$, 
and compare it with results from agent-based simulations. The heterogeneous pair approximation is in good agreement with simulations for low $p$, but discrepancies become 
very large as $p$ increases. The overall performance is surprisingly even worse than the homogeneous pair approximation. This might be due to the accruing of discrepancies 
in individual terms $[A_{k}B_{k'}]$. While the heterogeneous approximation certainly provides a more accurate description of networks with wide degree distribution, it does 
not suitably capture the correlations arising in the fragmentation transition. 

\subsection{Active Neighborhood Approach}\label{subsubsec:active_neigh}
Recently, an alternative heterogeneous moment-closure approximation, the active neighborhood approach \citep{Noel2009,Marceau2010}, was proposed to study the dynamics of the SIS model 
on adaptive networks \citep{Gross2006}. Here, nodes are grouped by their compartment, defined according to their state, total degree and the number of active links. Therefore, 
not only the heterogeneity of the network is taken into account, but also the state correlations between nearest neighbors and associated heterogeneities. In this formalism, 
there is no need of estimating the neighborhood of a node, because this information is already contained in its class.

The active neighborhood approach has been applied to other epidemics systems \citep{Lindquist2011,Gleeson2011,Taylor2012}, Glauber dynamics \citep{Gleeson2011}, and a voter-like model 
where the rewiring is state independent \citep{Durrett2012}. The approximation was found to reproduce the time evolution of both the states and the structure of the network with remarkable 
accuracy. A moment-generating function approach has been applied to mitigate the computational cost of the numerical integration \citep{Wieland2012,Wieland2012b}. 

We now follow the active neighborhood approach for the adaptive voter model. We place nodes with state A (B), degree $k$, and $n$ ($n=0,..,k$) neighbors in the opposite state B (A), 
into the compartment labeled as $[A,k,n]$ ($[B,k,n]$). For reasons of simplicity, we assume that the time is continuous, thus that opinion adoption and rewiring processes take place 
in parallel. Every node in the network transmits its state to its neighbors at rate $\beta$, and rewires the connection from each neighbor in the opposite state to a random node 
with the same state at rate $\gamma$. That is, in a small time interval $dt$ all links are updated with the same probability $(\beta + \gamma)dt$.  This ``link homogeneous'' dynamics 
is equivalent to the link update dynamics, in which links are chosen and updated with probability $1/L$, where $L$ is the number of links in the network. Therefore, stationary states 
obtained from numerical simulations of both dynamics are similar, as shown in Fig.\ref{fig:transition_compartment}. 

The evolution of $A_{k,n}$ is governed by the rate equation,  
\begin{eqnarray}
\frac{d A_{k,n}}{dt} &=& \beta \left[ (k-n) B_{k,k-n} - n A_{k,n} \right] 
\nonumber \\
&+& \beta \frac{A_{AB}}{A_A} \left[ (k-n+1) A_{k,n-1} - (k-n) A_{k,n}
\right] \nonumber \\
&+& \beta \frac{B_{AA}}{B_A} \left[ (n+1) A_{k,n+1} - n A_{k,n} \right]
\nonumber \\
&+& \gamma \left[ (n+1) A_{k,n+1} - n A_{k,n} \right] \nonumber \\ 
&+& \gamma \frac{A_B}{A} \left[ A_{k-1,n} - A_{k,n} \right] \nonumber \\ 
&+& \gamma \left[ (n+1) A_{k+1,n+1} - n A_{k,n} \right],
\label{dAkn}	
\end{eqnarray}
with the zeroth-order moments
\begin{equation}
A \equiv \sum_{k,n} A_{k,n} ~~ \mbox{and} ~~ B \equiv \sum_{k,n} B_{k,n},
\end{equation} 
first-order moments
\begin{eqnarray}
A_A \equiv \sum_{k,n} (k-n) A_{k,n}, ~~~~~~ A_B \equiv \sum_{k,n} n A_{k,n},
\nonumber \\
B_B \equiv \sum_{k,n} (k-n) A_{k,n}~~\mbox{and}~~B_A \equiv \sum_{k,n} n B_{k,n},
\label{first-mom}
\end{eqnarray} 
and second-order moments
\begin{eqnarray}	
A_{AB} \equiv \sum_{k,n} n(k-n) A_{k,n} ~~ \mbox{and} ~~ 
B_{AA} \equiv \sum_{k,n} n^2 B_{k,n} \nonumber \\
B_{BA} \equiv \sum_{k,n} n(k-n) B_{k,n} ~~ \mbox{and} ~~ 
A_{BB} \equiv \sum_{k,n} n^2 A_{k,n}.
\label{second-mom}
\end{eqnarray} 
These moments are related to the moments we defined in Sec.~\ref{sec:homogeneous} by $A=[A]$, $A_A= 2 [AA]$, $A_B = [AB]$, $A_{AB} = [AAB]$ and $A_{BB} = [AB] + 2[BAB]$.   

\begin{figure}[ht!]
\centering
 \includegraphics[width=3.4in,keepaspectratio]{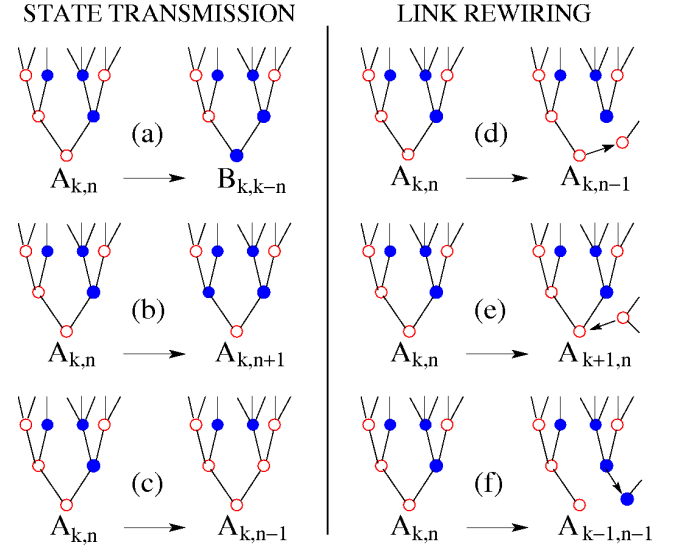}
  \caption{Schematic representation of the possible update events in node
	states (a), (b) and (c) and links (d), (e) and (f), for the active neighborhood
	approach to the voter dynamics.  Open and filled circles represent
	nodes in state A and B, respectively. }
  \label{fig:compart-rates}
\end{figure}

In Fig.~\ref{fig:compart-rates}, we illustrate the six possible transitions of nodes from compartment $[A,k,n]$ to other compartments (we denote the reference node in 
compartment $[A,k,n]$ as node i), which correspond to the six loss terms in brackets of Eq.~(\ref{dAkn}). The first of these describes the transition of node i from compartment 
$[A,k,n]$ to compartment $[B,k,k-n]$ at rate $\beta n$, when it adopts state B from an active neighbor (see Fig.~\ref{fig:compart-rates}-a). The second loss term describes 
the change of one of the $k-n$ A-neighbors of node i to B that happens at rate $\beta A_{AB}/A_A$, where $A_{AB}/A_A$ is the estimated number of the B-neighbors (see Fig.~\ref{fig:compart-rates}-b). 
This yields the transition of $i$ to $[A,k,n+1]$. The third term is analogous to the second term, but with the change $B \to A$ of one of the $n$ B-neighbors of node i at 
rate $\beta B_{AA}/B_A$ (see Fig.~\ref{fig:compart-rates}-c), that brings i to compartment $[A,k,n-1]$. The fourth term represents the replacement of a B-neighbor of node i 
by an A-node at rewiring rate $\gamma n$ (see Fig.~\ref{fig:compart-rates}-d), thus node i moves to $[A,k,n-1]$. Node i gains a link coming from an A-node due to a rewiring 
event which occurs at rate $\gamma A_B/A$ (see Fig.~\ref{fig:compart-rates}-e). Node i switches accordingly to $[A,k+1,n]$, represented by the fifth term. Finally, node i 
switches to the compartment $[A,k-1,n-1]$ when it loses a link due to the disconnection of one of its B-neighbors at rate $\gamma n$ (see Fig.~\ref{fig:compart-rates}-f).
The gain terms can be explained analogously.

\begin{figure}[ht!]
  \centering
  \includegraphics[width=3.2in,keepaspectratio]{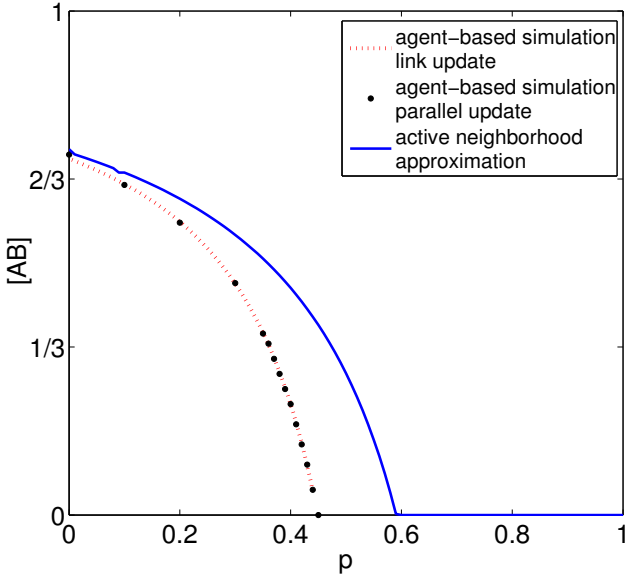}
  \caption{(Color online) Performance of active neighborhood and agent-based simulations with the link update and parallel update rules. To make agent-based simulations compatible with 
  the parallel update mentioned earlier, we used the following algorithm: In a time interval $dt = 0.01$, every node and link of the network is selected. Each node i with state A (B) 
  changes to B (A) with probability $\beta n dt$, where $n$ is the number of neighbors of $i$ in state B (A). Also, each link i--j is removed with probability $2 \gamma dt$ and replaced 
  either by a link i--k with probability $1/2$ or by a link j--k with the complementary probability $1/2$, where node k is randomly chosen within those nodes with state A (B).
  Parameters: $\beta = 0.01$, $\gamma = \beta p/q$, $N=10^5$, $\langle k \rangle = 4$.}
  \label{fig:transition_compartment}
\end{figure}

We obtain the equation for the density $[B,k,n]$ by exchanging A and B in Eq.~(\ref{dAkn}) using the symmetry of the model. This closes the system of equations. Numerical 
integration of the closed system of equations by standard numerical integration algorithms gives the values of the fractions $A_{k,n}$ for a given time, and therefore, 
it allows one to obtain the time evolution of macroscopic variables, such as the densities of active links. We numerically solve Eq.~(\ref{dAkn}) with initial conditions 
\begin{eqnarray}
A_{k,n}(0) &=& A_0 P_k {k \choose n} (1-A_0)^n A_0^{k-n} \nonumber \\
B_{k,n}(0) &=& B_0 P_k {k \choose n} A_0^n (1-A_0)^{k-n},
\end{eqnarray}                                                                                         
where $A_0=B_0= 1/2$ and $P_k=e^{-\langle k \rangle} \langle k \rangle ^k/k!$, and determine the asymptotic values of the density of AB-pairs $ A_B = \sum_{k,n} n A_{k,n}$ for different values of 
the ratio between the rewiring and state adoption dynamics $\gamma/ \beta = p/(1-p)$, where $p$ is the rewiring probability.  

Results from the active neighborhood approximation are compared with agent-based simulations in Fig.~\ref{fig:transition_compartment}. The agreement between the analytical approach 
and simulations is good for small values of $p$, but discrepancy increases with increasing $p$, such that also the active neighborhood approach fails to capture the fragmentation 
transition faithfully.  


\section{ACTIVE MOTIF APPROACH}\label{sec:percolation}
In conventional moment expansions of previous sections, moments are taken as densities of regular subgraphs, where subgraphs were characterized by a given number of links and prescribed node states, 
and degrees in case of heterogeneous moments. While such basis provide reasonable general purpose approaches, they do not take into account the specific dynamics of the system. The considerations 
presented in Sec.~\ref{sec:homogeneous} and the failure of the heterogeneous approximations in Sec.~\ref{sec:heterogeneous} convey a clear message: To capture the fragmentation transition faithfully 
it is essential to capture the very heterogeneous distribution of active links that appears close to the fragmentation transition. Even the very complex and sophisticated active neighborhood approximation 
is in essence only a first order approximation and thus fails to pick up the correlations in the active links.    

We can conclude that capturing the fragmentation transition faithfully requires tracking subgraphs of an order roughly up to the mean degree of the network. While the task of tracking all such subgraphs 
would be of enormous difficulty, it is greatly simplified, by tailoring the subgraph basis to the problem by using only those subgraphs that capture much information on the specific system. Our analysis 
above has shown that properties such as the density of ABB-triplets and even the degree distribution conforms very well to statistical expectations. By contrast the number of ABA-triplets and larger subgraphs 
comprising a number of active links attached to the same node defies statistical expectations close to the fragmentation point. 

The reasoning above suggests that we should use a basis consisting of subgraphs that contain different numbers of active links attached to the same node. Two such \emph{active motif} bases were proposed 
in \citet{Boehme2011}. In application to the adaptive voter model, it was shown that both basis allowed a precise prediction of the transition point \citep{Boehme2011}. The approach was subsequently 
extended to multi-state voter models \citep{Boehme2012}, where it likewise yielded good results. In the following, we briefly explain the approach and illustrate it by application to the adaptive voter model. 
Going beyond the previous works we extend this approach to the estimation of active link densities, which allows a direct comparison with the approaches discussed above.  

For computing the fragmentation point $p^*$ we consider the situation where two communities holding opposite opinions have formed that are only connected by few active links. In analogy to the approaches discussed so far, 
considering the possible updates affecting subgraphs in the basis we derive a system of differential equations, governing the evolution of the subgraphs in time. However, because we are only concerned with subgraphs 
containing active links and such links are rare close to the fragmentation point, we arrive at a linear system of equations. The time evolution is thus fully captured by the eigenvalues of the corresponding Jacobian 
matrix. If all eigenvalues of the Jacobian are negative the fragmented state is stable and the remaining active motifs will disappear over time. By contrast, if the Jacobian has an eigenvalue with positive real part, 
then the fragmented state is unstable and the network will remain connected. The transition point is thus marked by $\lambda(p,\langle k \rangle)=0$, where $\lambda(p,\langle k \rangle)$ is the leading eigenvalue.

To illustrate the approach in more detail let us assume that the network is \emph{degree-regular}, such that every node has the same degree. This assumption can be justified by our earlier observation that the degree 
distribution stays narrow for all values of $p$. For illustration let us further consider the specific case of $k=3$. Here, the dynamics of active motifs is illustrated in Fig.~\ref{bundleappr}. We start by considering 
a single active link (which we call a $1$-fan). In the next update, the link will be rewired with probability $p$ deactivating the motif. With probability $1-p$ one of the nodes connected by this 
link adopts the other's state. In the adoption event, the original active link becomes inert, but the $k-1$ other connections of the adopting agent become active. This leads to a $k-1$-fan, a motif of $k-1$ active links, 
connected by a \emph{base node}. If the next update, which affects the $k-1$-fan, is a rewiring event (probability $p$) the motif is turned into a $k-2$-fan. If the update is an adoption event, then either the base node 
changes its opinion or one of the fringe nodes adopts the base node's opinion, giving rise to a $1$-fan or to two fans, one containing $k-2$ and the other one $k-1$ active links. In the case of link update both processes 
occur with equal probability $(1-p)/2$. 

\begin{figure}
\begin{center}
\includegraphics[width=3.2in,keepaspectratio]{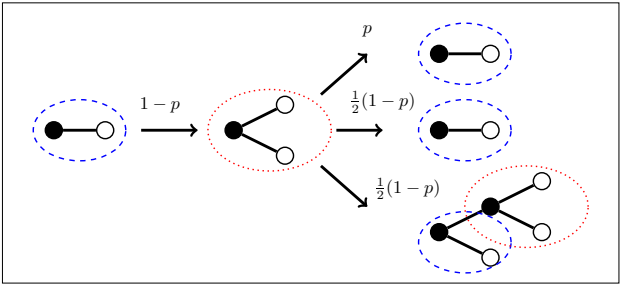}
\caption{(Illustration of the evolution of active links in a degree regular network with degree $k=3$ and link update rule. Shown is the network 
in the neighborhood of an active link connecting components of different opinions. Arrows correspond to dynamical 
updates and are labeled with the corresponding transition rate. Depending on the parameters the updates lead to 
proliferation or decline of active motifs containing one active link (encircled dotted) or two active links (encircled 
dashed).\label{bundleappr}}
 \end{center}
\end{figure}
For $k=3$, we obtain a closed system of two differential equations for the densities $\{q\}$ of $q$-fans:
\begin{eqnarray}\label{k2_rho0}\renewcommand{\arraystretch}{2}
  \displaystyle\frac{d\{1\}}{dt} & = &-\{1\}+2\{2\}, \nonumber \\
  \displaystyle\frac{d\{2\}}{dt}& = &-2\{2\}+(1-p)\{1\}+(1-p)\{2\}.
\end{eqnarray}
The corresponding Jacobian is
\begin{equation}
 {\rm \bf J}=\begin{pmatrix}-1&2\\1-p&-1-p\end{pmatrix},
\end{equation}
and the condition $\lambda(p,3)=0$ yields $p^*=1/3$ for the transition point. The described procedure can be generalized to arbitrary $k$. For $k=4$ (the degree considered here),
the predicted transition point is in good agreement with the value ($p^*\approx=0.445$) from agent-based simulations which violates the assumption of degree regularity. 
 
In order to account for a degree heterogeneous network, a basis set of $\{m,l\}$-spiders is used: a spider motif consists of one central base node which is connected to $m$ nodes 
of its own opinion and $l$ nodes of opposing opinion. The dynamical evolution equation for spider motifs is provided in \ref{sec:appendix I}, and leads to a very close 
approximation of the true transition point. 
\begin{figure}
\begin{center}
\includegraphics[width=3.2in,keepaspectratio]{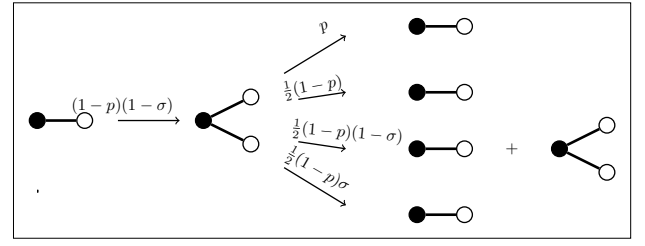}
\caption{Illustration of the evolution of active links in a degree regular network with degree $k=3$ and link update rule.
Updates which lead to transitions between different motifs are depicted as in Fig.~\ref{bundleappr}.
Now the transition rates depend on the probability $\sigma$ that a newly created fan is active. \label{include_rho}}
\end{center}
\end{figure}

In the estimation above for the calculation of the transition point, we assumed that in a $q$-fan motif, all neighbors of the fringe nodes (except for the base node) hold the 
same opinion as the fringe nodes. This assumption is valid for vanishing active link density at the symmetric state $\sigma=[AB]/(\langle k \rangle [A]) \to 0$. Now we consider 
the case $p<p^*$, where there is a finite density of active links. In our equations we therefore have to include the possibility that a fringe node already holds \emph{outer} 
active links (apart from the one connecting to the base node), which become inert when the fringe node adopts the state of the base node. By utilizing the observation that active links 
tend to gather, rather than distribute homogeneously over the whole system, we assume that whenever a new $k-1$-fan is created, this fan is either active, with probability $(1-\sigma)$, 
or inactive with probability $\sigma$.   

In Fig.~\ref{include_rho} an example for the degree regular case ($k=3$) is illustrated, where we include the active link density in the transition probabilities. As before, this can be 
summarized in a system of evolution equations for the motif densities
\begin{eqnarray}\renewcommand{\arraystretch}{2}
\frac{d\{1\}}{dt}&=&-\{1\}+2\{2\} \nonumber \\
\frac{d\{2\}}{dt}&=&-2\{2\}+(1-p)(1-\sigma) \left(\{1\}+\{2\}\right).
\end{eqnarray}
Note that for $\sigma=0$ we recover Eq.~\ref{k2_rho0}.

\begin{figure}[ht!]
  \centering
  \includegraphics[width=3.2in,keepaspectratio]{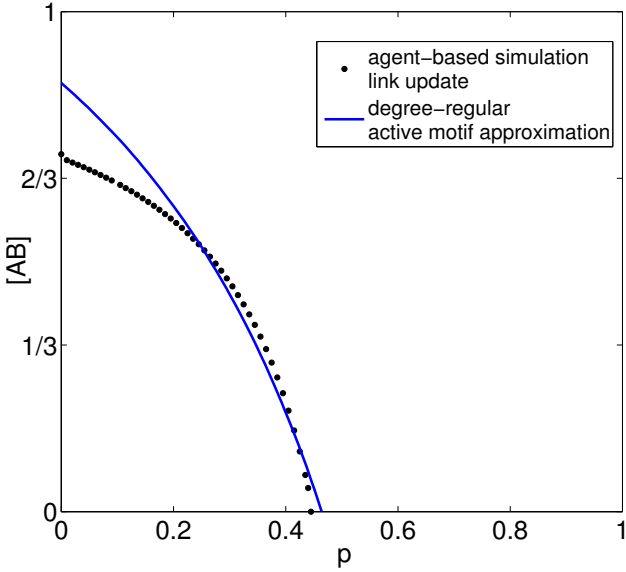}
  \caption{(Color online) Performance of active motif approach, compared to agent-based simulations for the link update rule.
	    Parameters: $N=10^5$, $\langle k \rangle = 4$.}  
\label{fig:transition_perc}
\end{figure}

The corresponding Jacobian now depends on $p$ and $\sigma$. Solving $\lambda(p,\sigma)=0$ yields 
\begin{equation}
 \sigma(p)=\frac{4(1-2p)-(1-p)^2}{8(1-p)-(1-p)^2}.
\end{equation}

In Fig.~\ref{fig:transition_perc} the resulting curve $[AB]=\sigma \langle k \rangle /2$ from the degree regular approach is shown for $\langle k \rangle =4$. Comparison 
to simulation results shows that, as expected, the approximation of the active link density works well in the vicinity of the fragmentation threshold, whereas for small 
rewiring rates it becomes very bad as active links are increasingly well mixed. 

\section{SUMMARY AND DISCUSSION} \label{sec:discussion}
In this paper we investigated the performance of moment-closure approximations for discrete adaptive networks. In particular we used the adaptive voter model as a benchmark 
model to assess different approaches. The comparison with agent-based simulations revealed that both homogeneous and heterogeneous moment-closure approximations capture 
qualitative properties of the fragmentation transition, but fail to provide good quantitative estimates close to the fragmentation point. Remarkably, even very sophisticated 
heterogeneous approaches can produce results that are worse than those from simple homogeneous schemes. Finally, we identified the active motif approximations as a class of 
approximations that were able to capture the behavior close to fragmentation point quantitatively. 

The present results are likely to hold in a much larger class of models. In the adaptive voter model conventional approaches fail close to the fragmentation point because some 
specific correlations appear. These correlations can be generally expected to arise in fragmentation transitions regardless of the specific model under consideration and should thus 
occur in a large variety of models \citep{Holme2006,Kimura2008,Durrett2012,Zanette2006,Vazquez2007,Gil2006,Boehme2012,Kozma2008a,Kozma2008b,Iniguez2009,Bryden2011}.  
We further expect that similar correlations could arise in networks that self-organize into specific topologies such as leader-follower networks, approximately bipartite nets, 
or complex topologies with other long-ranged state correlations \citep{Zimmermann2004,Holme2006a,Zimmermann2005}.

Perhaps the main message from the current work is that even the evolution of problematic models can be captured if an expansion is used that is tailored specifically to 
the system at hand. In this paper we have used extensive numerical and analytical investigations to identify the problematic correlations. However, in retrospect just considering 
a sketch of the situation close to fragmentation, such as the inset in Fig.~\ref{fig:ABA}, could have pointed us to these correlations and hence to a suitable approximation 
scheme. 

Anticipating the structures that are likely to emerge prominently in a given model should generally allow to identify a suitable approximation scheme. Using the spider or fan motifs 
of the active motif approximation will improve predictions in models close to fragmentation or related de-mixing transitions. By contrast, including such motifs in the subgraph basis 
of an approximation scheme could be cumbersome and even have an adverse effect in models that remain well-mixed. Models that are prone to evolve degree-state correlations or strong 
clustering will require specific approaches such as \citep{Matsuda1992,Keeling1999,Bauch2005,Eames2002,Bansal2007,Volz2007,Volz2009} (see \citep{House2011} for a review). By contrast, 
in models with strong random rewiring and sufficiently low degree, small cycles should be rare and thus subgraphs containing cycles can (and should) be ignored when selecting the 
approximation scheme. Clearly, for all models that tend to evolve very heterogeneous degree distributions, heterogeneous approximations such as the heterogeneous pair approximation 
or the active neighborhood approach are required. However, even in the case when degree distributions become broad but not exceedingly broad, such as in \citet{Gross2006}, it may be 
worth to consider homogeneous approximations as they may still provide relatively good results, at a significantly lower cost than the heterogeneous approximations.     

Finally, we note even for the deceptively simple adaptive voter model, we have not yet identified an approximation that works well over the whole parameter range. While the active 
motif approximation yields faithful results in the ordered states close to fragmentation it fails in well-mixed systems. Conversely, all other approaches studied here work reasonably 
well when far from fragmentation, but fail at the fragmentation point. In the future a scheme that works over the whole range of rewiring rates may be found, either as a reasonable 
interpolation between the existing approaches, or by the informed design of a suitable approximation. We hope that the information gathered in the present survey will contribute 
to reaching this goal.      

\section*{Acknowledgement}
The authors thank D. Kimura for the insightful discussions.

\appendix

\section{The origin of second-order terms in the first-order moment expansion} \label{sec:appendix A}

Here, we derive the triplet density terms in the moment expansion Eq.~\ref{eq:ode_link_triplet} from local selection events and combinatorics. We illustrate it on $Q(A|BA)$ which denotes the number of additional A-neighbors of the B-end of 
a randomly selected AB-link. 

The quantity $Q(A|BA)$ can be written as $Q(A|BA) = P(A|BA)$ $\langle q_{B} \rangle$, where $P(A|BA)$ denotes the probability that a random neighbor of the B-end of a randomly selected AB-link 
has state A and $\langle q_{B} \rangle$ denotes the mean excess degree, as derived in the main text. The probability  $P(A|BA)$ is expressed as a ratio of the corresponding triplet densities, i.e. 
$P(A|BA) = 2[ABA]/(2[ABA]+[ABB])$, where $[ABA] = N_{ABA} / N$ ($N_{ABA}$ is the number of ABA-triplets) and analogously $[ABB] = N_{ABB} / N$. 

Let $k_{i}$ and $n_{i}$ be the number of all (A and B) neighbors and A-neighbors of a B-node  selected at random (node i) respectively.  The number of triplets is obtained from a summation over all B-nodes: $N_{ABA} = \sum_{i\in\{B\}}^{}n_{i}(n_{i}-1)/2$ 
and $N_{ABB} = \sum_{i\in\{B\}}n_{i}k_{i}$, where $\{B\}$ is the set of B-nodes. We thus obtain $2N_{ABA}+N_{ABB} = \sum_{i\in\{B\}}n_{i}(k_{i}-1)$. Now, we proceed with the summation 
\begin{eqnarray*} \renewcommand{\arraystretch}{2}
\displaystyle \sum_{i\in\{B\}}n_{i}(k_{i}-1) & = & \sum_{i\in\{B\}}n_{i}k_{i} - \sum_{i\in\{B\}}n_{i}\ \\
				 & = & \sum_{\begin{array}{c} \mbox{\footnotesize $i\in{B}$} \\ \mbox{\footnotesize $j\in{A}$} \\ \mbox{\footnotesize $(i,j)\in Edges$} \end{array}} k_{i} - N [AB] \\
				& = & N [AB] \left(\sum_{k} k \frac{kP_{k}^{B}}{[B]\langle k_{B} \rangle} - 1\right) \\
				& = & N [AB] \sum_{k} \left(k \frac{kP_{k}^{B}}{[B]\langle k_{B} \rangle} - \frac{kP_{k}^{B}}{[B]\langle k_{B} \rangle}\right)\\
				& = & N [AB] \sum_{k}{(k-1)kP_{k}^{B}/([B]\langle k_{B} \rangle}) \\
				& = & N [AB] \langle q_{B} \rangle 
\end{eqnarray*}
and obtain $2[ABA]+[ABB] = (2N_{ABA} + N_{ABB})/N  = [AB] \langle q_{B} \rangle$. By this, we reach 
\begin{equation*}
Q(A|BA) =  \frac{2[ABA]}{[AB]}.
\end{equation*}

\section{Second order moment expansion for the link update rule} \label{sec:appendix B}

Since the right-hand side of Eq.~\ref{eq:ode_link_triplet} involves second-order moments, the first-order expansion is not closed. We now treat the second-order moments as dynamical variables and derive the corresponding rate equations. 
In Fig.~\ref{fig:secondorder}, we illustrate the complete set of possibilities and the corresponding rates where ABA-triplets are created or destroyed. The second-order expansion reads

\begin{eqnarray}\label{eq:ode_link_secondorder}\renewcommand{\arraystretch}{2}
\displaystyle \frac{\rm d}{\rm dt} [A] & = & 0, \nonumber \\
\displaystyle \frac{\rm d}{\rm dt} [AA]  & = & \displaystyle\frac{(1-p)}{2} \big( [AB] + 2 [ABA] - [AAB] \big) + \frac{p}{2} [AB], \nonumber \\
\displaystyle \frac{\rm d}{\rm dt} [BB] &=& \displaystyle \frac{(1-p)}{2}\big([AB] + 2 [BAB] - [ABB]\big) + \frac{p}{2} [AB], \nonumber \\ 
\displaystyle \frac{\rm d}{\rm dt} [AAA] & = & \displaystyle \frac{1-p}{2}  \bigg(2[ABA] + [AAB] + [ABAA]  \nonumber\\ 
					 &  &\displaystyle + 3 [(B)AAA] - [AAAB] - [(A)BAA] \bigg)  \nonumber\\
					 &  & \displaystyle + \frac{p}{2}  \bigg( [AAB] + 2 \frac{[AA][AB]}{[A]}\bigg), \nonumber\\
\displaystyle \frac{\rm d}{\rm dt} [AAB] & = & \displaystyle \frac{1-p}{2}  \bigg([ABB] + 2[BAB] -2 [AAB] \nonumber \\ 
					 &  & \displaystyle + [BAAA] + [ABAB] + 2 [(B)AAB] \nonumber\\ 
					 &  & \displaystyle - [ABAA] - 2[BAAB] - 2[(A)BBA] \bigg)  \nonumber\\ 
					 &  & \displaystyle + \frac{p}{2}  \bigg( 2[BAB] -2[AAB] + \frac{[AB]^{2}}{[A]}\bigg), \nonumber\\
\displaystyle \frac{\rm d}{\rm dt} [ABA] & = & \displaystyle \frac{1-p}{2} \bigg(-4[ABA] + 2[ABBA] + [(A)BAA] \nonumber\\ 
					 &  &\displaystyle  - [ABAB] - 3[(B)AAA] \bigg) - 4 p [ABA], \nonumber\\ 
\end{eqnarray}
where $[XYZW]$ denotes the density of chain-quadruplets constituted by a node of state X (node 1), a Y-neighbor of node 1 (node 2), a Z-neighbor of node 2 (node 3), and a W-neighbor of node 3 (node 4), and $[(X)YZW]$ denotes the density of 
star-quadruplets constituted by a node of state X at the center and its three neighbors of state Y, Z, and W with $\text{X, Y, Z, W}\in\{\text{A,B}\}$.

\begin{figure}[ht!]
  \centering \includegraphics[width=3.5in,keepaspectratio]{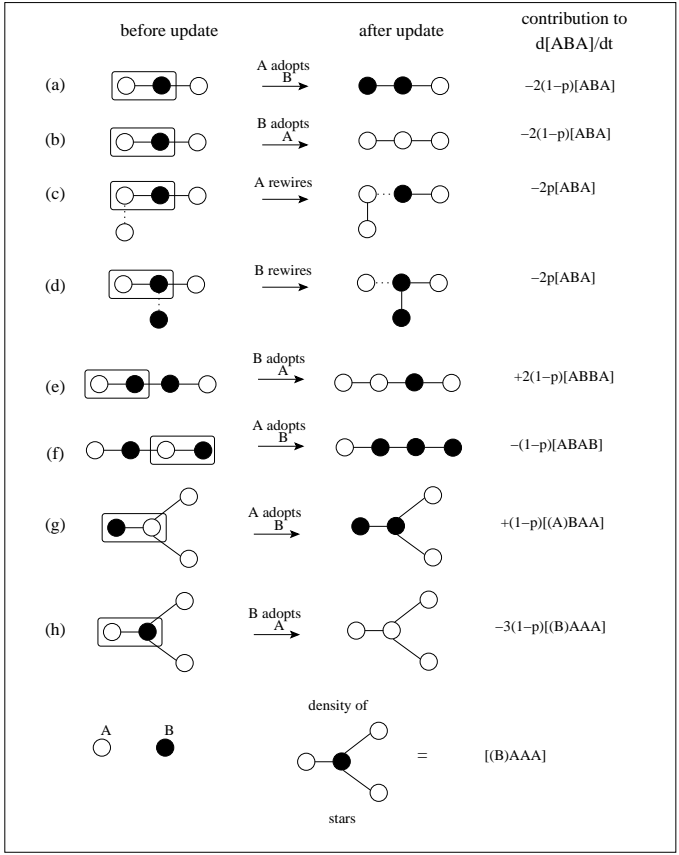}
  \caption{Second order moment expansion for $d[ABA]/dt$ in the link update rule.}
  \label{fig:secondorder}
\end{figure}

For illustration, we derive the corresponding contributions to $d[ABA]/dt$ for processes (e) and (h) in Fig.~\ref{fig:secondorder}.

In the process (e), an ABA-triplet is created per each AB-link connected through its B-node to the B-end of an AB-link on which B-node adopts state A. We denote the A (B) end of a randomly selected AB-link as node X (Y) and a random excess neighbor 
of node Y as node Z. The expected number of AB-links connected through its B-node to node Y is $Q(AB|BA)= P(AB|BA)K_{BA}$, where $P(AB|BA)$ is the probability that node Z has state B and a random neighbor of node Z has state A and $K_{BA}$ is 
the expected number of excess triplets attached to node Y. 

Node Y has degree $k$ with probability $kP_{k}^{B}/([B]\langle k_{B} \rangle)$. Assuming neutral mixing by degree, node Z has degree $k'$ with probability $k'P_{k'}/\langle k \rangle$. 
Thus, $K_{BA} = \left(\sum_{k}{(k-1)kP_{k}^{B}/([B]\langle k_{B} \rangle})\right)\times\left(\sum_{k'}{(k'-1)k'P_{k'}/\langle k \rangle}\right) = \langle q_{B} \rangle \langle q \rangle$.  In the symmetric state, $K_{BA} = \langle q \rangle^{2}$. 

Analogous to the expression of $P(A|BA)$ in \ref{sec:appendix A}, we write the conditional probability $P(AB|BA)$ as a fraction of appropriate network moments, i.e. $P(AB|BA)=2[ABBA]/$ $(2[ABBA]+[ABBB]+[ABAB]+[ABAA])$. In order to approximate 
the denominator, we use the second-order approximation for chain-quadruplets, which is derived in \ref{sec:appendix C} . Following this approximation, 
\begin{eqnarray*} \renewcommand{\arraystretch}{2}
& &  \hspace{-0.6in}2[ABBA] + [ABBB] + [ABAB]+[ABAA]  \\
& & \hspace{-0.3in} \simeq \frac{[ABB]([ABB]+2[BBB])}{2[BB]}+\frac{2[ABA]([AAB]+2[BAB])}{[AB]} \\
& & \hspace{-0.3in} = \langle q \rangle ([ABB]+2[ABA])\\
& & \hspace{-0.3in} = \langle q \rangle^{2} [AB].
\end{eqnarray*}
Thus, $P(AB|BA)=2[ABBA]/(\langle q \rangle^{2} [AB])$. Since  $K_{AB} = \langle q \rangle^{2}$, we reach 
\begin{equation}
 Q(AB|BA) = 2 [ABBA]/[AB].
\end{equation}

Similarly, 
\begin{eqnarray}\label{eq:indirect_linear}\renewcommand{\arraystretch}{2}
Q(AA|AB) & = & \frac{[AAAB]}{[AB]}, \nonumber\\ 
Q(AB|AB) & = & \frac{[ABAB]}{[AB]}, \nonumber\\ 
Q(BA|AB) & = & 2 \frac{[BAAB]}{[AB]}, \nonumber\\ 
Q(BB|AB) & = & \frac{[BBAB]}{[AB]}. 
\end{eqnarray}

Furthermore, in the process (h), an ABA-triplet becomes an AAA-triplet per every two A-neighbors of the B-end of a randomly selected AB-link. We call such subgraphs constituted by a node and its three neighbors as \emph{stars}. We again denote the A (B) end of a randomly selected AB-link as node X (Y). We denote the expected number of the configurations where 
node Y is connected to two additional A-nodes as $Q(A,A|BA) = P(A,A|BA)S_{B}$, where $P(A,A|BA)$ is the probability that two random neighbors of node Y both have state A, and $S_{B}$ is the average number of two-combinations of excess neighbors of node Y. 

Node Y has degree $k$ with probability $kP_{k}^{B}/([B]\langle k_{B} \rangle)$. The term $S_{B}$ is obtained from $S_{B} = \sum_{k}{(k-2)(k-1)kP_{k}^{B}/(2[B]\langle k_{B} \rangle})$. In terms of statistical moments of the degree distribution,
\begin{equation*}
S_{B} = \frac{\langle k_{B}^{3} \rangle -3 \langle k_{B}^{2} \rangle + 2\langle k_{B} \rangle}{2[B]\langle k_{B} \rangle}.
\end{equation*}

We write the conditional probability $P(A,A|BA)$ as a fraction of appropriate moments as before such that $P(A,A|BA)=3[(B)AAA]/(3[(B)AAA]+2[(B)AAB]+[(B)ABB])$. The density $[(B)AAA]$ is defined as $[(B)AAA] = N_{(B)AAA} / N$, where $N_{(B)AAA}$ is the total number of (B)AAA-stars (central B-node, three A-neighbors). 
The densities $[(B)AAB]=N_{(B)AAB}/N$ and $[(B)ABB]=N_{(B)ABB}/N$ are defined analogously. The number of stars are obtained from a summation over B-nodes such that $N_{(B)AAA} = \sum_{i\in\{B\}}^{}n_{i}(n_{i}-1)(n_{i}-2)/6$, $N_{(B)AAB} = \sum_{i\in\{B\}}^{}n_{i}(n_{i}-1)(k_{i}-n_{i})/2$, and 
$N_{(B)ABB} = \sum_{i\in\{B\}}^{}n_{i}(k_{i}-n_{i})(k_{i}-n_{i}-1)/2$, where $k_{i}$ and $n_{i}$ are the number of all neighbors and A-neighbors of node $i$ respectively. We reach the expression $3N_{(B)AAA} + 2N_{(B)AAB} + N_{(B)ABB}= \sum_{i\in\{B\}}n_{i}(k_{i}^{2}-3k_{i}+2)/2$ and
proceed with the summation
\begin{eqnarray*} \renewcommand{\arraystretch}{2}
\displaystyle \sum_{i\in\{B\}}n_{i}(k_{i}^{2}-3k_{i}+2) & = & \sum_{\begin{array}{c} \mbox{\footnotesize $i\in{B}$} \\ \mbox{\footnotesize $j\in{A}$} \\ \mbox{\footnotesize $(i,j)\in Edges$} \end{array}} (k_{i}^{2}-3k_{i}+2) \\
				& = & N [AB] \sum_{k}{\frac{(k-2)(k-1)kP_{k}^{B}}{2[B]\langle k_{B} \rangle}} \\
				& = & N [AB] \frac{\langle k_{B}^{3} \rangle -3 * \langle k_{B}^{2} \rangle + 2\langle k_{B} \rangle}{[B]\langle k_{B} \rangle}\\
				& = & 2 N [AB] S_{B}
\end{eqnarray*}
and obtain $3[(B)AAA] + 2[(B)AAB] + [(B)ABB] = [AB] S_{B}$. We reach 
\begin{equation}
Q(A,A|BA) = 3 [(B)AAA] / [AB].
\end{equation}

Similarly,
\begin{eqnarray}\label{eq:indirect_star}\renewcommand{\arraystretch}{2}
Q(A,A|AB) & = & \frac{[(A)BAA]}{[AB]}, \nonumber\\ 
Q(A,B|AB) & = & 2 \frac{[(A)BBA]}{[AB]}, \nonumber\\ 
Q(B,B|AB) & = & 3 \frac{[(A)BBB]}{[AB]}.
\end{eqnarray}

\section{Second-order moment closure approximation} \label{sec:appendix C}

In order to close the expansion Eq.~\ref{eq:ode_link_secondorder}, we need to express third-order moments in terms of lower order ones. 

We first derive the second-order approximation for the chain quadruplet density $[ABBA]$. We start with the BB-link at the center, which has density $[BB]$, and estimate the number of the A-nodes connected to the two ends, between which no correlation 
exists according to the second-order approximation. Each B-end has $[ABB]/2[BB]$ A-neighbors on average, as obtained in \ref{sec:appendix A}. We reach the following second-order approximation: 
$2[ABBA]\simeq0.5[ABB]^{2}/[BB]$. We note that $2$ appears on both sides, because we can't distinguish between the two B-nodes in the BB-link and the two ABB-triplets,

We derive the remaining terms for chain quadruplets analogously and obtain
\begin{eqnarray}\label{eq:secorderapprox_linear}\renewcommand{\arraystretch}{2}
{[AAAB]} & \simeq & \frac{[AAA][AAB]}{[AA]}, \nonumber\\
{[ABAA]} & \simeq & 2 \frac{[ABA][AAB]}{[AB]}, \nonumber\\
{[ABAB]} & \simeq & 4 \frac{[ABA][BAB]}{[AB]}, \nonumber\\
{[ABBA]} & \simeq & \frac{[ABB]^{2}}{4[BB]}. 
\end{eqnarray}

We now illustrate the second-order approximation for the star motif density $[(A)BBA]$, which is the density of subgraphs formed by an A-node in the center and its three (two B and one A) neighbors. We consider the star motif as an assembly of 
two triplets originating from the same end of a shared link. We start with locating an AB-link that appears at density $[AB]$. Each excess neighbor of the A-node has state A with probability $[AAB]/\langle q \rangle [AB]$ and state B with probability 
$2[BAB]/\langle q \rangle [AB]$. The expected number of star-quadruplets originating from one end of a random link is $1/2(\langle k^{3} \rangle -3 \langle k^{2} \rangle + 2 \langle k \rangle)/\langle k \rangle$, where $\langle k^{n} \rangle$ is the $n^{th}$ moment of the degree distribution. 
By using it, we obtain $2[(A)BBA] =  \langle k \rangle ((\langle k^{3} \rangle -3 \langle k^{2} \rangle + 2 \langle k \rangle)/( \langle k^{2} \rangle - \langle k \rangle)^{2})(2[BAB][AAB])/[AB]$. If we assume a Poissonian distribution, the second moment 
$\langle k^{2} \rangle = \langle k \rangle^{2} + \langle k \rangle$ and third moment $\langle k^{3} \rangle = \langle k \rangle ^{3} + 3 \langle k \rangle^{2} + \langle k \rangle$. Thence, we reach the approximation 
\begin{equation*}
{[(A)BBA]} \simeq \frac{[AAB][BAB]}{[AB]}.
\end{equation*}

However, approximations for star-quadruplet densities are not unique. It depends on which pair is considered to be shared by the two constituting triplets. Above we took the shared link to be an AB-link. We could also select the AA-link instead
as the shared link which leads to a different approximation. By considering such alternative ways of motif construction, star quadruplets can be approximated as
\begin{eqnarray}\label{eq:secorderapprox_star}\renewcommand{\arraystretch}{2}
{[(A)BAA]} & \simeq & \frac{[AAB]^{2}}{2[AB]} \  or \ \frac{[AAB][AAA]}{2[AA]}, \nonumber\\
{[(A)BBA]} & \simeq & \frac{[AAB]^{2}}{4[AA]} \ or \ \frac{[AAB][BAB]}{[AB]}, \nonumber\\
{[(A)BBB]} & \simeq & \frac{2[BAB]^{2}}{3[AB]}. 
\end{eqnarray}

\section{Reverse node update rule}\label{sec:appendix D}

Moment expansion equations for the reverse node update rule are very similar to those of the link update rule with the following changes in rates. 

An A-node adopts state B at rate $(1-p)[A][AB]/\langle k_{A} \rangle$ and  an AB-link is replaced by an BB-link through rewiring at rate $p[A][AB]/\langle k_{A} \rangle$. Analogously, a B-node adopts state A at rate $(1-p)[B][AB]/\langle k_{B} \rangle$ and 
an AB-link is replaced by an AA-link through rewiring at rate $p[B][AB]/\langle k_{B} \rangle$. 

The degree distributions of the nodes selected at an update event depend on the update rule. In the link update rule, both of the nodes at the two ends of a selected link have degree $k$ with the same probability $kP_{k}/\langle k \rangle$. In the reverse node update rule, 
the first selected node (node X) has degree $k'$ with probability $P_{k'}$ and its random neighbor (node Y) has degree $k$ with probability $kP_{k}/\langle k \rangle$. The degree distribution of node X does not show up in the first-order equation and the degree
distribution of node Y is the same as in the link update rule. Therefore, the first-order equation for the reverse node update rule differs from that of the link update rule due to the changes of rates of events only
\begin{eqnarray}
\label{eq:ode_reverse}\renewcommand{\arraystretch}{2}
\displaystyle \frac{\rm d}{\rm dt} [A] & =& \displaystyle (1-p)\left(\frac{[AB]}{\langle k_{A} \rangle} - \frac{[AB]}{\langle k_{B} \rangle}\right),  \nonumber\\
\displaystyle \frac{\rm d}{\rm dt} [AA]  & = & \displaystyle (1-p) \bigg(\frac{[AB]}{\langle k_{A} \rangle} + \frac{[AB]^{2}}{\langle k_{A} \rangle [A]} - \frac{2[AB][BB]}{\langle k_{B} \rangle [B]}\bigg) \nonumber\\
& & \displaystyle +  p \left(\frac{[AB]}{\langle k_{B} \rangle}\right) , \nonumber\\
\displaystyle \frac{\rm d}{\rm dt} [BB]  &=& \displaystyle (1-p) \bigg( \frac{[AB]}{\langle k_{B} \rangle} + \frac{[AB]^{2}}{\langle k_{A} \rangle [B]} - \frac{2[AB][AA]}{\langle k_{A} \rangle [A]} \bigg) \nonumber \\
& & \displaystyle + p \left( \frac{[AB]}{\langle k_{A} \rangle} \right).
\end{eqnarray}

At the symmetric state $[A]=[B]=1/2$, Eq.~\ref{eq:ode_reverse} is equivalent to Eq.~\ref{eq:ode_pair_apprx} with the rescaled time $t'=\langle k \rangle t$. 

\section{Direct node update rule} \label{sec:appendix E}

An A-node adopts state A at rate $(1-p)[AB]/\langle k_{A} \rangle$ and an AB-link is replaced by an AA-link through rewiring at rate $p [AB]/\langle k_{A} \rangle$. Similarly, a B-node adopts state A at rate $(1-p)[AB]/\langle k_{B} \rangle$ and an 
AB-link is replaced by a BB-link at rate $p [AB]/\langle k_{B} \rangle$.

To estimate the indirect contribution term, we use the quantity $Q(A|A_1 B_2)$ that represents the average number of A-neighbors of an A-node given that it already has a B-neighbor. Indices $1$ and $2$ indicate that the A-node 
was chosen first and the B-node was chosen after that. Given that node X  is chosen at random in the direct node update rule, it has degree $k$ with probability $P_{k}$. We also know that X has a B-neighbor, thus the probability that each of the $k-1$ remaining neighbors 
has state A can be estimated as $[AAB]/\langle q_{A} \rangle [AB]$.  Averaging over the entire network we obtain 
\begin{equation*}
\label{eq:directtripletappr}
Q(A|A_1 B_2)  = \sum_{k}  (k-1) P_k \frac{[AAB]}{\langle q_{A} \rangle [AB]}= \frac{\langle k_{A} \rangle - 1}{\langle q_{A} \rangle}\frac{[AAB]}{[AB]}.
\end{equation*}

Then, the following set of equations is reached for the direct node update rule:

\begin{eqnarray*}\renewcommand{\arraystretch}{2}
\displaystyle \frac{\rm d}{\rm dt} [A] & = & \displaystyle (1-p)\left(\frac{[AB]}{\langle k_{B} \rangle}-\frac{[AB]}{\langle k_{A} \rangle} \right), \\
\displaystyle \frac{\rm d}{\rm dt} [AA]  & = & \displaystyle (1-p) \Bigg\{ \frac{[AB]}{\langle k_{B} \rangle}+ \frac{2\left(\langle k_{B} \rangle -1 \right)[ABA]}{\langle q_{B} \rangle \langle k_{B} \rangle} \\
					& & - \frac{\left(\langle k_{A} \rangle -1\right) [AAB]}{\langle q_{A} \rangle   \langle k_{A} \rangle} \Bigg\} + p \frac{[AB]}{\langle k_{A} \rangle},
\end{eqnarray*}
\begin{eqnarray}\label{eq:ode_direct_triplet}\renewcommand{\arraystretch}{2}
\displaystyle \frac{\rm d}{\rm dt} [BB]  & = & \displaystyle (1-p) \Bigg\{ \frac{[AB]}{\langle k_{A} \rangle}+ \frac{2\left(\langle k_{A} \rangle -1 \right)[BAB]}{\langle q_{A} \rangle \langle k_{A} \rangle} \nonumber\\ 
					& & - \frac{\left(\langle k_{B} \rangle -1\right) [BBA]}{\langle q_{B} \rangle   \langle k_{B} \rangle} \Bigg\} + p \frac{[AB]}{\langle k_{B} \rangle}.
\end{eqnarray}

At the steady-state, the link density $[AB]$ can be expressed in terms of triplet density in Eq.~\ref{eq:ode_direct_triplet} 
\begin{equation}
[AB] = \frac{(1-p)( \langle k \rangle -1)}{\kappa \langle k \rangle} \left( [AAB] - 2 [ABA]\right).
\end{equation}

However, this requires numerical values of the triplet densities that should be measured from the agent-based simulations, which is undesirable as explained in the main text. Instead, we develop a first-order moment closure approximation. 
Since node X is selected without resorting to the information of node Y, node Y should be treated as a sample from the neighbors of node X. An alternative solution to this problem is to ignore the information from node Y and use a pair 
approximation. Then, any neighbor of node X has state A with probability $2[AA]/[A] \langle k \rangle_A$. The pair approximation formulation for $Q(A|A_1 B_2)$ is  

\begin{eqnarray*}
\displaystyle Q(A|A_1 B_2) \simeq Q(A|A_1) &=& \displaystyle \sum \limits_{k} P_k (k-1) \frac{2[AA]}{[A] \langle k_{A} \rangle} \nonumber \\
&=& \frac{ 2\left( \langle k_{A} \rangle - 1 \right)}{\langle k_{A} \rangle} \frac{[AA]}{[A]}.
\end{eqnarray*}

This is equivalent to using the pair approximation for the density of triplets, such as

\begin{equation*}
[AAB] \simeq 2\kappa_{A}\frac{ [AA] [AB]}{[A]}. 
\end{equation*}

Replacing this expression for $[AAB]$ and the analogous expressions for the other densities of triplets in Eq.~\ref{eq:ode_direct_triplet}, we arrive at

\begin{eqnarray}\renewcommand{\arraystretch}{2}
\displaystyle \frac{\rm d}{\rm dt} [A] & = & \displaystyle (1-p)\left(\frac{[AB]}{\langle k_{B} \rangle}-\frac{[AB]}{\langle k_{A} \rangle} \right), \nonumber\\
\displaystyle \frac{\rm d}{\rm dt} [AA]  & = & \displaystyle (1-p) \Bigg\{ \frac{[AB]}{\langle k_{B} \rangle}+ \frac{\left(\langle k_{B} \rangle -1 \right)[AB]^2}{[B] \langle k_{B} \rangle^2}\nonumber\\ 
					& & - \frac{2 \left(\langle k_{A} \rangle -1\right) [AA] [AB]} {[A] \langle k_{A} \rangle^2 } \Bigg\} + p \frac{[AB]}{\langle k_{A} \rangle}, \nonumber\\
\displaystyle \frac{\rm d}{\rm dt} [BB]  & = & \displaystyle (1-p) \Bigg\{ \frac{[AB]}{\langle k_{A} \rangle}+ \frac{\left(\langle k_{A} \rangle -1 \right)[AB]^2}{[A] \langle k_{A} \rangle^2} \nonumber\\ 
					& & - \frac{2 \left(\langle k_{B} \rangle -1\right) [BB] [AB]} {[B] \langle k_{B} \rangle^2 } \Bigg\} + p \frac{[AB]}{\langle k_{B} \rangle}.
\end{eqnarray}

\section{Derivation of pair approximation for the fraction of active triplets } \label{sec:appendix F}

Let's denote $B_{k}^{m}$ as the fraction of B-nodes with k total (A and B) and $m$ A neighbors with the normalization $\sum_{k}\sum_{m=0}^{k}B_{k}^{m} = 1$. The fraction $B_{k}^{m}$ is expressed as $B_{k}^{m} = P_{k}^{B}P(m,k)/[B]$, where $P_{k}^{B}/[B]$ is the fraction of 
B-nodes with degree $k$ and $P(m,k)$ is the probability that $m$ links out of $k$ are active. For an infinite Erd\H{o}s-Renyi random graph, $P_{k}^{B}/[B]$ is Poisson distributed, i.e. $P_{k}^{B}/[B]=\langle k \rangle ^{k} e^{-\langle k \rangle} / k!$. 
If  we neglect state correlations to second nearest-neighbors, then every link connected to a B-node is active with probability $\sigma_{p,\langle k \rangle} = [AB]_{p,\langle k \rangle}^{s}/([B] \langle k \rangle) = 2 [AB]_{p,\langle k \rangle}^{s}/\langle k \rangle $, 
at $[B]=1/2$, where we measure the average active link density $[AB]_{p,\langle k \rangle}^{s}$ at a specific $p$ and $\langle k \rangle$. When second nearest-neighbor correlations are ignored, $P(m,k)$ is a binomial distribution, i.e. 
$P(m,k)= (m!(k-m)!/k!)\sigma_{p,\langle k \rangle}^{m}(1-\sigma_{p,\langle k \rangle})^{k-m}$. Therefore, in the uncorrelated infinite-size Erd\H{o}s-Renyi case $B_{k}^{m}$ obeys 
\begin{equation}
B_{k}^{m} =  \frac{ m!(k-m)!e^{- \langle k \rangle}\langle k \rangle ^{k}}{(k!)^{2}} \sigma_{p,\langle k \rangle}^{m}(1-\sigma_{p,\langle k \rangle})^{k-m}. 
\end{equation}

The quantity $T_{ABA}^{m}$ is the fraction of ABA-triplets that have a B-node with $m$ A-neighbors. By definition,  
\begin{equation}
T_{ABA}^{m} = \sum_{k\geq m} 1/2 m(m-1)B_{k}^{m}. 
\end{equation}

\section{Derivation of the heterogeneous pair approximation} \label{sec:appendix G}

Here, we first derive the rate equation for the average change in the density of active links $[A_k B_{k'}]$ when a node with state A is chosen. We rewrite Eq.~(\ref{dABkk'dt}) as

\begin{equation}
\left. \frac{d [A_k B_{k'}]}{dt} \right|_A = \sum_{l=1}^{\overline k} \frac{[A]_l}{l} \sum_{\mathcal N_1=0}^l ... \sum_{\mathcal N_{\overline k}=0}^l ~ M(\mathcal N_1,..,\mathcal N_{\overline k};l)~ \mathcal S(\mathcal N_1,..,\mathcal N_{\overline k}), 
\label{dABkk'dt3}
\end{equation}
where 
\begin{eqnarray}\label{S}
\hspace{-1cm} \mathcal S(\mathcal N_1,..,\mathcal N_{\overline k}) & \equiv & \sum_{n_1=0}^{\mathcal N_1} ...  \sum_{n_{\overline k}=0}^{\mathcal N_{\overline k}} ~ \prod_{m=1}^{\overline k} B(n_{m};\mathcal N_{m}) \nonumber \\
&\Bigg\{&  (1-p) \bigg[ \mathcal N_{k} n_{k} - n_{k}^2 + (\mathcal N_{k}- n_{k}) \sum_{m\neq k}^{\overline k} n_{m} \bigg] \delta_{l,k'} \nonumber \\  
&-& (1-p) \bigg[ n_{k'}^2 + n_{k'} \sum_{m \neq k'} n_m \bigg] \delta_{l,k}- p n_{k'} \delta_{l,k} \nonumber \\
&+& p n_{k'+1} N(A_k|B_{k'+1 }A_l) - p\, n_{k'} N(A_k|B_{k'} A_l) \nonumber \\ 
&+& p \left[ \frac{[A]_{k-1} N(B_{k'}|A_{k-1})}{[A]} - \frac{[A]_k N(B_{k'}|A_k)}{[A]} \right] \nonumber \\
&& \ \ \ \times \sum_{m=1}^{\overline k} n_{m}\Bigg\}. 
\end{eqnarray}

We have also replaced the number of active links $n$ around the chosen node by $\sum_{m=1}^{\overline k} n_{m}$.  To carry out the summations in Eqs.~(\ref{dABkk'dt3}) and (\ref{S}), we assume that the network has no degree correlations, thus 
the probability that a given node has a neighbor of degree $m$ is $Q_{m} = m\,P_{m}/\langle k \rangle$. Then, the probabilities $M$ and B become the multinomial and binomial distributions, respectively, 
\begin{eqnarray}
&& \hspace{-1cm} M\left(\mathcal N_1,...,\mathcal N_{\overline k};l;Q_1,...,Q_{\overline k} \right) = \nonumber \\ 
&&\begin{cases} 
\frac{l!}{\mathcal N_1!...\mathcal N_{\overline k!}} \, Q_1^{\mathcal N_1}\,..\, Q_{\overline k}^{\mathcal N_{\overline k}} & \text{when $\sum_{m=1}^{\overline k} \mathcal N_{m}=l$;} \nonumber \\
0 & \text{otherwise,} 
\end{cases} \\
\label{Multi}
\end{eqnarray} 
and
\begin{eqnarray}
B(n_{m};\mathcal N_{m}) =
\frac{\mathcal N_{m} ! ~ q_{m|l}^{n_{m}} (1-q_{m|l})^{\mathcal
N_{m}-n_{m}} }{n_{m} ! (\mathcal N_{m}-n_{m})!}.
\label{Bino}
\end{eqnarray}

Here, $q_{m|l} = P(B | m;A,l) \simeq [A_l B_{m}]/ l Q_{m}[A]_l$ is the conditional probability that a neighbor of a node in class $(A,l)$ that has degree $m$ is in state B. This probability is estimated as the ratio $N(A,l \to B,m)/N(A,l \to m)$ 
between the number of links $N(A,l \to B,m)= [A_l B_{m}] N$ from nodes of class $(A,l)$ to nodes of class $(B,m)$, and the number of links $N(A,l \to m)= [A]_l N l Q_{m}$ from nodes of class $(A,l)$ to nodes of degree $m$ and state A or B. 
The multiple summation in Eq.~(\ref{S}), weighted by the product of the binomials, leads to the first and second moments, $\langle n_{m} \rangle = \sum_{m} B(n_{m};\mathcal N_{m}) n_{m}$ and $\langle n_{m}^2 \rangle = \sum_{m} B(n_{m};\mathcal N_{m}) n_{m}^2$,
respectively, obtaining 
\begin{eqnarray}\label{S1}
& & \hspace{-2.0cm} \mathcal S(\mathcal N_1,..,\mathcal N_{\overline k}) = (1-p) \bigg[ \mathcal N_{k} \langle n_{k} \rangle - \langle n_{k}^2 \rangle + (\mathcal N_{k}- \langle n_{k} \rangle ) \sum_{m\neq k}^{\overline k} \langle n_{m} \rangle \bigg] \delta_{l,k'} \nonumber \\
&-& (1-p) \bigg[ \langle n_{k'}^2 \rangle + \langle n_{k'} \rangle \sum_{m \neq k'} \langle n_m \rangle \bigg] \delta_{l,k} \nonumber \\ 
&-& p \langle n_{k'} \rangle \delta_{l,k} + p \langle n_{k'+1} \rangle N(A_k|B_{k'+1 }A_l) \nonumber \\
&-& p \langle n_{k'} \rangle N(A_k|B_{k'} A_l) \nonumber \\
&+&p \left[ \frac{[A]_{k-1} N(B_{k'}|A_{k-1})}{[A]} - \frac{[A]_k N(B_{k'}|A_k)}{[A]} \right] \nonumber \\
& & \ \ \ \ \ \ \times \ \sum_{m=1}^{\overline k} \langle n_{m} \rangle \nonumber \\
&=&  (1-p) \bigg[ q_{k|l} (q_{k|l}-1) \mathcal N_{k} + (1-q_{k|l}) \mathcal N_{k} \sum_{m}^{\overline k} q_{m|l} \mathcal N_{m} \bigg] \nonumber \\
& & \ \ \ \ \ \ \ \ \  \ \ \ \ \ \times \ \delta_{l,k'} \nonumber \\
&-& (1-p) \bigg[ (1-q_{k'|l}) q_{k'|l} \mathcal N_{k'} + q_{k'|l} \mathcal N_{k'} \sum_m q_{m|l} \mathcal N_m \bigg] \delta_{l,k} \nonumber \\  
&-& p q_{k'|l} \mathcal N_{k'} \delta_{l,k} + p N(A_k|B_{k'+1 }A_l) q_{k'+1|l} \mathcal N_{k'+1} \nonumber \\
&-& p N(A_k|B_{k'} A_l) q_{k'|l} \mathcal N_{k'}  \nonumber \\
&+& p \left[ \frac{[A]_{k-1} N(B_{k'}|A_{k-1})}{[A]} - \frac{[A]_k N(B_{k'}|A_k)}{[A]} \right] \sum_{m=1}^{\overline k} q_{m|l} \mathcal N_{m}, \nonumber \\    
\end{eqnarray}                 
where we have used the expression for the moments $\langle n_{m} \rangle =  q_{m|l} \mathcal N_{m}$ and $\langle n_{m}^2 \rangle = q_{m|l} \mathcal N_{m} + q_{m|l}^2 \mathcal N_{m} (\mathcal N_{m} -1)$ of the binomials defined in Eq.~(\ref{Bino}).  
Now, inserting expression~(\ref{S1}) for $\mathcal S$ in Eq.~(\ref{dABkk'dt3}) we obtain

\begin{eqnarray}\label{dABkk'dt4}
& & \hspace{-1.9cm} \left. \frac{d [A_k B_{k'}]}{dt} \right|_A = \sum_{l=1}^{\overline k} \frac{[A]_l}{l} \Bigg\{ (1-p) \bigg[ (q_{k|l}-1) q_{k,l} \langle \mathcal N_{k} \rangle \nonumber \\
&&  \ \ \ \ \ \ \ \ \ \ \ \ \ \ \ \ \ \ \ \ \ \ \ \ \ \ \ \ \ \ + (1- q_{k|l}) \sum_{m=1}^{\overline k}  q_{m|l} \langle \mathcal N_{k} \mathcal N_{m} \rangle \bigg] \delta_{l,k'} \nonumber \\ 
&-&(1-p) \bigg[ (1-q_{k'|l}) q_{k'|l} \langle \mathcal N_{k'} \rangle + q_{k'|l} \sum_m q_{m|l} \langle \mathcal N_{k'} \mathcal N_m \rangle \bigg] \nonumber \\
& &   \ \ \ \ \ \ \ \ \ \ \ \ \ \ \times \ \delta_{l,k} \nonumber \\
&-& p q_{k'|l} \langle \mathcal N_{k'} \rangle \delta_{l,k} + p N(A_k|B_{k'+1 }A_l) q_{k'+1|l} \langle \mathcal N_{k'+1} \rangle \nonumber \\ 
&-& p N(A_k|B_{k'} A_l) q_{k'|l} \langle \mathcal N_{k'} \rangle  \nonumber \\
&+& p \left[ \frac{[A]_{k-1} N(B_{k'}|A_{k-1})}{[A]} - \frac{[A]_k N(B_{k'}|A_k)}{[A]} \right] \nonumber \\
& &   \ \ \ \ \ \ \ \ \ \ \times \ \sum_{m=1}^{\overline k} q_{m|l} \langle \mathcal N_{m} \rangle \Bigg\}. 
\end{eqnarray}

Then, using the pair approximation to estimate the number of nodes in class $(B,k')$ connected to a node in class $(A,k)$ as $N(B_{k'}|A_k) \simeq k Q_{k'} q_{k'|k}$ and the additional number of nodes in class $(A,k)$ attached to the B-node of an 
$A_l B_{k'}$-link as $N(A_k | B_{k'} A_l) \simeq N(A_k | B_{k'}) \simeq (k'-1) Q_k r_{k|k'}$, with $r_{k|k'} = P(A | k;B,k')\simeq [A_k B_{k'}]/$ $(k' Q_{k}[B]_{k'})$, we arrive to
\begin{eqnarray}\label{dABkk'dt5}
&& \hspace{-1.9cm} \left. \frac{d [A_k B_{k'}]}{dt} \right|_A = \sum_{l=1}^{\overline k} [A]_l \Bigg\{ (1-p) (l-1) \sum_{m=1}^{\overline k} q_{m|l} Q_{m} \nonumber \\
& &   \ \ \ \  \ \ \ \ \ \ \ \ \ \ \ \times \bigg[ (1- q_{k|l}) Q_{k} \delta_{l,k'} - q_{k'|l} Q_{k'} \delta_{l,k} \bigg] \nonumber \\ 
&-& q_{k'|l} Q_{k'} \delta_{l,k} + p k'r_{k|k'+1}  q_{k'+1|l} Q_k Q_{k'+1|l} \nonumber \\
&-& p (k'-1)r_{k|k'} q_{k'|l} Q_k Q_{k'} \nonumber \\
&+& p \left[ \frac{(k-1) q_{k'|k-1} Q_{k'} [A]_{k-1}}{[A]} - \frac{k q_{k'|k} Q_{k'} [A]_k}{[A]} \right] \nonumber \\
& &   \ \ \ \ \ \times \ \sum_{m=1}^{\overline k} q_{m|l} Q_{m} \Bigg\},
\end{eqnarray}

where we have used the following expressions for the moments of the multinomial distribution $M$, defined in Eq.~(\ref{Multi}):
\begin{eqnarray}
\langle \mathcal N_{k'} \rangle &\equiv& \sum_{N_1=0}^l ... \sum_{N_{\overline k}=0}^l ~ M(\mathcal N_1,..,\mathcal N_{\overline k};l)~\mathcal N_{k'} = Q_{k'} l  \nonumber \\
\langle \mathcal N_{k'} \mathcal N_{m} \rangle &\equiv& \sum_{N_1=0}^l ... \sum_{N_{\overline k}=0}^l ~ M(\mathcal N_1,..,\mathcal N_{\overline k};l)~ \mathcal N_{k'} \mathcal N_{m} \nonumber \\
&=& \begin{cases}
 Q_{k'} Q_{m} l (l-1)  & \text{for $k' \neq m$;} \nonumber \\
Q_{k'} l + Q_{k'}^2 l (l-1) & \text{for $k'=m$}.
\end{cases} \\
\end{eqnarray}

Using in Eq.~(\ref{dABkk'dt5}) the expression for the probabilities $q_{k'|l}$  and $r_{k|k'}$, and expressing the sum $\sum_{m} q_{m|l} Q_{m}$ as $[A_l B]/ l [A]_l$, with $[A_l B] \equiv \sum_{m=1}^{\overline k} [A_l B_{m}]$, we arrive at the
expression Eq.~(\ref{dABkk'dt2}).
\begin{eqnarray}
\label{dABkk'dt2}
\hspace{-1.0cm} \left. \frac{d [A_k B_{k'}]}{dt} \right|_A & = & (1-p)\frac{k'-1}{k'} Q_{k} \left( 1 - \frac{[A_{k'} B_{k}]}{k' Q_{k} [A]_{k'}} \right) [A_{k'} B] \nonumber \\
&-& (1-p) \frac{k-1}{k^2} \frac{[A_k B_{k'}] [A_k B]}{[A]_k}-\frac{[A_k B_{k'}]}{k} \nonumber \\
&+& p \Bigg\{ \frac{k'}{k'+1} \frac{[A_k B_{k'+1}]\{ A B_{k'+1} \}}{[B]_{k'+1}} \nonumber \\
&& \ \ \ \ \ - \frac{k'-1}{k'} \frac{[A_k B_{k'}]\{A B_{k'} \} }{[B]_{k'}} \nonumber \\
&& \ \ \ \ \ + \frac{ [A_{k-1} B_{k'}] - [A_k B_{k'}]}{[A]}  \{ A B \}	   
\Bigg\},  
\end{eqnarray}

where $Q_k \equiv k P_k /\langle k \rangle$, $[A_k B] \equiv \sum_{m=1}^{\overline k} [A_k B_{m}]$, \hspace{0.1cm} $\{A B_{k'}\} \equiv \sum_{l=1}^{\overline k} [A_{l} B_{k'}]/l$, and $\{A B \} \equiv \sum_{l=1}^{\overline k} [A_{l} B]/l$.

Given that $[A_k B_{k'}]$ may also change when a B-node is chosen, the evolution of $[A_k B_{k'}]$ is given by 
\begin{eqnarray*}
\left. \frac{d [A_k B_{k'}]}{dt} = \frac{d [A_k B_{k'}]}{dt} \right|_A + 
\left. \frac{d [A_k B_{k'}]}{dt} \right|_B.
\end{eqnarray*}

The second term on the right hand side can be obtained from Eq.~({\ref{dABkk'dt2}), by interchanging A and $k$ by B and $k'$, respectively. Adding the two contributions leads to 

\begin{eqnarray}
\hspace{-1.0cm} \frac{d [A_k B_{k'}]}{dt} &=& (1-p) \Bigg\{ Q_{k'} \frac{k-1}{k} [A B_k] + Q_{k} \frac{k'-1}{k'} [A_{k'} B] \nonumber \\
&& \ \ \ \ \  - \left( \frac{ [A B_k]}{[B]_k}\,\frac{k-1}{k^2} + \frac{[A_{k'} B]}{[A]_{k'}}\,\frac{k'-1}{k'^2} \right) [A_{k'} B_k] \nonumber \\ 
&& \ \ \ \ \ - \left( \frac{[A_k B]}{[A]_k} \frac{k-1}{k^2} + \frac{[A B_{k'}]}{[B]_{k'}} \frac{k'-1}{k'^2} \right) [A_k B_{k'}] \Bigg\} \nonumber \\
&+& \left( \frac{1}{k} + \frac{1}{k'} \right) [A_k B_{k'}] \nonumber \\
&+& p \Bigg\{ \frac{k'}{k'+1} \frac{[A_k B_{k'+1}] \{A B_{k'+1} \} }{[B]_{k'+1}} \nonumber \\
&& \ \ \ \ \ + \frac{k}{k+1} \frac{[A_{k+1} B_{k'}] \{ A_{k+1} B \}}{[A]_{k+1}} \nonumber \\
&& \ \ \ \ \  - \left( \frac{k'-1}{k'} \frac{\{A B_{k'}\}}{[B]_{k'}} + \frac{k-1}{k} \frac{\{A_k B\}}{[A]_k} \right) [A_k B_{k'}] \nonumber \\ 
&& \ \ \ \ \  + \left(  \frac{[A_{k-1} B_{k'}]}{[A]} + \frac{[A_k B_{k'-1}]}{[B]} \right) \{A B\} \nonumber \\
&& \ \ \ \ \  - \frac{[A_k B_{k'}] \{A B\}}{[A][B]}
\Bigg\}.
\end{eqnarray}

Finally, given that $[AB]_{k,k'}=[A_k B_{k'}]+[B_k A_{k'}]$, and using the symmetry between A and B states and the fact that in a time of order unity, a quasi-stationary state is established, in which the fraction of nodes in different 
degree classes $[A]_k/P_k$ ($[B]_k/P_k$) reach the value corresponding to the global density $[A]$ ($[B]$), we reach the final equation Eq.~(\ref{d[AB]kk'dt}).

\section{Derivation of the homogeneous moment expansion from active neighborhood approach} \label{sec:appendix H}

Summing over $k$ and $n$ and using the constraint $A_B = B_A$, Eq.~(\ref{dAkn})  leads to 
\begin{equation}
\frac{d A}{dt} = 0, ~~~\mbox{and}
\label{dAdt}
\end{equation}
\begin{equation}
\frac{d A_B}{dt} = \beta \left[ A_{AB} + B_{BA} - A_{BB} - B_{AA} \right] - 
2 \gamma A_B.
\label{dABdt}
\end{equation}
Equation~(\ref{dAdt}) expresses the conservation of the global density of
nodes in a given state, that is a well-known property of the voter model
under link update dynamics.  Equation~({\ref{dABdt}) can be associated to the
evolution of the density of AB-pairs, by writing the first and second
moments in terms of densities of pairs and triplets as we already defined
after Eq.~(\ref{second-mom}).
\begin{eqnarray}
\frac{d [AB]}{dt} &=& \beta \Big\{ [AAB] + [BBA] - 2 [AB] - 2 [ABA] 
\nonumber \\ 
&-& 2 [BAB] \Big\}- 2 \gamma [AB]
\label{dABdt-2} 
\end{eqnarray}
Eq.~(\ref{dABdt-2}) is the same as the equation derived for link update Eq.~(\ref{eq:ode_link_triplet}), 
showing the equivalence between parallel and link update dynamics. 

\section{Active motif approach: equations for spider motifs} \label{sec:appendix I}

The rate equations for spider densities $\{m,l\}$ for link update are given by
\begin{align*}
\frac{d}{dt}{\{m,l\}}=&-l\{m,l\}+\tfrac{1}{2}(l+1)\{m-1,l+1\}\\
+&\tfrac{1}{2}(1-p)m\{l,m\}+\tfrac{p}{2}(l+1)\{m,l+1\},
\end{align*}
where $m\neq1$, and
\begin{align*}
\frac{d}{dt}{\{1,l\}}=&-l\{1,l\}
+\tfrac{1}{2}(1-p)\{l,1\}+
\tfrac{p}{2}(l+1)\{1,l+1\}\\
&+\tfrac{1}{2}(1-p)\frac{e^{-\langle k\rangle}\langle k\rangle^{l+1}}{(l+1)!}\sum_{\{x,y\}}y\{x,y\}
\end{align*}
for $m=1$. 
Here we use a Poissonian degree distribution $P(k)=e^{-\langle k\rangle}\langle k\rangle^{k}/k!$ with mean degree $\langle k\rangle$. In order to obtain a Jacobian of finite size we consider only motifs with $m+l \leq k_{\rm max}$. 

The transition probabilities for node update rules differ from those for link update, depending on the degree of the chosen nodes. The corresponding equations for direct node update are given by
\begin{align*}
\frac{d}{dt}{\{m,l\}}=&-l\{m,l\}\\
+&(1-p)(l+1)\{m-1,l+1\}\sum_gP_{\langle k\rangle}(g+1)\alpha_{m,l}(g)\\
+&p(l+1)\{m-1,l+1\}\sum_gP_{\langle k\rangle}(g+1)\beta_{m,l}(g)\\
\end{align*}
\begin{align*}
+&(1-p)m\{l,m\}\sum_gP_{\langle k\rangle}(g+1)\beta_{m,l}(g)\\
+&p(l+1)\{m,l+1\}\sum_gP_{\langle k\rangle}(g+1)\alpha_{m,l}(g),\\
\frac{d}{dt}{\{1,l\}}=&-l\{1,l\}\\
+&(1-p)\{l,1\}\sum_gP_{\langle k\rangle}(g+1)\beta_{1,l}(g)\\
+&p(l+1)\{1,l+1\}\sum_gP_{\langle k\rangle}(g+1)\alpha_{1,l}(g)\\
+&(1-p)P_{\langle k\rangle}(l+1)\alpha_{1,l}(l)\sum_{\{x,y\}}y\{x,y\},
\end{align*}
where $\alpha_{m,l}(g)=(l+m)/(l+m+g+1)$ and $\beta_{m,l}(g)=(g+1)/(l+m+g+1)$. 

The equations look similar for reverse node update.

\bibliographystyle{model1-num-names}
\bibliography{refs}

\begin{thebibliography}{124}
\expandafter\ifx\csname natexlab\endcsname\relax\def\natexlab#1{#1}\fi
\providecommand{\url}[1]{\texttt{#1}}
\providecommand{\href}[2]{#2}
\providecommand{\path}[1]{#1}
\providecommand{\DOIprefix}{doi:}
\providecommand{\ArXivprefix}{arXiv:}
\providecommand{\URLprefix}{URL: }
\providecommand{\Pubmedprefix}{pmid:}
\providecommand{\doi}[1]{\href{http://dx.doi.org/#1}{\path{#1}}}
\providecommand{\Pubmed}[1]{\href{pmid:#1}{\path{#1}}}
\providecommand{\bibinfo}[2]{#2}
\ifx\xfnm\relax \def\xfnm[#1]{\unskip,\space#1}\fi
\bibitem[{Albert and Barab{\'a}si(2002)}]{Albert2002}
\bibinfo{author}{R.~Albert}, \bibinfo{author}{A.-L. Barab{\'a}si},
\newblock \bibinfo{title}{Statistical mechanics of complex networks},
\newblock \bibinfo{journal}{Rev. Mod. Phys.} \bibinfo{volume}{74}
  (\bibinfo{year}{2002}) \bibinfo{pages}{47--97}.
\bibitem[{Newman(2003)}]{Newman2003}
\bibinfo{author}{M.~E.~J. Newman},
\newblock \bibinfo{title}{The structure and function of complex networks},
\newblock \bibinfo{journal}{SIAM Review} \bibinfo{volume}{45}
  (\bibinfo{year}{2003}) \bibinfo{pages}{167--256}.
\bibitem[{Newman et~al.(2006)Newman, Barab{\'a}si, and Watts}]{Newman2006}
\bibinfo{editor}{M.~E.~J. Newman}, \bibinfo{editor}{A.-L. Barab{\'a}si},
  \bibinfo{editor}{D.~J. Watts} (Eds.), \bibinfo{title}{The Structure and
  Dynamics of Networks}, \bibinfo{publisher}{Princeton University Press},
  \bibinfo{address}{Princeton, NJ}, \bibinfo{year}{2006}.
\bibitem[{Boccaletti et~al.(2006)Boccaletti, Latora, Moreno, Chavez, and
  Hwang}]{Boccaletti2006}
\bibinfo{author}{S.~Boccaletti}, \bibinfo{author}{V.~Latora},
  \bibinfo{author}{Y.~Moreno}, \bibinfo{author}{M.~Chavez},
  \bibinfo{author}{D.-U. Hwang},
\newblock \bibinfo{title}{Complex networks: structure and dynamics},
\newblock \bibinfo{journal}{Physics Reports} \bibinfo{volume}{424}
  (\bibinfo{year}{2006}) \bibinfo{pages}{175--308}.
\bibitem[{Dorogovtsev et~al.(2008)Dorogovtsev, Goltsev, and
  Mendes}]{Dorogovtsev2008}
\bibinfo{author}{S.~N. Dorogovtsev}, \bibinfo{author}{A.~V. Goltsev},
  \bibinfo{author}{J.~F.~F. Mendes},
\newblock \bibinfo{title}{Critical phenomena in complex networks},
\newblock \bibinfo{journal}{Rev. Mod. Phys.} \bibinfo{volume}{80}
  (\bibinfo{year}{2008}) \bibinfo{pages}{1275--1335}.
\bibitem[{Castellano et~al.(2009)Castellano, Fortunato, and
  Loreto}]{Castellano2009}
\bibinfo{author}{C.~Castellano}, \bibinfo{author}{S.~Fortunato},
  \bibinfo{author}{V.~Loreto},
\newblock \bibinfo{title}{Statistical physics of social dynamics},
\newblock \bibinfo{journal}{Rev. Mod. Phys.} \bibinfo{volume}{81}
  (\bibinfo{year}{2009}) \bibinfo{pages}{591--646}.
\bibitem[{Barab{\'a}si and Albert(1999)}]{Barabasi1999}
\bibinfo{author}{A.-L. Barab{\'a}si}, \bibinfo{author}{R.~Albert},
\newblock \bibinfo{title}{Emergence of scaling in random networks},
\newblock \bibinfo{journal}{Science} \bibinfo{volume}{286}
  (\bibinfo{year}{1999}) \bibinfo{pages}{509--512}.
\bibitem[{Watts and Strogatz(1998)}]{Watts1998}
\bibinfo{author}{D.~J. Watts}, \bibinfo{author}{S.~H. Strogatz},
\newblock \bibinfo{title}{Collective dynamics of 'small-world' networks},
\newblock \bibinfo{journal}{Nature} \bibinfo{volume}{393}
  (\bibinfo{year}{1998}) \bibinfo{pages}{440--442}.
\bibitem[{Pastor-Satorras and Vespignani(2001)}]{Pastor-Satorras2001}
\bibinfo{author}{R.~Pastor-Satorras}, \bibinfo{author}{A.~Vespignani},
\newblock \bibinfo{title}{Epidemic spreading in scale-free networks},
\newblock \bibinfo{journal}{Phys. Rev. Lett.} \bibinfo{volume}{86}
  (\bibinfo{year}{2001}) \bibinfo{pages}{3200--3203}.
\bibitem[{Gross and Blasius(2008)}]{Gross2008}
\bibinfo{author}{T.~Gross}, \bibinfo{author}{B.~Blasius},
\newblock \bibinfo{title}{Adaptive coevolutionary networks: a review},
\newblock \bibinfo{journal}{J. R. Soc. Interface} \bibinfo{volume}{5}
  (\bibinfo{year}{2008}) \bibinfo{pages}{259--271}.
\bibitem[{Gross and Sayama(2009)}]{Gross2009}
\bibinfo{editor}{T.~Gross}, \bibinfo{editor}{H.~Sayama} (Eds.),
  \bibinfo{title}{Adaptive Networks. Theory, Models and Applications},
  \bibinfo{publisher}{Springer Verlag}, \bibinfo{address}{New York},
  \bibinfo{year}{2009}.
\bibitem[{Holme and Newman(2006)}]{Holme2006}
\bibinfo{author}{P.~Holme}, \bibinfo{author}{M.~E.~J. Newman},
\newblock \bibinfo{title}{Nonequilibrium phase transition in the coevolution of
  networks and opinions},
\newblock \bibinfo{journal}{Phys. Rev. E} \bibinfo{volume}{74}
  (\bibinfo{year}{2006}) \bibinfo{pages}{056108}.
\bibitem[{Vazquez et~al.(2008)Vazquez, Eguiluz, and Miguel}]{Vazquez2008}
\bibinfo{author}{F.~Vazquez}, \bibinfo{author}{V.~M. Eguiluz},
  \bibinfo{author}{M.~S. Miguel},
\newblock \bibinfo{title}{Generic absorbing transition in coevolution
  dynamics},
\newblock \bibinfo{journal}{Phys. Rev. Lett.} \bibinfo{volume}{100}
  (\bibinfo{year}{2008}) \bibinfo{pages}{108702}.
\bibitem[{Nardini et~al.(2008)Nardini, Kozma, and Barrat}]{Nardini2008}
\bibinfo{author}{C.~Nardini}, \bibinfo{author}{B.~Kozma},
  \bibinfo{author}{A.~Barrat},
\newblock \bibinfo{title}{Who's talking first? consensus or lack thereof in
  coevolving opinion formation models},
\newblock \bibinfo{journal}{Phys. Rev. Lett.} \bibinfo{volume}{100}
  (\bibinfo{year}{2008}) \bibinfo{pages}{158701}.
\bibitem[{Kimura and Hayakawa(2008)}]{Kimura2008}
\bibinfo{author}{D.~Kimura}, \bibinfo{author}{Y.~Hayakawa},
\newblock \bibinfo{title}{Coevolutionary networks with homophily and
  heterophily},
\newblock \bibinfo{journal}{Phys. Rev. E} \bibinfo{volume}{78}
  (\bibinfo{year}{2008}) \bibinfo{pages}{016103}.
\bibitem[{Zschaler et~al.(2012)Zschaler, B{\"o}hme, Sei{\ss}inger, Huepe, and
  Gross}]{Zschaler2012}
\bibinfo{author}{G.~Zschaler}, \bibinfo{author}{G.~A. B{\"o}hme},
  \bibinfo{author}{M.~Sei{\ss}inger}, \bibinfo{author}{C.~Huepe},
  \bibinfo{author}{T.~Gross},
\newblock \bibinfo{title}{Early fragmentation in the adaptive voter model on
  directed networks},
\newblock \bibinfo{journal}{Phys. Rev. E} \bibinfo{volume}{85}
  (\bibinfo{year}{2012}) \bibinfo{pages}{046107}.
\bibitem[{Durrett et~al.(2012)Durrett, Gleeson, Lloyd, Mucha, Shi, Sivakoff,
  Socolar, and Varghese}]{Durrett2012}
\bibinfo{author}{R.~Durrett}, \bibinfo{author}{J.~P. Gleeson},
  \bibinfo{author}{A.~L. Lloyd}, \bibinfo{author}{P.~J. Mucha},
  \bibinfo{author}{F.~Shi}, \bibinfo{author}{D.~Sivakoff},
  \bibinfo{author}{J.~Socolar}, \bibinfo{author}{C.~Varghese},
\newblock \bibinfo{title}{Graph fission in an evolving voter model},
\newblock \bibinfo{journal}{Proc. Natl. Acad. Sci. USA} \bibinfo{volume}{109}
  (\bibinfo{year}{2012}) \bibinfo{pages}{3682}.
\bibitem[{B{\"o}hme and Gross(2012)}]{Boehme2012}
\bibinfo{author}{G.~A. B{\"o}hme}, \bibinfo{author}{T.~Gross},
\newblock \bibinfo{title}{Fragmentation transitions in multi-state voter
  models},
\newblock \bibinfo{journal}{Phys. Rev. E} \bibinfo{volume}{85}
  (\bibinfo{year}{2012}) \bibinfo{pages}{066117}.
\bibitem[{Gross et~al.(2006)Gross, D'Lima, and Blasius}]{Gross2006}
\bibinfo{author}{T.~Gross}, \bibinfo{author}{C.~J.~D. D'Lima},
  \bibinfo{author}{B.~Blasius},
\newblock \bibinfo{title}{Epidemic dynamics on an adaptive network},
\newblock \bibinfo{journal}{Phys. Rev. Lett.} \bibinfo{volume}{96}
  (\bibinfo{year}{2006}) \bibinfo{pages}{208701}.
\bibitem[{Shaw and Schwartz(2008)}]{Shaw2008}
\bibinfo{author}{L.~B. Shaw}, \bibinfo{author}{I.~B. Schwartz},
\newblock \bibinfo{title}{Fluctuating epidemics on adaptive networks},
\newblock \bibinfo{journal}{Phys. Rev. E} \bibinfo{volume}{77}
  (\bibinfo{year}{2008}) \bibinfo{pages}{066101}.
\bibitem[{Risau-Gusman and Zanette(2009)}]{Risau-Gusman2009}
\bibinfo{author}{S.~Risau-Gusman}, \bibinfo{author}{D.~H. Zanette},
\newblock \bibinfo{title}{Contact switching as a control strategy for epidemic
  outbreaks},
\newblock \bibinfo{journal}{Journal of Theoretical Biology}
  \bibinfo{volume}{257} (\bibinfo{year}{2009}) \bibinfo{pages}{52--60}.
\bibitem[{Shaw and Schwartz(2010)}]{Shaw2010}
\bibinfo{author}{L.~B. Shaw}, \bibinfo{author}{I.~B. Schwartz},
\newblock \bibinfo{title}{Enhanced vaccine control of epidemics in adaptive
  networks},
\newblock \bibinfo{journal}{Phys. Rev. E} \bibinfo{volume}{81}
  (\bibinfo{year}{2010}) \bibinfo{pages}{046120}.
\bibitem[{Marceau et~al.(2010)Marceau, N{\"o}el, Hebert-Dufresne, Allard, and
  Dube}]{Marceau2010}
\bibinfo{author}{V.~Marceau}, \bibinfo{author}{P.-A. N{\"o}el},
  \bibinfo{author}{L.~Hebert-Dufresne}, \bibinfo{author}{A.~Allard},
  \bibinfo{author}{L.~Dube},
\newblock \bibinfo{title}{Adaptive networks: coevolution of disease and
  topology},
\newblock \bibinfo{journal}{Phys. Rev. E} \bibinfo{volume}{82}
  (\bibinfo{year}{2010}) \bibinfo{pages}{036116}.
\bibitem[{Gr{\"a}ser et~al.(2011)Gr{\"a}ser, Hui, and Xu}]{Graeser2011}
\bibinfo{author}{O.~Gr{\"a}ser}, \bibinfo{author}{P.~M. Hui},
  \bibinfo{author}{C.~Xu},
\newblock \bibinfo{title}{Separatrices between healthy and endemic states in an
  adaptive epidemic model},
\newblock \bibinfo{journal}{Physica A} \bibinfo{volume}{390}
  (\bibinfo{year}{2011}) \bibinfo{pages}{906--913}.
\bibitem[{Lagorio et~al.(2011)Lagorio, Dickison, Vazquez, Braunstein, Macri,
  Migueles, Havlin, and Stanley}]{Lagorio2011}
\bibinfo{author}{C.~Lagorio}, \bibinfo{author}{M.~Dickison},
  \bibinfo{author}{F.~Vazquez}, \bibinfo{author}{L.~A. Braunstein},
  \bibinfo{author}{P.~A. Macri}, \bibinfo{author}{M.~V. Migueles},
  \bibinfo{author}{S.~Havlin}, \bibinfo{author}{H.~E. Stanley},
\newblock \bibinfo{title}{Quarantine-generated phase transition in epidemic
  spreading},
\newblock \bibinfo{journal}{Phys. Rev. E} \bibinfo{volume}{83}
  (\bibinfo{year}{2011}) \bibinfo{pages}{026102}.
\bibitem[{Wang et~al.(2011)Wang, Cao, Suzuki, and Aihara}]{Wang2011}
\bibinfo{author}{B.~Wang}, \bibinfo{author}{L.~Cao},
  \bibinfo{author}{H.~Suzuki}, \bibinfo{author}{K.~Aihara},
\newblock \bibinfo{title}{Epidemic spread in adaptive networks with multitype
  agents},
\newblock \bibinfo{journal}{J. Phys. A} \bibinfo{volume}{44}
  (\bibinfo{year}{2011}) \bibinfo{pages}{035101}.
\bibitem[{Juher et~al.(2012)Juher, Ripoll, and Saldana}]{Juher2012}
\bibinfo{author}{D.~Juher}, \bibinfo{author}{J.~Ripoll},
  \bibinfo{author}{J.~Saldana},
\newblock \bibinfo{title}{Outbreak analysis of an sis epidemic model with
  rewiring},
\newblock \bibinfo{journal}{J. Math. Biol.}  (\bibinfo{year}{2012}).
\bibitem[{Zimmermann et~al.(2000)Zimmermann, Eguiluz, Miguel, and
  Spadaro}]{Zimmermann2000}
\bibinfo{author}{M.~Zimmermann}, \bibinfo{author}{V.~Eguiluz},
  \bibinfo{author}{M.~S. Miguel}, \bibinfo{author}{A.~Spadaro},
\newblock \bibinfo{title}{Cooperation in an adaptive network},
\newblock in: \bibinfo{editor}{G.~Ballot}, \bibinfo{editor}{G.~Weisbuch}
  (Eds.), \bibinfo{booktitle}{Application of Simulations to Social Sciences},
  \bibinfo{publisher}{Hermes Science Publications}, \bibinfo{address}{Oxford,
  UK}, \bibinfo{year}{2000}, pp. \bibinfo{pages}{283--297}.
\bibitem[{Skyrms and Pemantle(2000)}]{Skyrms2000}
\bibinfo{author}{B.~Skyrms}, \bibinfo{author}{R.~Pemantle},
\newblock \bibinfo{title}{A dynamic model of social network formation},
\newblock \bibinfo{journal}{Proc. Natl. Acad. Sci. USA} \bibinfo{volume}{97}
  (\bibinfo{year}{2000}) \bibinfo{pages}{9340--9346}.
\bibitem[{Zimmermann et~al.(2004)Zimmermann, Eguiluz, and
  Miguel}]{Zimmermann2004}
\bibinfo{author}{M.~G. Zimmermann}, \bibinfo{author}{V.~M. Eguiluz},
  \bibinfo{author}{M.~S. Miguel},
\newblock \bibinfo{title}{Coevolution of dynamical states and interactions in
  dynamic networks},
\newblock \bibinfo{journal}{Phys. Rev. E} \bibinfo{volume}{69}
  (\bibinfo{year}{2004}) \bibinfo{pages}{065102(R)}.
\bibitem[{Pacheco et~al.(2006)Pacheco, Traulsen, and Nowak}]{Pacheco2006}
\bibinfo{author}{J.~M. Pacheco}, \bibinfo{author}{A.~Traulsen},
  \bibinfo{author}{M.~A. Nowak},
\newblock \bibinfo{title}{Coevolution of strategy and structure in complex
  networks with dynamical linking},
\newblock \bibinfo{journal}{Phys. Rev. Lett.} \bibinfo{volume}{97}
  (\bibinfo{year}{2006}) \bibinfo{pages}{258103}.
\bibitem[{van Segbroeck et~al.(2009)van Segbroeck, Santos, Lenaerts, and
  Pacheco}]{Segbroeck2009}
\bibinfo{author}{S.~van Segbroeck}, \bibinfo{author}{F.~Santos},
  \bibinfo{author}{T.~Lenaerts}, \bibinfo{author}{J.~Pacheco},
\newblock \bibinfo{title}{Reacting differently to adverse ties promotes
  cooperation in social networks},
\newblock \bibinfo{journal}{Phys. Rev. Lett.} \bibinfo{volume}{102}
  (\bibinfo{year}{2009}) \bibinfo{pages}{058105}.
\bibitem[{Poncela et~al.(2009)Poncela, G{\'o}mez-Garde{\~n}es, Traulsen, and
  Moreno}]{Poncela2009}
\bibinfo{author}{J.~Poncela}, \bibinfo{author}{J.~G{\'o}mez-Garde{\~n}es},
  \bibinfo{author}{A.~Traulsen}, \bibinfo{author}{Y.~Moreno},
\newblock \bibinfo{title}{Evolutionary game dynamics in a growing structured
  population},
\newblock \bibinfo{journal}{New J. Phys.} \bibinfo{volume}{11}
  (\bibinfo{year}{2009}) \bibinfo{pages}{083031}.
\bibitem[{Szolnoki and Perc(2009)}]{Szolnoki2009}
\bibinfo{author}{A.~Szolnoki}, \bibinfo{author}{M.~Perc},
\newblock \bibinfo{title}{Emergence of multilevel selection in the prisoner's
  dilemma game on coevolving random networks},
\newblock \bibinfo{journal}{New J. Phys.} \bibinfo{volume}{11}
  (\bibinfo{year}{2009}) \bibinfo{pages}{093033}.
\bibitem[{Do et~al.(2010)Do, Rudolf, and Gross}]{Do2010}
\bibinfo{author}{A.-L. Do}, \bibinfo{author}{L.~Rudolf},
  \bibinfo{author}{T.~Gross},
\newblock \bibinfo{title}{Patterns of cooperation},
\newblock \bibinfo{journal}{New J. Phys.} \bibinfo{volume}{12}
  (\bibinfo{year}{2010}) \bibinfo{pages}{063023}.
\bibitem[{Zschaler et~al.(2010)Zschaler, Traulsen, and Gross}]{Zschaler2010}
\bibinfo{author}{G.~Zschaler}, \bibinfo{author}{A.~Traulsen},
  \bibinfo{author}{T.~Gross},
\newblock \bibinfo{title}{A homoclinic route to asymptotic full cooperation in
  adaptive networks and its failure},
\newblock \bibinfo{journal}{New J. Phys.} \bibinfo{volume}{12}
  (\bibinfo{year}{2010}) \bibinfo{pages}{093015}.
\bibitem[{van Segbroeck et~al.(2011)van Segbroeck, Santos, Lenaerts, and
  Pacheco}]{Segbroeck2011}
\bibinfo{author}{S.~van Segbroeck}, \bibinfo{author}{F.~Santos},
  \bibinfo{author}{T.~Lenaerts}, \bibinfo{author}{J.~Pacheco},
\newblock \bibinfo{title}{Selection pressure transforms the nature of social
  dilemmas in adaptive networks},
\newblock \bibinfo{journal}{New J. Phys.} \bibinfo{volume}{13}
  (\bibinfo{year}{2011}) \bibinfo{pages}{013007}.
\bibitem[{Lee et~al.(2011)Lee, Holme, and Wu}]{Lee2011}
\bibinfo{author}{S.~Lee}, \bibinfo{author}{P.~Holme}, \bibinfo{author}{Z.-X.
  Wu},
\newblock \bibinfo{title}{Emergent hierarchical structures in multiadaptive
  games},
\newblock \bibinfo{journal}{Phys. Rev. Lett.} \bibinfo{volume}{106}
  (\bibinfo{year}{2011}) \bibinfo{pages}{028702}.
\bibitem[{Fehl et~al.(2011)Fehl, van~der Post, and Semmann}]{Fehl2011}
\bibinfo{author}{K.~Fehl}, \bibinfo{author}{D.~J. van~der Post},
  \bibinfo{author}{D.~Semmann},
\newblock \bibinfo{title}{Co-evolution of behaviour and social network
  structure promotes human cooperation},
\newblock \bibinfo{journal}{Ecology Letters} \bibinfo{volume}{14}
  (\bibinfo{year}{2011}) \bibinfo{pages}{546--551}.
\bibitem[{Santos et~al.(2012)Santos, Pinheiro, Lenaerts, and
  Pacheco}]{Santos2012}
\bibinfo{author}{F.~C. Santos}, \bibinfo{author}{F.~L. Pinheiro},
  \bibinfo{author}{T.~Lenaerts}, \bibinfo{author}{J.~M. Pacheco},
\newblock \bibinfo{title}{The role of diversity in the evolution of
  cooperation},
\newblock \bibinfo{journal}{J. Theor. Biol.} \bibinfo{volume}{299}
  (\bibinfo{year}{2012}) \bibinfo{pages}{88--96}.
\bibitem[{Zhou and Kurths(2006)}]{Zhou2006}
\bibinfo{author}{C.~Zhou}, \bibinfo{author}{J.~Kurths},
\newblock \bibinfo{title}{Dynamical weights and enhanced synchronization in
  adaptive complex networks},
\newblock \bibinfo{journal}{Phys. Rev. Lett.} \bibinfo{volume}{96}
  (\bibinfo{year}{2006}) \bibinfo{pages}{164102}.
\bibitem[{Sorrentino and Ott(2008)}]{Sorrentino2008}
\bibinfo{author}{F.~Sorrentino}, \bibinfo{author}{E.~Ott},
\newblock \bibinfo{title}{Adaptive synchronization of dynamics on evolving
  complex networks},
\newblock \bibinfo{journal}{Phys. Rev. Lett.} \bibinfo{volume}{100}
  (\bibinfo{year}{2008}) \bibinfo{pages}{114101}.
\bibitem[{Aoki and Aoyagi(2009)}]{Aoki2009}
\bibinfo{author}{T.~Aoki}, \bibinfo{author}{T.~Aoyagi},
\newblock \bibinfo{title}{Co-evolution of phases and connection strengths in a
  network of phase oscillators},
\newblock \bibinfo{journal}{Phys. Rev. Lett.} \bibinfo{volume}{102}
  (\bibinfo{year}{2009}) \bibinfo{pages}{034101}.
\bibitem[{Assenza et~al.(2011)Assenza, Guti{\'e}rrez, G{\'o}mez-Garde{\~n}es,
  Latora, and Boccaletti}]{Assenza2011}
\bibinfo{author}{S.~Assenza}, \bibinfo{author}{R.~Guti{\'e}rrez},
  \bibinfo{author}{J.~G{\'o}mez-Garde{\~n}es}, \bibinfo{author}{V.~Latora},
  \bibinfo{author}{S.~Boccaletti},
\newblock \bibinfo{title}{Emergence of structural patterns out of
  synchronization in networks with competitive interactions},
\newblock \bibinfo{journal}{Scientific Reports} \bibinfo{volume}{1}
  (\bibinfo{year}{2011}) \bibinfo{pages}{99}.
\bibitem[{Botella-Soler and Glendinning(2012)}]{Botella2012}
\bibinfo{author}{V.~Botella-Soler}, \bibinfo{author}{P.~Glendinning},
\newblock \bibinfo{title}{Emergence of hierarchical networks and
  polysynchronous behaviour in simple adaptive systems},
\newblock \bibinfo{journal}{Europhys. Lett.} \bibinfo{volume}{97}
  (\bibinfo{year}{2012}) \bibinfo{pages}{1332--1355}.
\bibitem[{Bornholdt and Rohlf(2000)}]{Bornholdt2000}
\bibinfo{author}{S.~Bornholdt}, \bibinfo{author}{T.~Rohlf},
\newblock \bibinfo{title}{Topological evolution of dynamical networks: global
  criticality from local dynamics},
\newblock \bibinfo{journal}{Phys. Rev. Lett.} \bibinfo{volume}{84}
  (\bibinfo{year}{2000}) \bibinfo{pages}{6114--6117}.
\bibitem[{Bornholdt and R{\"o}hl(2003)}]{Bornholdt2003}
\bibinfo{author}{S.~Bornholdt}, \bibinfo{author}{T.~R{\"o}hl},
\newblock \bibinfo{title}{Self-organized critical neural networks},
\newblock \bibinfo{journal}{Phys. Rev. E} \bibinfo{volume}{67}
  (\bibinfo{year}{2003}) \bibinfo{pages}{066118}.
\bibitem[{Levina et~al.(2007{\natexlab{a}})Levina, Herrmann, and
  Geisel}]{Levina2007}
\bibinfo{author}{A.~Levina}, \bibinfo{author}{J.~M. Herrmann},
  \bibinfo{author}{T.~Geisel},
\newblock \bibinfo{title}{Dynamical synapses causing self-organized criticality
  in neural networks},
\newblock \bibinfo{journal}{Nature Physics} \bibinfo{volume}{3}
  (\bibinfo{year}{2007}{\natexlab{a}}) \bibinfo{pages}{857--860}.
\bibitem[{Levina et~al.(2007{\natexlab{b}})Levina, Herrmann, and
  Geisel}]{Levina2009}
\bibinfo{author}{A.~Levina}, \bibinfo{author}{J.~M. Herrmann},
  \bibinfo{author}{T.~Geisel},
\newblock \bibinfo{title}{Phase transitions towards criticality in a neural
  system with adaptive interactions},
\newblock \bibinfo{journal}{Phys. Rev. Lett.} \bibinfo{volume}{102}
  (\bibinfo{year}{2007}{\natexlab{b}}) \bibinfo{pages}{118110}.
\bibitem[{Jost and Kolwankar(2009)}]{Jost2009}
\bibinfo{author}{J.~Jost}, \bibinfo{author}{K.~M. Kolwankar},
\newblock \bibinfo{title}{Evolution of network structure by temporal learning},
\newblock \bibinfo{journal}{Physica A} \bibinfo{volume}{388}
  (\bibinfo{year}{2009}) \bibinfo{pages}{1959--1966}.
\bibitem[{Meisel and Gross(2009)}]{Meisel2009}
\bibinfo{author}{C.~Meisel}, \bibinfo{author}{T.~Gross},
\newblock \bibinfo{title}{Adaptive self-organization in a realistic neural
  network model},
\newblock \bibinfo{journal}{Phys. Rev. E} \bibinfo{volume}{80}
  (\bibinfo{year}{2009}) \bibinfo{pages}{061917}.
\bibitem[{Ren et~al.(2010)Ren, Kolwankar, Samal, and Jost}]{Ren2010}
\bibinfo{author}{Q.~Ren}, \bibinfo{author}{K.~M. Kolwankar},
  \bibinfo{author}{A.~Samal}, \bibinfo{author}{J.~Jost},
\newblock \bibinfo{title}{Stdp-driven networks and the c. elegans neuronal
  network},
\newblock \bibinfo{journal}{Physica A} \bibinfo{volume}{389}
  (\bibinfo{year}{2010}) \bibinfo{pages}{3900--3914}.
\bibitem[{Meisel et~al.(2012)Meisel, Storch, Hallmeyer-Elgner, Bullmore, and
  Gross}]{Meisel2012}
\bibinfo{author}{C.~Meisel}, \bibinfo{author}{A.~Storch},
  \bibinfo{author}{S.~Hallmeyer-Elgner}, \bibinfo{author}{E.~Bullmore},
  \bibinfo{author}{T.~Gross},
\newblock \bibinfo{title}{Failure of adaptive self-organized criticality during
  epileptic seizure attacks},
\newblock \bibinfo{journal}{PLoS Comput. Biol.} \bibinfo{volume}{8(1)}
  (\bibinfo{year}{2012}) \bibinfo{pages}{e1002312}.
\bibitem[{Droste et~al.(2012)Droste, Do, and Gross}]{Droste2012}
\bibinfo{author}{F.~Droste}, \bibinfo{author}{A.-L. Do},
  \bibinfo{author}{T.~Gross},
\newblock \bibinfo{title}{Analytical investigation of self-organized
  criticality in neural networks},
\newblock \bibinfo{journal}{J. R. Soc. Interface}  (\bibinfo{year}{2012}).
\bibitem[{Huepe et~al.(2011)Huepe, Zschaler, Do, and Gross}]{Huepe2011}
\bibinfo{author}{C.~Huepe}, \bibinfo{author}{G.~Zschaler},
  \bibinfo{author}{A.-L. Do}, \bibinfo{author}{T.~Gross},
\newblock \bibinfo{title}{Adaptive-network models of swarm dynamics},
\newblock \bibinfo{journal}{New J. Phys.} \bibinfo{volume}{13}
  (\bibinfo{year}{2011}) \bibinfo{pages}{073022}.
\bibitem[{Couzin et~al.(2011)Couzin, Ioannou, Demirel, Gross, Torney, Hartnett,
  Conradt, Levin, and Leonard}]{Couzin2011}
\bibinfo{author}{I.~D. Couzin}, \bibinfo{author}{C.~C. Ioannou},
  \bibinfo{author}{G.~Demirel}, \bibinfo{author}{T.~Gross},
  \bibinfo{author}{C.~J. Torney}, \bibinfo{author}{A.~Hartnett},
  \bibinfo{author}{L.~Conradt}, \bibinfo{author}{S.~A. Levin},
  \bibinfo{author}{N.~E. Leonard},
\newblock \bibinfo{title}{Uninformed individuals promote democratic consensus
  in animal groups},
\newblock \bibinfo{journal}{Science} \bibinfo{volume}{334}
  (\bibinfo{year}{2011}) \bibinfo{pages}{1578--1580}.
\bibitem[{Peixoto and Bornholdt(2012)}]{Peixoto2012}
\bibinfo{author}{T.~P. Peixoto}, \bibinfo{author}{S.~Bornholdt},
\newblock \bibinfo{title}{No need for conspiracy: self-organized cartel
  formation in a modified trust game},
\newblock \bibinfo{journal}{Phys. Rev. Lett.} \bibinfo{volume}{108}
  (\bibinfo{year}{2012}) \bibinfo{pages}{218702}.
\bibitem[{Kim and Noh(2008)}]{Kim2008}
\bibinfo{author}{S.-W. Kim}, \bibinfo{author}{J.~D. Noh},
\newblock \bibinfo{title}{Instability in a network coevolving with a particle
  system},
\newblock \bibinfo{journal}{Phys. Rev. Lett.} \bibinfo{volume}{100}
  (\bibinfo{year}{2008}) \bibinfo{pages}{118702}.
\bibitem[{Tomita et~al.(2009)Tomita, Kurokawa, and Muarata}]{Tomita2009}
\bibinfo{author}{K.~Tomita}, \bibinfo{author}{H.~Kurokawa},
  \bibinfo{author}{S.~Muarata},
\newblock \bibinfo{title}{Graph-rewiring automata as a natural extension of
  cellular automata},
\newblock in: \bibinfo{editor}{T.~Gross}, \bibinfo{editor}{H.~Sayama} (Eds.),
  \bibinfo{booktitle}{Adaptive Networks. Theory, Models and Applications},
  \bibinfo{publisher}{Springer Verlag}, \bibinfo{address}{New York},
  \bibinfo{year}{2009}, pp. \bibinfo{pages}{291--310}.
\bibitem[{Atay and Jost(2004)}]{Atay2004}
\bibinfo{author}{F.~M. Atay}, \bibinfo{author}{J.~Jost},
\newblock \bibinfo{title}{Delays, connection topology, and synchronization of
  coupled chaotic maps},
\newblock \bibinfo{journal}{Phys. Rev. Lett.} \bibinfo{volume}{92}
  (\bibinfo{year}{2004}) \bibinfo{pages}{144101}.
\bibitem[{Do et~al.(2012)Do, Boccaletti, and Gross}]{Do2012}
\bibinfo{author}{A.-L. Do}, \bibinfo{author}{S.~Boccaletti},
  \bibinfo{author}{T.~Gross},
\newblock \bibinfo{title}{Graphical notation reveals topological stability
  criteria for collective dynamics in complex networks},
\newblock \bibinfo{journal}{Phys. Rev. Lett.} \bibinfo{volume}{108}
  (\bibinfo{year}{2012}) \bibinfo{pages}{194102}.
\bibitem[{Bethe(1935)}]{Bethe1935}
\bibinfo{author}{H.~A. Bethe},
\newblock \bibinfo{title}{Statistical theory of superlattices},
\newblock \bibinfo{journal}{Proc. R. Soc. B} \bibinfo{volume}{150}
  (\bibinfo{year}{1935}) \bibinfo{pages}{552--575}.
\bibitem[{Peierls(1936)}]{Peierls1936}
\bibinfo{author}{R.~Peierls},
\newblock \bibinfo{title}{On ising's model of ferromagnetism},
\newblock \bibinfo{journal}{Proc. Cambridge Philos. Soc.} \bibinfo{volume}{32}
  (\bibinfo{year}{1936}) \bibinfo{pages}{477--481}.
\bibitem[{Matsuda et~al.(1992)Matsuda, Ogita, Sasaki, and Sato}]{Matsuda1992}
\bibinfo{author}{H.~Matsuda}, \bibinfo{author}{N.~Ogita},
  \bibinfo{author}{A.~Sasaki}, \bibinfo{author}{K.~Sato},
\newblock \bibinfo{title}{Statistical mechanics of population: the lattice
  lotka-volterra model},
\newblock \bibinfo{journal}{Prog. Theor. Phys.} \bibinfo{volume}{88}
  (\bibinfo{year}{1992}) \bibinfo{pages}{1035--1049}.
\bibitem[{Bauch(2005)}]{Bauch2005}
\bibinfo{author}{C.~T. Bauch},
\newblock \bibinfo{title}{The spread of infectious diseases in spatially
  structured populations: an invasory pair approximation},
\newblock \bibinfo{journal}{Mathematical Biosciences} \bibinfo{volume}{198}
  (\bibinfo{year}{2005}) \bibinfo{pages}{217--237}.
\bibitem[{Szab{\'o} et~al.(2004)Szab{\'o}, Szolnoki, and Izsak}]{Szabo2004}
\bibinfo{author}{G.~Szab{\'o}}, \bibinfo{author}{A.~Szolnoki},
  \bibinfo{author}{R.~Izsak},
\newblock \bibinfo{title}{Rock-scissors-paper game on regular small-world
  networks},
\newblock \bibinfo{journal}{J. Phys. A} \bibinfo{volume}{37}
  (\bibinfo{year}{2004}) \bibinfo{pages}{2599}.
\bibitem[{Peyrard et~al.(2008)Peyrard, Dieckmann, and Franc}]{Peyrard2008}
\bibinfo{author}{N.~Peyrard}, \bibinfo{author}{U.~Dieckmann},
  \bibinfo{author}{A.~Franc},
\newblock \bibinfo{title}{Long-range correlations improve understanding of the
  influence of network structure on contact dynamics},
\newblock \bibinfo{journal}{Theor. Popul. Biol.} \bibinfo{volume}{73}
  (\bibinfo{year}{2008}) \bibinfo{pages}{383--394}.
\bibitem[{Gross and Kevrekidis(2008)}]{Gross2008a}
\bibinfo{author}{T.~Gross}, \bibinfo{author}{I.~G. Kevrekidis},
\newblock \bibinfo{title}{Robust oscillations in sis epidemics on adaptive
  networks: coarse graining by automated moment closure},
\newblock \bibinfo{journal}{Europhys. Lett.} \bibinfo{volume}{82}
  (\bibinfo{year}{2008}) \bibinfo{pages}{38004--38006}.
\bibitem[{Jolad et~al.(2011)Jolad, Liu, Schmittmann, and Zia}]{Jolad2011}
\bibinfo{author}{S.~Jolad}, \bibinfo{author}{W.~Liu},
  \bibinfo{author}{B.~Schmittmann}, \bibinfo{author}{R.~K.~P. Zia},
  \bibinfo{title}{Epidemic spreading on preferred degree adaptive networks},
  \bibinfo{year}{2011}. \bibinfo{note}{ArXiv:1109.5440}.
\bibitem[{Wieland et~al.(2012{\natexlab{a}})Wieland, Aquino, and
  Nunes}]{Wieland2012}
\bibinfo{author}{S.~Wieland}, \bibinfo{author}{T.~Aquino},
  \bibinfo{author}{A.~Nunes},
\newblock \bibinfo{title}{The structure of coevolving infection networks},
\newblock \bibinfo{journal}{Europhys. Lett.} \bibinfo{volume}{97}
  (\bibinfo{year}{2012}{\natexlab{a}}) \bibinfo{pages}{18003}.
\bibitem[{Wieland et~al.(2012{\natexlab{b}})Wieland, Parisi, and
  Nunes}]{Wieland2012b}
\bibinfo{author}{S.~Wieland}, \bibinfo{author}{A.~Parisi},
  \bibinfo{author}{A.~Nunes},
\newblock \bibinfo{title}{Detecting and describing dynamic equilibria in
  adaptive networks},
\newblock \bibinfo{journal}{Eur. Phys. J. Special Topics} \bibinfo{volume}{212}
  (\bibinfo{year}{2012}{\natexlab{b}}) \bibinfo{pages}{99--113}.
\bibitem[{Demirel and Gross(2012)}]{Demirel2012b}
\bibinfo{author}{G.~Demirel}, \bibinfo{author}{T.~Gross},
  \bibinfo{title}{Absence of epidemic thresholds in a growing adaptive
  network}, \bibinfo{year}{2012}. \bibinfo{note}{ArXiv:1209.2541}.
\bibitem[{Gr{\"a}ser et~al.(2009)Gr{\"a}ser, Xu, and Hui}]{Graeser2009}
\bibinfo{author}{O.~Gr{\"a}ser}, \bibinfo{author}{C.~Xu},
  \bibinfo{author}{P.~M. Hui},
\newblock \bibinfo{title}{Disconnected-connected network transitions and phase
  separation driven by co-evolving dynamics},
\newblock \bibinfo{journal}{Europhys. Lett.} \bibinfo{volume}{87}
  (\bibinfo{year}{2009}) \bibinfo{pages}{38003}.
\bibitem[{Fu et~al.(2009)Fu, Wu, and Wang}]{Fu2009}
\bibinfo{author}{F.~Fu}, \bibinfo{author}{T.~Wu}, \bibinfo{author}{L.~Wang},
\newblock \bibinfo{title}{Partner switching stabilizes cooperation in
  coevolutionary prisoner’s dilemma},
\newblock \bibinfo{journal}{Phys. Rev. E} \bibinfo{volume}{79}
  (\bibinfo{year}{2009}) \bibinfo{pages}{036101}.
\bibitem[{Zanette and Gil(2006)}]{Zanette2006}
\bibinfo{author}{D.~H. Zanette}, \bibinfo{author}{S.~Gil},
\newblock \bibinfo{title}{Opinion spreading and agent segregation on evolving
  networks},
\newblock \bibinfo{journal}{Physica D} \bibinfo{volume}{224}
  (\bibinfo{year}{2006}) \bibinfo{pages}{156--165}.
\bibitem[{Demirel et~al.(2011)Demirel, Prizak, Reddy, and Gross}]{Demirel2011}
\bibinfo{author}{G.~Demirel}, \bibinfo{author}{R.~Prizak},
  \bibinfo{author}{P.~N. Reddy}, \bibinfo{author}{T.~Gross},
\newblock \bibinfo{title}{Cyclic dominance in adaptive networks},
\newblock \bibinfo{journal}{Eur. Phys. J. B} \bibinfo{volume}{84}
  (\bibinfo{year}{2011}) \bibinfo{pages}{541--548}.
\bibitem[{Holley and Liggett(1975)}]{Holley1975}
\bibinfo{author}{R.~Holley}, \bibinfo{author}{T.~Liggett},
\newblock \bibinfo{title}{Ergodic theorems for weakly interacting infinite
  systems and the voter model},
\newblock \bibinfo{journal}{Ann. Probab.} \bibinfo{volume}{3}
  (\bibinfo{year}{1975}) \bibinfo{pages}{643--663}.
\bibitem[{Sood and Redner(2005)}]{Sood2005}
\bibinfo{author}{V.~Sood}, \bibinfo{author}{S.~Redner},
\newblock \bibinfo{title}{Voter model on heterogeneous graphs},
\newblock \bibinfo{journal}{Phys. Rev. Lett.} \bibinfo{volume}{94}
  (\bibinfo{year}{2005}) \bibinfo{pages}{178701}.
\bibitem[{Sood et~al.(2008)Sood, Antal, and Redner}]{Sood2008}
\bibinfo{author}{V.~Sood}, \bibinfo{author}{T.~Antal},
  \bibinfo{author}{S.~Redner},
\newblock \bibinfo{title}{Voter models on heterogeneous networks},
\newblock \bibinfo{journal}{Phys. Rev. E} \bibinfo{volume}{77}
  (\bibinfo{year}{2008}) \bibinfo{pages}{041121}.
\bibitem[{Vazquez and Eguiluz(2008)}]{Vazquez2008a}
\bibinfo{author}{F.~Vazquez}, \bibinfo{author}{V.~M. Eguiluz},
\newblock \bibinfo{title}{Analytical solution of the voter model on
  uncorrelated networks},
\newblock \bibinfo{journal}{New J. Phys.} \bibinfo{volume}{10}
  (\bibinfo{year}{2008}) \bibinfo{pages}{063011}.
\bibitem[{Castellano et~al.(2003)Castellano, Vilone, and
  Vespignani}]{Castellano2003}
\bibinfo{author}{C.~Castellano}, \bibinfo{author}{D.~Vilone},
  \bibinfo{author}{A.~Vespignani},
\newblock \bibinfo{title}{Incomplete ordering of the voter model on small-world
  networks},
\newblock \bibinfo{journal}{Europhys. Lett.} \bibinfo{volume}{63}
  (\bibinfo{year}{2003}) \bibinfo{pages}{153--158}.
\bibitem[{Vilone and Castellano(2004)}]{Vilone2004}
\bibinfo{author}{D.~Vilone}, \bibinfo{author}{C.~Castellano},
\newblock \bibinfo{title}{Solution of voter model dynamics on annealed
  small-world networks},
\newblock \bibinfo{journal}{Phys. Rev. E} \bibinfo{volume}{69}
  (\bibinfo{year}{2004}) \bibinfo{pages}{016109}.
\bibitem[{Suchecki et~al.(2005)Suchecki, Eguiluz, and Miguel}]{Suchecki2005a}
\bibinfo{author}{K.~Suchecki}, \bibinfo{author}{V.~M. Eguiluz},
  \bibinfo{author}{M.~S. Miguel},
\newblock \bibinfo{title}{Conservation laws for the voter model in complex
  networks},
\newblock \bibinfo{journal}{Europhys. Lett.} \bibinfo{volume}{69}
  (\bibinfo{year}{2005}) \bibinfo{pages}{228--234}.
\bibitem[{Castellano et~al.(2005)Castellano, Loreto, Barrat, Cecconi, and
  Parisi}]{Castellano2005}
\bibinfo{author}{C.~Castellano}, \bibinfo{author}{V.~Loreto},
  \bibinfo{author}{A.~Barrat}, \bibinfo{author}{F.~Cecconi},
  \bibinfo{author}{D.~Parisi},
\newblock \bibinfo{title}{Comparison of voter and glauber ordering dynamics on
  networks},
\newblock \bibinfo{journal}{Phys. Rev. E} \bibinfo{volume}{71}
  (\bibinfo{year}{2005}) \bibinfo{pages}{066107}.
\bibitem[{Suchecki et~al.(2005)Suchecki, Eguiluz, and Miguel}]{Suchecki2005b}
\bibinfo{author}{K.~Suchecki}, \bibinfo{author}{V.~M. Eguiluz},
  \bibinfo{author}{M.~S. Miguel},
\newblock \bibinfo{title}{Voter model dynamics in complex networks: role of
  dimensionality, disorder, and degree distribution},
\newblock \bibinfo{journal}{Phys. Rev. E} \bibinfo{volume}{72}
  (\bibinfo{year}{2005}) \bibinfo{pages}{036132}.
\bibitem[{Castell{\'o} et~al.(2007)Castell{\'o}, Toivonen, Eguiluz,
  Saram{\"a}ki, Kaski, and Miguel}]{Castello2007a}
\bibinfo{author}{X.~Castell{\'o}}, \bibinfo{author}{R.~Toivonen},
  \bibinfo{author}{V.~M. Eguiluz}, \bibinfo{author}{J.~Saram{\"a}ki},
  \bibinfo{author}{K.~Kaski}, \bibinfo{author}{M.~S. Miguel},
\newblock \bibinfo{title}{Anomalous lifetime distributions and topological
  traps in ordering dynamics},
\newblock \bibinfo{journal}{Europhys. Lett.} \bibinfo{volume}{79}
  (\bibinfo{year}{2007}) \bibinfo{pages}{66006}.
\bibitem[{B{\"o}hme and Gross(2011)}]{Boehme2011}
\bibinfo{author}{G.~A. B{\"o}hme}, \bibinfo{author}{T.~Gross},
\newblock \bibinfo{title}{Analytical calculation of fragmentation transitions
  in adaptive networks},
\newblock \bibinfo{journal}{Phys. Rev. E} \bibinfo{volume}{83}
  (\bibinfo{year}{2011}) \bibinfo{pages}{035101(R)}.
\bibitem[{Gil and Zanette(2006)}]{Gil2006}
\bibinfo{author}{S.~Gil}, \bibinfo{author}{D.~H. Zanette},
\newblock \bibinfo{title}{Coevolution of agents and networks: Opinion spreading
  and community disconnection},
\newblock \bibinfo{journal}{Phys. Lett. A} \bibinfo{volume}{356}
  (\bibinfo{year}{2006}) \bibinfo{pages}{89--94}.
\bibitem[{Benczik et~al.(2008)Benczik, Benczik, Schmittmann, and
  Zia}]{Benczik2008}
\bibinfo{author}{I.~J. Benczik}, \bibinfo{author}{S.~Z. Benczik},
  \bibinfo{author}{B.~Schmittmann}, \bibinfo{author}{R.~K.~P. Zia},
\newblock \bibinfo{title}{Lack of consensus in social systems},
\newblock \bibinfo{journal}{Europhys. Lett.} \bibinfo{volume}{82}
  (\bibinfo{year}{2008}) \bibinfo{pages}{48006}.
\bibitem[{Fu and Wang(2008)}]{Fu2008}
\bibinfo{author}{F.~Fu}, \bibinfo{author}{L.~Wang},
\newblock \bibinfo{title}{Coevolutionary dynamics of opinions and networks:
  From diversity to uniformity},
\newblock \bibinfo{journal}{Phys. Rev. E} \bibinfo{volume}{78}
  (\bibinfo{year}{2008}) \bibinfo{pages}{016104}.
\bibitem[{Benczik et~al.(2009)Benczik, Benczik, Schmittmann, and
  Zia}]{Benczik2009}
\bibinfo{author}{I.~J. Benczik}, \bibinfo{author}{S.~Z. Benczik},
  \bibinfo{author}{B.~Schmittmann}, \bibinfo{author}{R.~K.~P. Zia},
\newblock \bibinfo{title}{Opinion dynamics on an adaptive random network},
\newblock \bibinfo{journal}{Phys. Rev. E} \bibinfo{volume}{79}
  (\bibinfo{year}{2009}) \bibinfo{pages}{046104}.
\bibitem[{Sobkowicz(2009)}]{Sobkowicz2009}
\bibinfo{author}{P.~Sobkowicz},
\newblock \bibinfo{title}{Studies of opinion stability for small dynamic
  networks with opportunistic agents},
\newblock \bibinfo{journal}{Int. J. Mod. Phys. C} \bibinfo{volume}{20}
  (\bibinfo{year}{2009}) \bibinfo{pages}{1645--1662}.
\bibitem[{Zhong et~al.(2010)Zhong, Ren, Qiu, Xu, Chen, and Liu}]{Zhong2010}
\bibinfo{author}{L.-X. Zhong}, \bibinfo{author}{F.~Ren},
  \bibinfo{author}{T.~Qiu}, \bibinfo{author}{J.-R. Xu}, \bibinfo{author}{B.-H.
  Chen}, \bibinfo{author}{C.-F. Liu},
\newblock \bibinfo{title}{Effects of attachment preferences on coevolution of
  opinions and networks},
\newblock \bibinfo{journal}{Physica A} \bibinfo{volume}{389}
  (\bibinfo{year}{2010}) \bibinfo{pages}{2557--2565}.
\bibitem[{Herrera et~al.(2011)Herrera, Cosenza, Tucci, and
  Gonz{\'a}lez-Avella}]{Herrera2011}
\bibinfo{author}{J.~L. Herrera}, \bibinfo{author}{M.~G. Cosenza},
  \bibinfo{author}{K.~Tucci}, \bibinfo{author}{J.~C. Gonz{\'a}lez-Avella},
\newblock \bibinfo{title}{General coevolution of topology and dynamics in
  networks},
\newblock \bibinfo{journal}{Europhys. Lett.} \bibinfo{volume}{95}
  (\bibinfo{year}{2011}) \bibinfo{pages}{58006}.
\bibitem[{Gleeson et~al.(2012)Gleeson, Melnik, Ward, Porter, and
  Mucha}]{Gleeson2012}
\bibinfo{author}{J.~P. Gleeson}, \bibinfo{author}{S.~Melnik},
  \bibinfo{author}{J.~Ward}, \bibinfo{author}{M.~A. Porter},
  \bibinfo{author}{P.~J. Mucha},
\newblock \bibinfo{title}{Accuracy of mean-field theory for dynamics on
  real-world networks},
\newblock \bibinfo{journal}{Phys. Rev. E} \bibinfo{volume}{85}
  (\bibinfo{year}{2012}) \bibinfo{pages}{026106}.
\bibitem[{Pastor-Satorras and Vespignani(2001)}]{Pastor-Satorras2001a}
\bibinfo{author}{R.~Pastor-Satorras}, \bibinfo{author}{A.~Vespignani},
\newblock \bibinfo{title}{Epidemic dynamics and endemic states in complex
  networks},
\newblock \bibinfo{journal}{Phys. Rev. E} \bibinfo{volume}{63}
  (\bibinfo{year}{2001}) \bibinfo{pages}{066117}.
\bibitem[{Moreno et~al.(2002)Moreno, Pastor-Satorras, and
  Vespignani}]{Moreno2002}
\bibinfo{author}{Y.~Moreno}, \bibinfo{author}{R.~Pastor-Satorras},
  \bibinfo{author}{A.~Vespignani},
\newblock \bibinfo{title}{Epidemic outbreaks in complex heterogeneous
  networks},
\newblock \bibinfo{journal}{Eur. Phys. J. B} \bibinfo{volume}{26}
  (\bibinfo{year}{2002}) \bibinfo{pages}{521--529}.
\bibitem[{Pugliese and Castellano(2009)}]{Pugliese2009}
\bibinfo{author}{E.~Pugliese}, \bibinfo{author}{C.~Castellano},
\newblock \bibinfo{title}{Heterogeneous pair approximation for voter models on
  networks},
\newblock \bibinfo{journal}{Europhys. Lett.} \bibinfo{volume}{88}
  (\bibinfo{year}{2009}) \bibinfo{pages}{58004}.
\bibitem[{N{\"o}el et~al.(2009)N{\"o}el, Davoudi, Brunham, Dube, and
  Pourbohloul}]{Noel2009}
\bibinfo{author}{P.-A. N{\"o}el}, \bibinfo{author}{B.~Davoudi},
  \bibinfo{author}{R.~C. Brunham}, \bibinfo{author}{L.~J. Dube},
  \bibinfo{author}{B.~Pourbohloul},
\newblock \bibinfo{title}{Time evolution of epidemic disease on finite and
  infinite networks},
\newblock \bibinfo{journal}{Phys. Rev. E} \bibinfo{volume}{79}
  (\bibinfo{year}{2009}) \bibinfo{pages}{026101}.
\bibitem[{Gleeson(2011)}]{Gleeson2011}
\bibinfo{author}{J.~P. Gleeson},
\newblock \bibinfo{title}{High-accuracy approximation of binary-state dynamics
  on networks},
\newblock \bibinfo{journal}{Phys. Rev. Lett.} \bibinfo{volume}{107}
  (\bibinfo{year}{2011}) \bibinfo{pages}{068701}.
\bibitem[{Keeling(1999)}]{Keeling1999}
\bibinfo{author}{M.~J. Keeling},
\newblock \bibinfo{title}{The effects of local spatial structure on
  epidemiological invasions},
\newblock \bibinfo{journal}{Proc. R. Soc. B} \bibinfo{volume}{266}
  (\bibinfo{year}{1999}) \bibinfo{pages}{859--867}.
\bibitem[{House and Keeling(2011)}]{House2011}
\bibinfo{author}{T.~House}, \bibinfo{author}{M.~J. Keeling},
\newblock \bibinfo{title}{Insights from unifying modern approximations to
  infections on networks},
\newblock \bibinfo{journal}{J. R. Soc. Interface} \bibinfo{volume}{8}
  (\bibinfo{year}{2011}) \bibinfo{pages}{67--73}.
\bibitem[{Zanette and Risau-Gusman(2008)}]{Zanette2008}
\bibinfo{author}{D.~H. Zanette}, \bibinfo{author}{S.~Risau-Gusman},
\newblock \bibinfo{title}{Infection spreading in a population with evolving
  contacts},
\newblock \bibinfo{journal}{J. Biol. Phys.} \bibinfo{volume}{34}
  (\bibinfo{year}{2008}) \bibinfo{pages}{135--148}.
\bibitem[{Kuehn(2011)}]{Kuehn2011}
\bibinfo{author}{C.~Kuehn},
\newblock \bibinfo{title}{A mathematical framework for critical transitions:
  Bifurcations, fast–slow systems and stochastic dynamics},
\newblock \bibinfo{journal}{Physica D} \bibinfo{volume}{240}
  (\bibinfo{year}{2011}) \bibinfo{pages}{1020--1035}.
\bibitem[{Katsura and Takizawa(1974)}]{Katsura1973}
\bibinfo{author}{S.~Katsura}, \bibinfo{author}{M.~Takizawa},
\newblock \bibinfo{title}{Bethe lattice and bethe approximation},
\newblock \bibinfo{journal}{Prog. Theor. Phys.} \bibinfo{volume}{51}
  (\bibinfo{year}{1974}) \bibinfo{pages}{82--98}.
\bibitem[{Burley(1972)}]{Burley1972}
\bibinfo{author}{D.~M. Burley},
\newblock \bibinfo{title}{Closed form approximations for lattice systems},
\newblock in: \bibinfo{editor}{C.~Domb}, \bibinfo{editor}{M.~Green} (Eds.),
  \bibinfo{booktitle}{Phase Transitions and Critical Phenomena},
  \bibinfo{publisher}{Academic Press}, \bibinfo{address}{London, UK},
  \bibinfo{year}{1972}, pp. \bibinfo{pages}{329--374}.
\bibitem[{Yedidia et~al.(2005)Yedidia, Freeman, and Weiss}]{Yedidia2005}
\bibinfo{author}{J.~Yedidia}, \bibinfo{author}{W.~Freeman},
  \bibinfo{author}{Y.~Weiss},
\newblock \bibinfo{title}{Constructing free energy approximations and
  generalized belief propagation algorithms},
\newblock \bibinfo{journal}{IEEE Trans. Inform. Theory} \bibinfo{volume}{51}
  (\bibinfo{year}{2005}) \bibinfo{pages}{2282--2312}.
\bibitem[{Law et~al.(2003)Law, Murrell, and Dieckmann}]{Law2003}
\bibinfo{author}{R.~Law}, \bibinfo{author}{D.~J. Murrell},
  \bibinfo{author}{U.~Dieckmann},
\newblock \bibinfo{title}{Population growth in space and time: spatial logistic
  equations},
\newblock \bibinfo{journal}{Ecology} \bibinfo{volume}{84}
  (\bibinfo{year}{2003}) \bibinfo{pages}{252--262}.
\bibitem[{Kikuchi(1951)}]{Kikuchi1951}
\bibinfo{author}{R.~Kikuchi},
\newblock \bibinfo{title}{A theory of cooperative phenomena},
\newblock \bibinfo{journal}{Phys. Rev.} \bibinfo{volume}{81}
  (\bibinfo{year}{1951}) \bibinfo{pages}{988--1003}.
\bibitem[{Kevrekidis et~al.(2004)Kevrekidis, Gear, and Hummer}]{Kevrekidis2004}
\bibinfo{author}{I.~G. Kevrekidis}, \bibinfo{author}{C.~W. Gear},
  \bibinfo{author}{G.~Hummer},
\newblock \bibinfo{title}{Equation-free: The computer-aided analysis of complex
  multiscale systems},
\newblock \bibinfo{journal}{AIChE Journal} \bibinfo{volume}{50}
  (\bibinfo{year}{2004}) \bibinfo{pages}{1346--1355}.
\bibitem[{Rogers(2011)}]{Rogers2011}
\bibinfo{author}{T.~Rogers},
\newblock \bibinfo{title}{Maximum-entropy moment-closure for stochastic systems
  on networks},
\newblock \bibinfo{journal}{J. Stat. Mech.}  (\bibinfo{year}{2011})
  \bibinfo{pages}{P05007}.
\bibitem[{Lindquist et~al.(2011)Lindquist, Ma, van~den Driessche, and
  Willeboordse}]{Lindquist2011}
\bibinfo{author}{J.~Lindquist}, \bibinfo{author}{J.~Ma},
  \bibinfo{author}{P.~van~den Driessche}, \bibinfo{author}{F.~H. Willeboordse},
\newblock \bibinfo{title}{Effective degree network disease models},
\newblock \bibinfo{journal}{J. Math. Biol.} \bibinfo{volume}{62}
  (\bibinfo{year}{2011}) \bibinfo{pages}{143--164}.
\bibitem[{Taylor et~al.(2012)Taylor, Taylor, and Kiss}]{Taylor2012}
\bibinfo{author}{M.~Taylor}, \bibinfo{author}{T.~J. Taylor},
  \bibinfo{author}{I.~Z. Kiss},
\newblock \bibinfo{title}{Epidemic threshold and control in a dynamic network},
\newblock \bibinfo{journal}{Phys. Rev. E} \bibinfo{volume}{85}
  (\bibinfo{year}{2012}) \bibinfo{pages}{016103}.
\bibitem[{Vazquez et~al.(2007)Vazquez, Gonz{\'a}lez-Avella, Eguiluz, and
  Miguel}]{Vazquez2007}
\bibinfo{author}{F.~Vazquez}, \bibinfo{author}{J.~C. Gonz{\'a}lez-Avella},
  \bibinfo{author}{V.~M. Eguiluz}, \bibinfo{author}{M.~S. Miguel},
\newblock \bibinfo{title}{Time scale competition leading to fragmentation and
  recombination transition in the coevolution of network and states},
\newblock \bibinfo{journal}{Phys. Rev. E} \bibinfo{volume}{76}
  (\bibinfo{year}{2007}) \bibinfo{pages}{046120}.
\bibitem[{Kozma and Barrat(2008{\natexlab{a}})}]{Kozma2008a}
\bibinfo{author}{B.~Kozma}, \bibinfo{author}{A.~Barrat},
\newblock \bibinfo{title}{Consensus formation on adaptive networks},
\newblock \bibinfo{journal}{Phys. Rev. E} \bibinfo{volume}{77}
  (\bibinfo{year}{2008}{\natexlab{a}}) \bibinfo{pages}{016102}.
\bibitem[{Kozma and Barrat(2008{\natexlab{b}})}]{Kozma2008b}
\bibinfo{author}{B.~Kozma}, \bibinfo{author}{A.~Barrat},
\newblock \bibinfo{title}{Consensus formation on coevolving networks: groups'
  formation and structure},
\newblock \bibinfo{journal}{J. Phys. A: Math. Theor.} \bibinfo{volume}{41}
  (\bibinfo{year}{2008}{\natexlab{b}}) \bibinfo{pages}{224020}.
\bibitem[{I{\~n}iguez et~al.(2009)I{\~n}iguez, Kert{\'e}sz, Kaski, and
  Barrio}]{Iniguez2009}
\bibinfo{author}{G.~I{\~n}iguez}, \bibinfo{author}{J.~Kert{\'e}sz},
  \bibinfo{author}{K.~K. Kaski}, \bibinfo{author}{R.~A. Barrio},
\newblock \bibinfo{title}{Opinion and community formation in coevolving
  networks},
\newblock \bibinfo{journal}{Phys. Rev. E} \bibinfo{volume}{80}
  (\bibinfo{year}{2009}) \bibinfo{pages}{066119}.
\bibitem[{Bryden et~al.(2011)Bryden, Funk, Geard, Bullock, and
  Jansen}]{Bryden2011}
\bibinfo{author}{J.~Bryden}, \bibinfo{author}{S.~Funk},
  \bibinfo{author}{N.~Geard}, \bibinfo{author}{S.~Bullock},
  \bibinfo{author}{V.~A.~A. Jansen},
\newblock \bibinfo{title}{Stability in flux: community structure in dynamic
  networks},
\newblock \bibinfo{journal}{J. R. Soc. Interface} \bibinfo{volume}{8}
  (\bibinfo{year}{2011}) \bibinfo{pages}{1031--1040}.
\bibitem[{Holme and Ghoshal(2006)}]{Holme2006a}
\bibinfo{author}{P.~Holme}, \bibinfo{author}{G.~Ghoshal},
\newblock \bibinfo{title}{Dynamics of networking agents competing for high
  centrality and low degree},
\newblock \bibinfo{journal}{Phys. Rev. Lett.} \bibinfo{volume}{96}
  (\bibinfo{year}{2006}) \bibinfo{pages}{098701}.
\bibitem[{Zimmermann and Eguiluz(2005)}]{Zimmermann2005}
\bibinfo{author}{M.~G. Zimmermann}, \bibinfo{author}{V.~M. Eguiluz},
\newblock \bibinfo{title}{Cooperation, social networks and the emergence of
  leadership in a prisoners dilemma with adaptive local interactions},
\newblock \bibinfo{journal}{Phys. Rev. E} \bibinfo{volume}{72}
  (\bibinfo{year}{2005}) \bibinfo{pages}{056118}.
\bibitem[{Eames and Keeling(2002)}]{Eames2002}
\bibinfo{author}{K.~T.~D. Eames}, \bibinfo{author}{M.~J. Keeling},
\newblock \bibinfo{title}{Modeling dynamic and network heterogeneities in the
  spread of sexually transmitted diseases},
\newblock \bibinfo{journal}{Proc. Natl. Acad. Sci. USA} \bibinfo{volume}{99}
  (\bibinfo{year}{2002}) \bibinfo{pages}{13330--13335}.
\bibitem[{Bansal et~al.(2007)Bansal, Grenfell, and Meyers}]{Bansal2007}
\bibinfo{author}{S.~Bansal}, \bibinfo{author}{B.~T. Grenfell},
  \bibinfo{author}{L.~A. Meyers},
\newblock \bibinfo{title}{When individual behaviour matters: homogeneous and
  network models in epidemiology},
\newblock \bibinfo{journal}{J. R. Soc. Interface} \bibinfo{volume}{4}
  (\bibinfo{year}{2007}) \bibinfo{pages}{879--891}.
\bibitem[{Volz and Meyers(2007)}]{Volz2007}
\bibinfo{author}{E.~Volz}, \bibinfo{author}{L.~A. Meyers},
\newblock \bibinfo{title}{Susceptible–infected–recovered epidemics in
  dynamic contact networks},
\newblock \bibinfo{journal}{Proc. R. Soc. B} \bibinfo{volume}{274}
  (\bibinfo{year}{2007}) \bibinfo{pages}{2925--2934}.
\bibitem[{Volz and Meyers(2009)}]{Volz2009}
\bibinfo{author}{E.~Volz}, \bibinfo{author}{L.~A. Meyers},
\newblock \bibinfo{title}{Epidemic thresholds in dynamic contact networks},
\newblock \bibinfo{journal}{J. R. Soc. Interface} \bibinfo{volume}{6}
  (\bibinfo{year}{2009}) \bibinfo{pages}{233--241}.

\end{thebibliography}

\end{document}